\documentclass[a4paper,fleqn,twocolumn,margin=1in]{aastex63}
\pdfoutput=1
\usepackage{float}
\usepackage{graphicx}
\usepackage{ae,aecompl}
\usepackage{enumitem}
\usepackage{amssymb}
\usepackage{amsmath}
\usepackage{hyperref}
\usepackage[lofdepth,lotdepth,caption=false]{subfig}

\def\muhz{\mu\mathrm{Hz}}
\def\numax{\nu_{\mathrm{max}}}
\def\numaxdetect{\nu_{\mathrm{max, detect}}}
\def\dnu{\Delta \nu}

\def\teff{T_{\mathrm{eff}}}

\def\teffsun{T_{\mathrm{eff,} \odot}}

\def\rsun{R_{\odot}}

\def\msun{M_{\odot}}
\def\numaxsun{\nu_{\mathrm{max,} \odot}}
\def\dnusun{\Delta\nu_{\odot}}
\def\numaxsunpip{\nu_{\mathrm{max,} \odot, {\rm PIP}}}
\def\dnusunpip{\Delta\nu_{\odot, {\rm PIP}}}

\def\Ks{K_{\mathrm{s}}}

\def\numaxmean{\langle \numax' \rangle}
\def\dnumean{\langle \dnu' \rangle}
\def\kapparmean{\langle \kappa'_R \rangle}
\def\kappammean{\langle \kappa'_M \rangle}
\graphicspath {{plots/}}

\floatstyle{boxed}
\newfloat{program}{thp}{lop}
\floatname{program}{Figure}

\newenvironment{mycode}{\ttfamily}{\par}
\DeclareTextFontCommand{\textmycode}{\mycode}

\begin{document}

\title{THE {\it K2} GALACTIC ARCHAEOLOGY PROGRAM DATA RELEASE 2:
  ASTEROSEISMIC RESULTS FROM CAMPAIGNS 4, 6, \& 7}
\author{Joel C. Zinn}
\affiliation{Department of Astrophysics, American Museum of Natural
  History, Central Park West at 79th Street, New York, NY 10024, USA}
\affiliation{School of Physics, University of New South Wales, Barker
  Street, Sydney, NSW 2052, Australia}
\affiliation{Department of Astronomy, The Ohio State University, 140
  West 18th Avenue, Columbus, OH 43210, USA}
\author{Dennis Stello}
\affiliation{School of Physics, University of New South Wales, Barker
  Street, Sydney, NSW 2052, Australia}
\affiliation{Sydney Institute for Astronomy (SIfA), School of Physics,
  University of Sydney, NSW 2006, Australia}
\affiliation{Stellar Astrophysics Centre, Department of Physics and
Astronomy, Aarhus University, Ny Munkegade 120, DK-8000
Aarhus C, Denmark}
\affiliation{Center of Excellence for Astrophysics in Three Dimensions
(ASTRO-3D), Australia}
\author{Yvonne Elsworth}
\affiliation{School of Physics and Astronomy, University of Birmingham, Edgbaston, Birmingham, B15 2TT, UK}
\affiliation{Stellar Astrophysics Centre, Department of Physics and Astronomy, Aarhus University, Ny Munkegade 120, DK-8000 Aarhus C, Denmark}
\author{Rafael A. Garc{\'i}a}
\affiliation{AIM, CEA, CNRS, Universit{\'e} Paris-Saclay, Universit{\'e} Paris Diderot, Sorbonne Paris Cit{\'e}, F-91191 Gif-sur-Yvette, France}
\author{Thomas Kallinger}
\affiliation{Institute of Astrophysics, University of Vienna, T{\"u}rkenschanzstrasse 17, Vienna 1180, Austria}
\author{Savita Mathur}
\affiliation{Space Science Institute, 4750 Walnut Street Suite \#205, Boulder, CO 80301, USA}
\affiliation{Instituto de Astrof\'{\i}sica de Canarias, La Laguna, Tenerife, Spain}
\affiliation{Dpto. de Astrof\'{\i}sica, Universidad de La Laguna, La Laguna, Tenerife, Spain}
\author{Beno{\^i}t Mosser}
\affiliation{LESIA, Observatoire de Paris, PSL Research University, CNRS, Sorbonne Universit{\'e}, Universit{\'e} de Paris Diderot, 92195 Meudon, France}
\author{Lisa Bugnet}
\affiliation{IRFU, CEA, Universit{\'e} Paris-Saclay, 91191 Gif-sur-Yvette, France}
\affiliation{AIM, CEA, CNRS, Universit{\'e} Paris-Saclay, Universit{\'e} Paris Diderot, Sorbonne Paris Cit{\'e}, F-91191 Gif-sur-Yvette, France}
\author{Caitlin Jones}
\affiliation{School of Physics and Astronomy, University of Birmingham, Edgbaston, Birmingham, B15 2TT, UK}
\author{Marc Hon}
\affiliation{School of Physics, University of New South Whales, Barker Street, Sydney, NSW 2052, Australia}
\author{Sanjib Sharma}
\affiliation{Sydney Institute for Astronomy (SIfA), School of Physics,
  University of Sydney, NSW 2006, Australia}
\affiliation{Center of Excellence for Astrophysics in Three Dimensions
(ASTRO-3D), Australia}
\author{Ralph Sch{\"o}nrich}
\affiliation{Mullard Space Science Laboratory, University College London, Holmbury St Mary, Dorking RH5 6NT, UK}
\author{Jack T. Warfield}
\affiliation{Department of Astronomy, The Ohio State University, 140 West
  18th Avenue, Columbus OH 43210, USA}
\affiliation{Department of Physics,
The Ohio State University, 191 West
Woodruff Avenue, Columbus OH 43210}
\author{Rodrigo Luger}
\affiliation{Center for Computational Astrophysics, Flatiron Institute, New York, NY, USA}
\affiliation{Virtual Planetary Laboratory, University of Washington, Seattle, WA, USA}
\author{Marc H. Pinsonneault}
\affiliation{Department of Astronomy, The Ohio State University, 140 West
  18th Avenue, Columbus OH 43210, USA}
\author{Jennifer A. Johnson}
\affiliation{Department of Astronomy, The Ohio State University, 140 West
  18th Avenue, Columbus OH 43210, USA}
\author{Daniel Huber}
\affiliation{ Institute for Astronomy, University of Hawai`i, 2680 Woodlawn Drive, Honolulu, HI 96822, USA}
\author{Victor Silva Aguirre}
\affiliation{Stellar Astrophysics Centre, Department of Physics and Astronomy, Aarhus University, Ny Munkegade 120, DK-8000 Aarhus C, Denmark}
\author{William J. Chaplin}
\affiliation{School of Physics and Astronomy, University of Birmingham, Edgbaston, Birmingham, B15 2TT, UK}
\affiliation{Stellar Astrophysics Centre, Department of Physics and Astronomy, Aarhus University, Ny Munkegade 120, DK-8000 Aarhus C, Denmark}
\author{Guy R. Davies}
\affiliation{School of Physics and Astronomy, University of Birmingham, Edgbaston, Birmingham, B15 2TT, UK}
\affiliation{Stellar Astrophysics Centre, Department of Physics and Astronomy, Aarhus University, Ny Munkegade 120, DK-8000 Aarhus C, Denmark}
\author{Andrea Miglio}
\affiliation{Stellar Astrophysics Centre, Department of Physics and Astronomy, Aarhus University, Ny Munkegade 120, DK-8000 Aarhus C, Denmark}
\affiliation{School of Physics and Astronomy, University of Birmingham, Edgbaston, Birmingham, B15 2TT, UK}

\correspondingauthor{Joel C. Zinn}
\email{jzinn@amnh.org}

\begin{abstract}
Studies of Galactic structure and evolution have benefitted enormously from \textit{Gaia} kinematic information, though additional, intrinsic stellar parameters like age are required to best constrain Galactic models. Asteroseismology is the most precise method of providing such information for field star populations \textit{en masse}, but existing samples
for the most part have been limited to a few narrow fields of view by the \textit{CoRoT} and \textit{Kepler} missions. In an effort to provide well-characterized stellar parameters across a wide range in
Galactic position, we present the second data release of red giant asteroseismic parameters for the \textit{K2} Galactic Archaeology Program (GAP). We provide $\numax$ and $\dnu$ based on six independent pipeline analyses; first-ascent red giant branch (RGB) and red clump (RC) evolutionary state classifications from machine learning; and ready-to-use radius \& mass coefficients, $\kappa_R$ \& $\kappa_M$, which, when appropriately
multiplied by a solar-scaled effective temperature factor, yield
physical stellar radii and masses. In total, we report 4395 radius and mass coefficients, with typical uncertainties of $3.3\% \mathrm{\ (stat.)} \pm 1\% \mathrm{\ (syst.)}$ for $\kappa_R$ and $7.7\% \mathrm{\ (stat.)} \pm 2\% \mathrm{\ (syst.)}$ for $\kappa_M$ among RGB stars, and $5.0\% \mathrm{\ (stat.)} \pm 1\% \mathrm{\ (syst.)}$ for $\kappa_R$ and $10.5\% \mathrm{\ (stat.)} \pm 2\% \mathrm{\ (syst.)}$ for $\kappa_M$ among RC stars. We verify that
the sample is nearly complete --- except for a dearth of stars with $\numax
\lesssim 10-20\muhz$ --- by comparing to Galactic models and
visual inspection. Our asteroseismic radii agree with radii derived from \textit{Gaia} Data
Release 2 parallaxes to within $2.2 \pm 0.3\%$ for RGB stars and $2.0 \pm 0.6\%$ for RC stars.
\end{abstract}

\section{Introduction}
\label{sec:introduction}
The Galactic Archaeology Program (GAP; \citealt{stello+2015}) has taken
advantage of the multidirectional view of the Galaxy offered by the re-purposed {\it Kepler}
mission, {\it K2}. With hundreds of thousands of
stars observed, {\it K2}'s potential for
studying the Galaxy is significant. Instead of a single snapshot of the Galaxy with {\it Kepler} \citep{borucki+2008}, {\it K2} \citep{howell+2014}
observed along the ecliptic, including the local disk, the bulge,
and even distant regions of the halo. Importantly for this
work, the \textit{K2} mission has delivered the quality of data necessary for asteroseismic analysis.

The {\it K2} GAP aims to provide fundamental stellar parameters for red giants across the Galaxy. In
combination with temperature and metallicity information,
asteroseismology can provide stellar radii, masses, and, when combined with stellar models, ages. \textit{Kepler} red giant asteroseismology has yielded important findings for Galactic archaeology, including verifying the presence of a vertical age gradient in the Galactic disc \citep{miglio+2013,casagrande+2016}; testing Galactic chemical evolution models \citep[e.g.,][]{spitoni+2020}; and confirming an age difference between chemically- and kinematically-defined thin and thick discs \citep{silva-aguirre+2018a}. Nevertheless, the \textit{Kepler} asteroseismic sample was not curated for Galactic studies, and so GAP's deliberate and well-understood target selection for Galactic archaeology purposes sets up \textit{K2} to be a more useful tool for Galactic archaeology, particularly in light of its expanded view of the Galaxy. Indeed, the \textit{K2} data is providing interesting insights into the relative ages of chemically-defined stellar populations beyond the solar vicinity (\citealt{rendle+2019}, Warfield et al., in preparation). 

\textit{K2}'s potential is tempered, however, by a decreased photometric precision compared to \textit{Kepler} and a $\sim 80$d dwell time per campaign instead of up to $\sim 4$yr for \textit{Kepler}. These two limitations mean that \textit{K2} is mostly suited for giants with logg above $\sim 1.4$; probes one to two magnitudes `shallower' than \textit{Kepler}; and yields less precise asteroseismic measurements compared to \textit{Kepler} \citep{stello+2017}. And, although the accuracy of stellar parameters derived through asteroseismology is at the percent level \citep[e.g.,][]{silva_aguirre+2012,huber+2012,zinn+2019a}, at this level, there are measurement systematics that need to be corrected for \citep{pinsonneault+2018}. We therefore devote special attention in what follows to understanding the statistical and systematic uncertainties in our asteroseismic quantities.

We have previously released a collection of $\numax$ and $\dnu$ values for
1210 {\it K2} GAP red giants in \cite{stello+2017}. The
present release covers campaigns 4, 6, \& 7, and comprises 4395 stars.
In addition to the global
asteroseismic parameters, $\numax$ and $\dnu$, we
also provide scaling-relation quantities that, when combined with an
effective temperature, yield radii and masses. We also provide
estimates of systematic and statistical errors on the
asteroseismic quantities, and establish the
completeness of observed targets in order to ensure a well-defined
selection function.

\section{Data}
\label{sec:data}

\subsection{Target selection}
\label{sec:selection}
In the context of the GAP, analysis of the
campaigns presented here were prioritized due to their coverage of the sky: the
Galactic center (C7) the Galactic anti-center (C4) and out of the
Galactic plane (C6). These results will ultimately be joined with
a forthcoming analysis of the rest of the {\it K2} campaigns for which
GAP targets have been observed. The GAP targets red giants because they are bright (probing far into the Galaxy) and because their oscillations are detectable from the \textit{K2} long-cadence data, which has a
Nyquist frequency of $\sim 280 \muhz$. All GAP targets for campaigns 4, 6 \& 7 were selected from 2MASS
\citep{skrutskie+2006} to have $J-K > 0.5$ and good photometric
quality based on 2MASS flags.\footnote{The 2MASS \texttt{qflg}
  photometric quality flag was required to be $A$ or $B$ for $J$, $H$,
  and $\Ks$, which ensures, among other things, that the
  signal-to-noise ratio is greater than 7. Additional flags ensured
  that the photometry did not suffer from confusion from nearby
  objects ($\texttt{cflg} == 0$); was a single, unblended source
  ($\texttt{bflg} == 1$); was not extended ($\texttt{xflg} == 0$); was
  not a known solar system object ($\texttt{aflg} == 0$); and had no
  neighbors within 6'' ($\texttt{prox} > 6.0$). See
  http:\/\/vizier.u-strasbg.fr\/cgi-bin\/VizieR?-source=B\/2mass for
  more details.} The proposed targets passing these selection criteria
were prioritized based on a rank ordering in
$V$-band magnitude from bright to faint.\footnote{At the time of targeting, it was typically not well-known which stars were giants or dwarfs, but generally the giant fraction was expected to be close to 100\% at the bright end and down to as low as 20\% ,depending on the campaign, at the faint end of the selection.} C4 and C6 targets were chosen to have $9 < V <
15$, and C7 targets were chosen to have $9 < V <
14.5$, with some exceptions to the
prioritization on a campaign-to-campaign basis, as follows: One giant
with existing RAVE data was prioritized in C4.  In C6, priority was given to 129 giants with existing APOGEE \citep{majewski+2010} spectra, 607
with existing RAVE spectra, and 5 low-metallicity giants chosen from
the literature to be giants with [Fe/H]$ < -3$. The highest priority to GAP
targets in C7 was given to 222 known giants with existing
spectroscopic data from APOGEE, and 23 targets
in NGC 6717 (but see below).

\begin{figure*}[htp]
\centering
\includegraphics[width=0.9\textwidth]{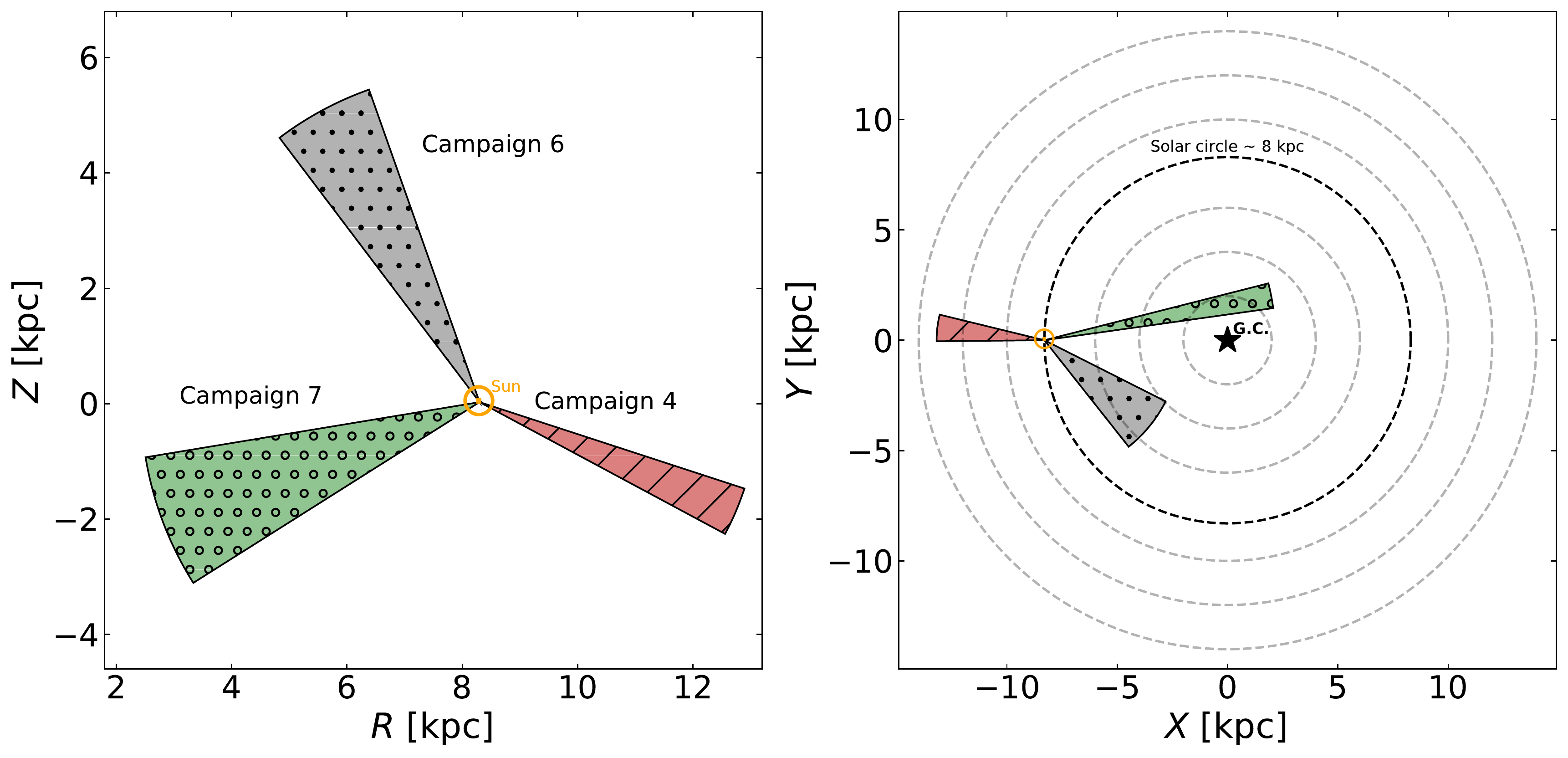}
\caption{Left: approximate Z-R distribution of stars in \textit{K2} GAP DR2, for each campaign. Right: approximate X-Y distribution of stars in \textit{K2} GAP DR2, for each campaign.}
\label{fig:loc}
\end{figure*}

\begin{figure*}[htp]
\centering
\includegraphics[width=0.9\textwidth]{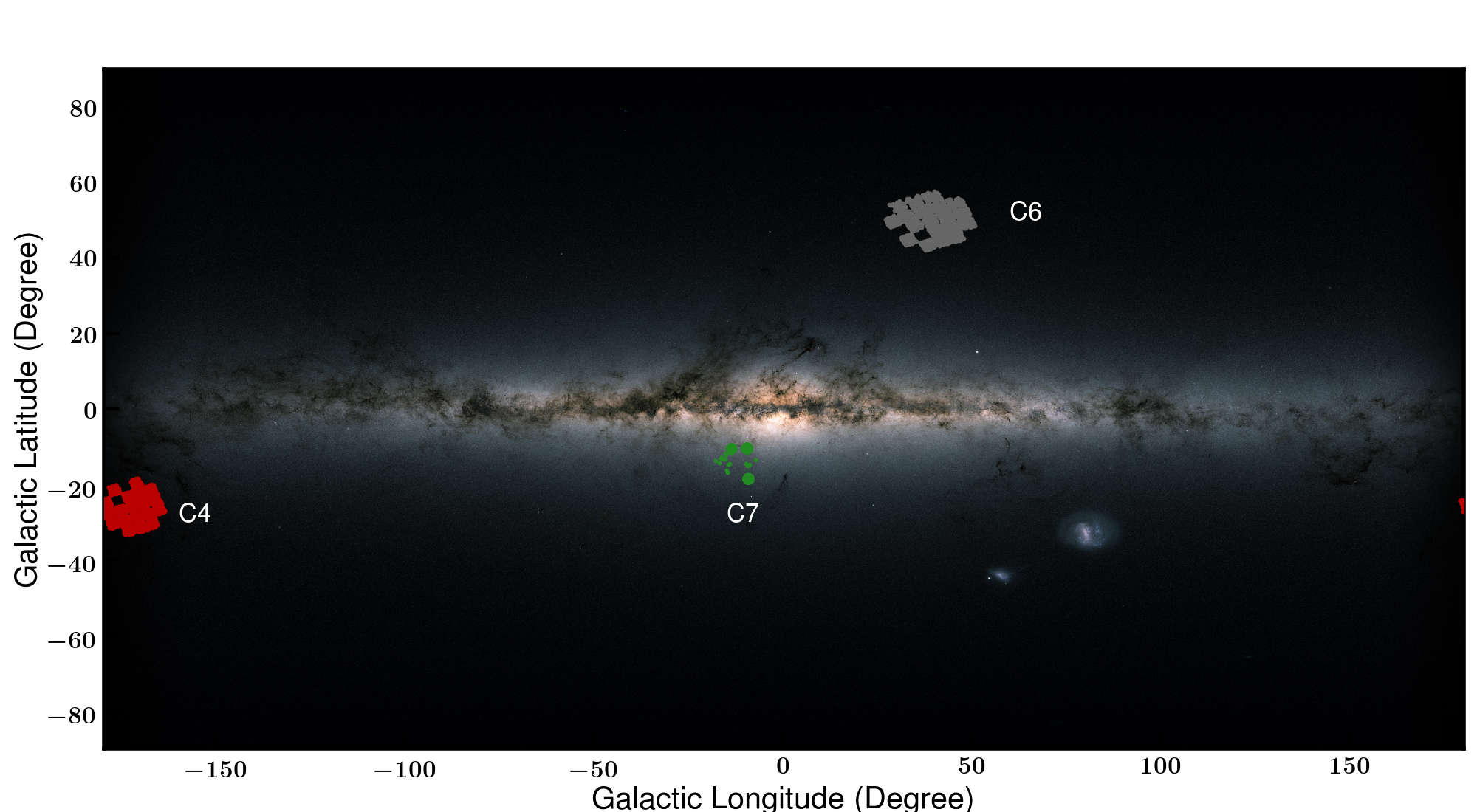}
\caption{Positions of observed \textit{K2} GAP stars in C4, C6, and C7, relative to the Galactic plane. Background image modified from ESA/{\it Gaia}/DPAC.}
\label{fig:radec}
\end{figure*}

Because observed targets were selected in a linear way from the target priority list, the selection functions for each of the campaigns are well-defined. Nearly all of the C6 targets were observed, and so the observed targets conform to the selection function $9 < V <
15$. Our C4 targets were observed down to $V = 13.447$, and therefore follow the selection function down to that magnitude limit (this magnitude limit approximately corresponds to the top 5000 GAP targets). In addition to these GAP-selected targets, there are fainter stars on the GAP target list that were serendipitously observed by {\it K2} through other non-GAP target proposals. Those stars do not follow the GAP selection function but are still analyzed in this paper. However, we caution their use for population studies. The observed C7 targets were accidentally chosen from an inverted priority list during the mission-wide target list consolidation before upload to the spacecraft. As a result, only two of the higher priority APOGEE C7
targets were observed and none from NGC 6717. As discussed in
\cite{sharma+2019}, the resulting effective selection function is one
of either $9 < V < 14.5$ or $14.276 < V < 14.5$, depending on the
position on the sky. Approximately, the 3500 lowest-priority GAP
targets follow the full selection function. There is,
however, an additional population of $\sim 600$ stars in C7 that can
be used for scientific purposes, with the understanding that its
selection function of $14.276 < V < 14.5$ is different than that of the rest of the campaign ($9 < V < 14.5$). For additional details, see \cite{sharma+2019}. We account for these selection functions when comparing to models in our analysis by making sure to compare to the subset of observed stars that follow reproducible GAP target criteria. An accounting of observed and targeted stars is given in Table~\ref{tab:stats}. The approximate lines of sight for these campaigns are shown in Figure~\ref{fig:loc}. An on-sky map of the campaigns is shown in Figure~\ref{fig:radec}.

\begin{deluxetable}{cccc}
  \tablecaption{Targeted and observed stars as a function of campaign \label{tab:stats}}
  \tabletypesize{\footnotesize}
\tablehead{Campaign & {\it K2} GAP targeted & {\it K2} GAP observed &
  GAP selected observed}
\startdata
C4 & 17410 & 6357 & 5000 \\ \hline 
C6 & 8371 & 8313 & 8300 \\ \hline 
C7 & 18698 & 4362 & 3500 \\ \hline
\enddata
\tablecomments{`\textit{K2} GAP targeted' refers to the number of
  proposed targets and `\textit{K2} GAP observed' are those that were
  observed, including targets not following the GAP selection function
  (see text). `GAP selected observed' refers to the approximate number of GAP observed stars that follow the GAP target selection function from the bright end of $V=9$ to a limiting magnitude of 13.447, 15, and 14.5 for C4, C6, and C7, respectively.}
\end{deluxetable}

\subsection{{\it K2} light curve pre-processing}
We used the light curves generated by the EVEREST pipeline
\citep{luger+2018}, which uses campaign-specific basis vectors to
remove noise correlated across pixels. Though the {\it K2} GAP DR1 \citep{stello+2017} used
\cite{vanderburg&johnson2014} (K2SFF) light curves, we found that the
EVEREST pipeline removes at least as much of the sawtooth-like systematic flux variation induced by the {\it K2} six-hour
thruster firings as K2SFF does, while furthermore generating lower
levels of white noise in the giant spectra than K2SFF. Because the {\it K2}
GAP DR1 used K2SFF light curves, we attempted to reconcile
systematic differences in the amplitudes of flux variations in the
EVEREST and K2SFF light curves for consistency between this release
and DR1. To do so, for each star, we multiply the EVEREST light curve flux by a scalar factor such that its power is equal to the K2SFF power. We
then applied a boxcar high-pass filter with a width of 4 days to the light curve to
remove most of the non-oscillation variability
\citep{stello+2015}, and outliers $4\sigma$ in the time series were removed. Seven targets were labeled as extended in the EPIC,
and were not included in the present analysis.\footnote{The
  following EPIC IDs are affected in C4: 210344244, 210489346, 210766860,
  210497173, 210608879; and in C6: 212680904, 212708542.}

\section{Methods}
\label{sec:methods}
\subsection{Extraction of asteroseismic parameters}
\label{sec:extraction}
In this data release, we focus on two asteroseismic quantities, $\numax$ and $\dnu$, which we describe in
turn.

The frequency at maximum acoustic power, $\numax$, scales with the acoustic
cutoff frequency at the stellar atmosphere
\citep{brown+1991,kjeldsen&bedding1995,chaplin+2008,belkacem+2011}, so that
\begin{equation}
\frac{\numax}{\numaxsun} \approx \frac{M/\msun}{(R/\rsun)^2\sqrt{(\teff/\teffsun)}}.
\label{eq:scaling1}
\end{equation}

The frequency separation between modes of the same degree but
consecutive radial order, $\dnu$, scales with the stellar density
\citep{ulrich1986,kjeldsen&bedding1995} according to
\begin{equation}
\frac{\dnu}{\dnusun} \approx \sqrt{\frac{M/\msun}{(R/\rsun)^3}}.
\label{eq:scaling2}
\end{equation}

The $\numaxsun$ and $\dnusun$ solar reference values are not set in stone, and
are themselves measured quantities from asteroseismic data of the Sun.  All the pipelines have an internal set of recommended
solar reference values, $\numaxsunpip$ and $\dnusunpip$. In addition to those pipeline-specific solar reference
values, we adopt a solar reference temperature of $\teffsun = 5772
K$ \citep{mamajek+2015a}.

Like \cite{stello+2017}, we analyze the data using multiple asteroseismic pipelines, whose descriptions
are found therein. We briefly revisit each pipeline below.

A2Z+ (hereafter A2Z) is based on the A2Z pipeline described in \cite{mathur+2010}. $\Delta \nu$ is computed using two methods: the autocorrelation function of the time series and the Power Spectrum of the Power Spectrum. Then, the results from both methods are compared, and a $\dnu$ value is only kept when both methods agree to within 10\%.  The Power Density Spectra (PDS) of those stars are checked to select the ones where modes are present to high confidence. The FliPer metric \citep{bugnet+2018} is used to check stars where the amplitude of the convective background does not agree with the seismic detection. Finally, the $\Delta \nu$ value is refined by cross-correlating a template of the radial modes around the region of the modes where $\Delta \nu$ is varied. The value reported corresponds to the one obtained for the highest correlation coefficient. After fitting the convective background with two Harvey laws \citep{mathur+2011,kallinger+2014} and subtracting it from the PDS, a Gaussian function is fit to the modes to estimate the frequency of maximum power, $\nu_{\rm max}$.

BHM is based on the OCT pipeline \citep{hekker+2010}, and performs hypothesis testing for solar-like oscillations above the granulation background. If there is a frequency window that has significant signal, a $\numax$ and $\dnu$ is computed according to \cite{hekker+2010}, with increased, {\it K2}-specific quality control to ensure $\dnu$ is not an alias of the true value.

CAN returns $\numax$ and $\dnu$ for stars whose autocorrelation functions have characteristic time-scales and rms variability that accord with a relation expected from solar-like oscillators \citep{kallinger+2016}. A Bayesian evidence is computed to determine whether there are solar-like oscillations near the detected autocorrelation time-scale (and therefore near $\numax$). A threshold for the evidence is determined based on visual inspection of a test set of power spectra. $\dnu$ is then calculated by fitting individual radial modes.

COR first fits for $\dnu$ using the autocorrelation of the light curve. COR \citep{mosser&appourchaux2009} returns results for stars that have $\dnu$, FWHM, amplitude of power, and the granulation at $\numax$ that follow relations from \cite{mosser+2010}, with stricter, {\it K2}-specific requirements for stars with a possible $\numax$ detected near the {\it K2} thruster firing frequency, or alias thereof.

SYD computes $\numax$ and $\dnu$ according to \cite{huber+2009}. Results are returned only for stars that are classified as having solar-like oscillations by a machine-learning algorithm described in \cite{hon+2018b}. $\dnu$ results are provided only for stars that are classified as having reliable $\dnu$ values by a machine-learning algorithm. The $\dnu$-vetting algorithm is a convolutional neural network trained on {\it K2} C1 data that were classified by eye as having detectable $\dnu$ values. The algorithm takes as an input the autocorrelation of a granulation background--corrected spectrum in a window with a width of $\pm 0.59\times \numax ^{0.9}$ centered around $\numax$. This latter window size is taken from the observed width of detected stellar oscillations in giants \citep{mosser+2010}. The algorithm takes the autocorrelation, passes it through a convolutional neural network, and outputs a score varying from 0 (not a valid $\dnu$ detection) to 1 (100\% certain valid $\dnu$ detection), but which is not a linear mapping to percent confidence in the detection. For this reason, the score was calibrated using visually-verified $\dnu$ detections, which showed that a score of 0.8 corresponds to a completeness of about 70\% (i.e., this score rejects 30\% stars that have valid $\dnu$ detections) and near total purity (i.e., all stars have visually-verified $\dnu$ detections); $\dnu$ detections were considered valid for stars with a score above this 0.8 value.

BAM calculates $\numax$ and $\dnu$ according to \cite{zinn+2019bam}. $\dnu$ values for this data release are provided based on the autocorrelation method described therein, which, in turn, is based on that of SYD. Stars are classified as oscillators or not based on the Bayesian evidence for the presence of solar-like oscillations (see \citealt{zinn+2019bam} for details), and results in this data release are returned for stars with $3.5\muhz < \numax < 250 \muhz$. $\dnu$ values are only provided for those that satisfy the SYD $\dnu$-vetting algorithm described above.

\begin{figure*}[htp]
\centering
  \includegraphics[width=0.7\textwidth]{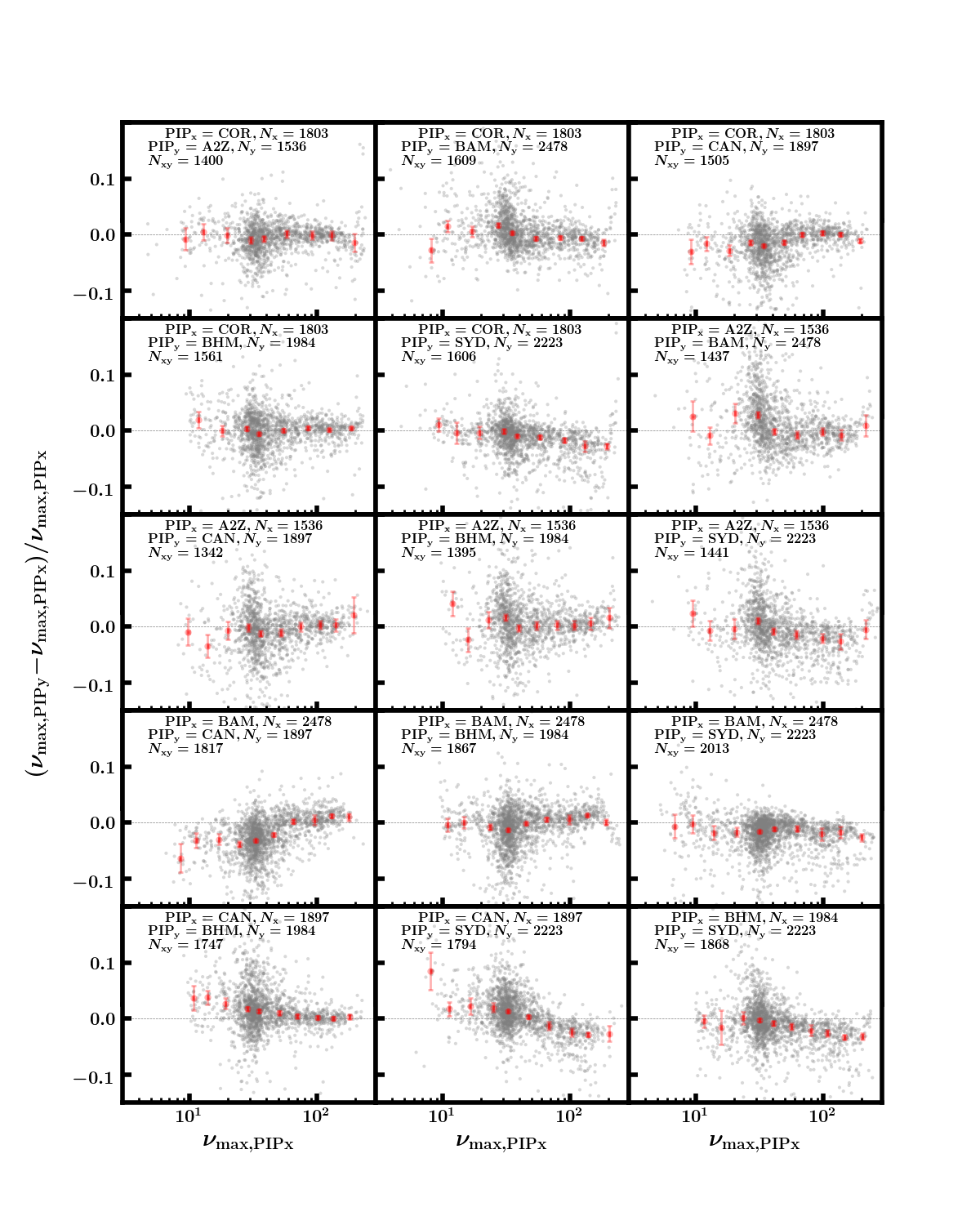}
  \caption{Pairwise comparisons between asteroseismology pipeline
    $\numax$ values for {\it K2} C4. The red error bars represent binned medians and uncertainty on those binned medians. $N_x$ and $N_y$ indicate the number of stars with asteroseismic values returned for pipeline $x$ and pipeline $y$, with the number of stars returned by both pipelines indicated by $N_{xy}$. Trends seen here are present in C6 and C7, as well.}
  \label{fig:numax_twod_c4}
\end{figure*}

\begin{figure*}[htp]
\centering
  \includegraphics[width=0.7\textwidth]{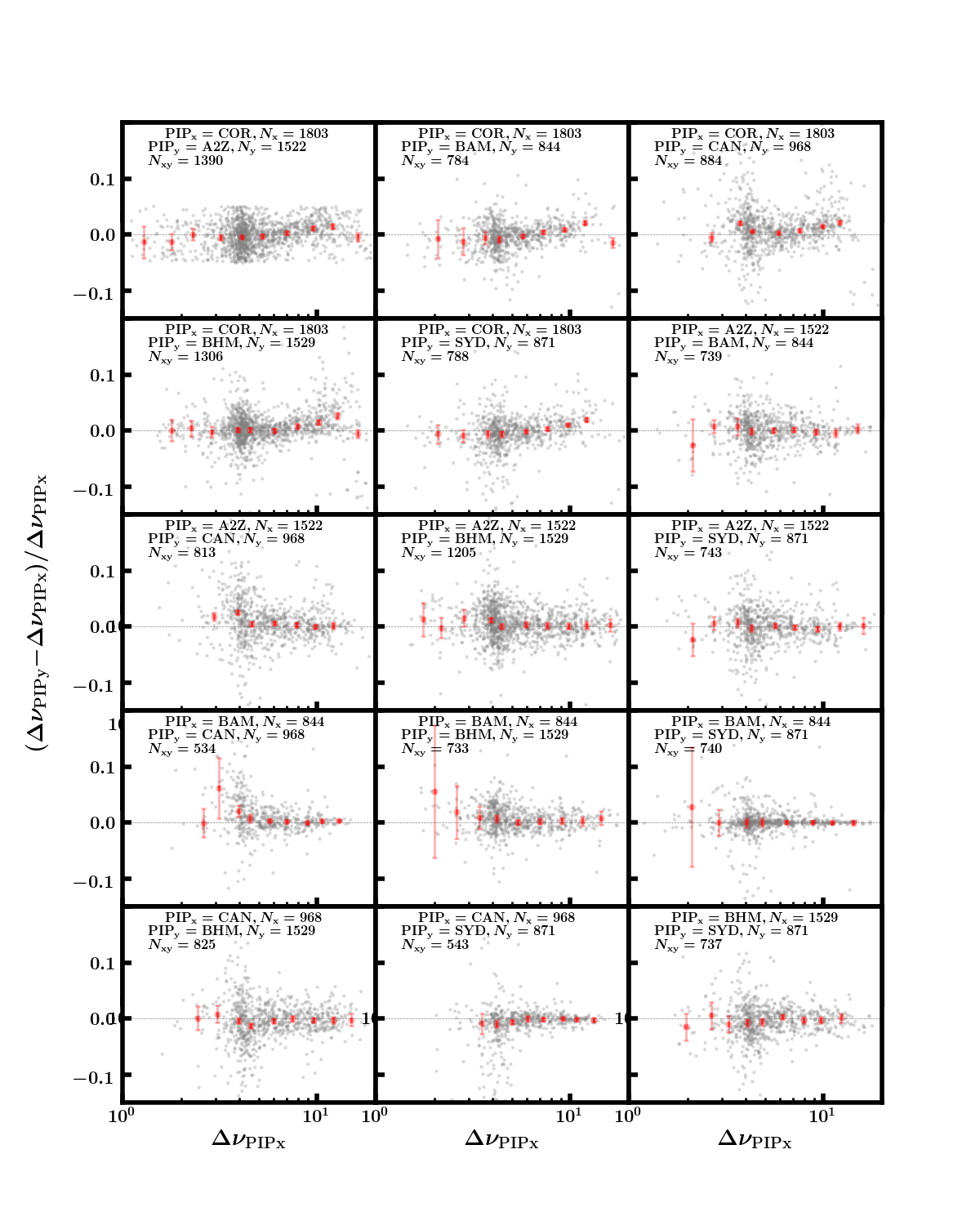}
  \caption{Pairwise comparisons between asteroseismology pipeline
    $\dnu$ values for {\it K2} C4. The red error bars represent binned medians and uncertainty on those binned medians. $N_x$ and $N_y$ indicate the number of stars with asteroseismic values returned for pipeline $x$ and pipeline $y$, with the number of stars returned by both pipelines indicated by $N_{xy}$. Trends seen here are present in C6 and C7, as well.}
  \label{fig:dnu_twod_c4}
\end{figure*}

\begin{figure}
  \plotone{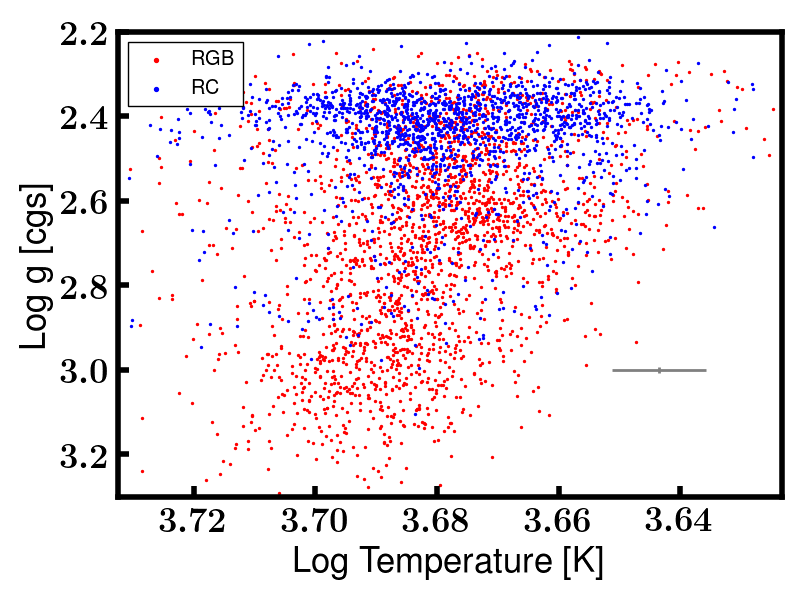}
  \caption{Kiel diagram for the C4, C6, and C7 observed {\it K2} GAP DR2 stars, colored by evolutionary state, and using EPIC effective temperatures. Surface gravities are computed using $\numaxmean$, and so only stars with consensus values from multiple asteroseismic pipelines are shown.}
  \label{fig:selection}
\end{figure}

\begin{figure*}[htp]
\centering
\subfloat{\includegraphics[width=0.37\textwidth]{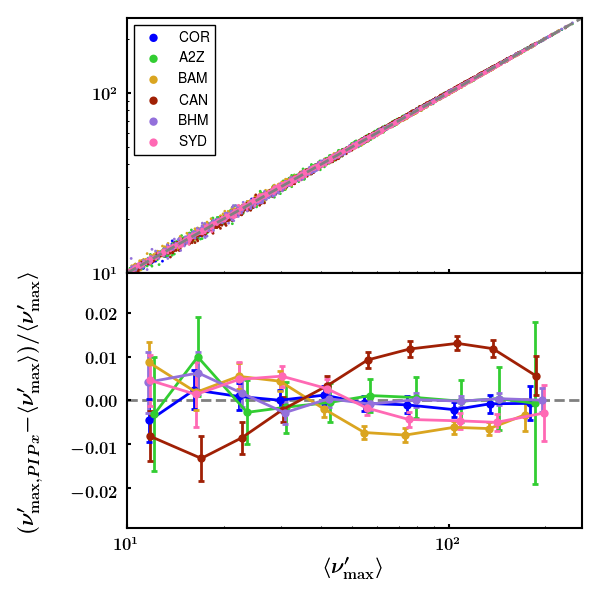}}
\subfloat{\includegraphics[width=0.37\textwidth]{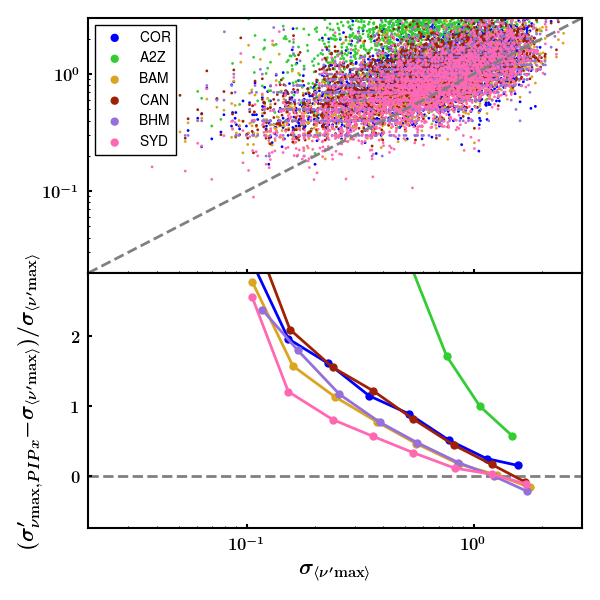}}
\caption{Left: comparison of RGB $\numax$ among pipelines, showing the re-scaled $\numax'$
  from each pipeline versus the mean $\numaxmean$ across pipelines in the top panel,
  and the fractional difference between $\numaxmean$ and $\numax'$ in the bottom panel, with error bars showing
  binned errors on the median fractional difference, assuming the uncertainty
  on $\numaxmean$ to be the standard deviation among the
  re-scaled pipeline $\numax'$ values, $\sigma_{\numaxmean}$ (this quantity is described in \S\ref{sec:zeropoint}~\&~\S\ref{sec:unc}). Right: $\sigma_{\numaxmean}$ is plotted
  against the reported uncertainty on $\numax'$ for each pipeline. The
  fractional difference between the two uncertainties is shown in the bottom panel.}
\label{fig:numax_rgb_mean}
\end{figure*}

Any pipeline may return a $\numax$ and/or a $\dnu$ for a given
star. In what follows, however, we consider a ``detection'' for
a given pipeline to be a target for which a $\numax$ is returned, unless
otherwise noted. A fraction of $\numax$-detected stars will not have a reliable $\dnu$ measurement reported. The raw $\numax$ and $\dnu$ and the uncertainties reported by each pipeline are
provided in Table~\ref{tab:raw}, while the number of $\numax$ and $\dnu$ values returned are provided in Table~\ref{tab:num}. 

\begin{figure*}[htp]
\centering
\label{fig:numax_rc_mean}
  \subfloat{\includegraphics[width=0.37\textwidth]{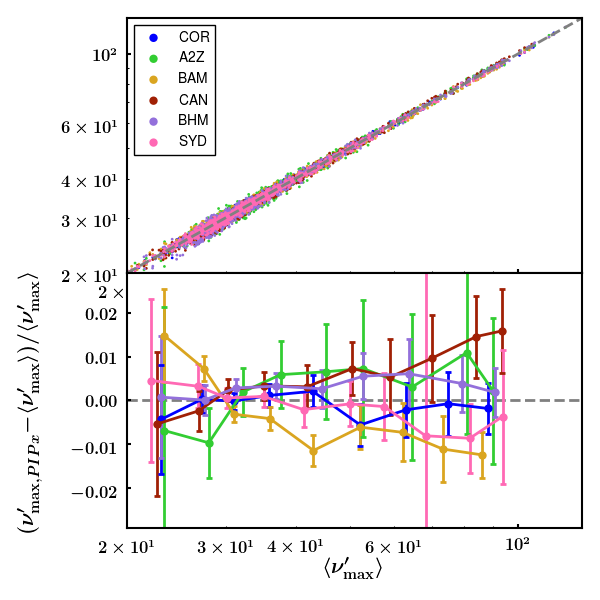}}
  \subfloat{\includegraphics[width=0.37\textwidth]{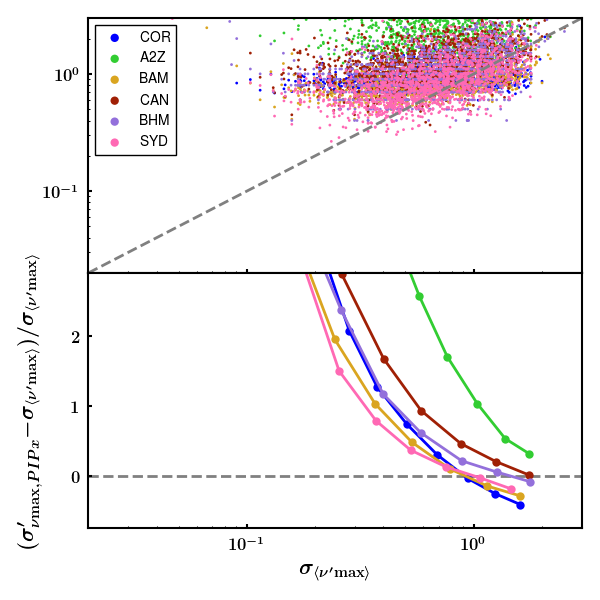} }
\caption{Same as Figure~\ref{fig:numax_rgb_mean}, but for RC stars.}
\end{figure*}

\subsection{Derived quantities}
\subsubsection{Mean asteroseismic parameters, $\langle \numax'\rangle$ and $\langle \dnu' \rangle$}
\label{sec:zeropoint}

Consolidating results from multiple pipelines, as we do here, has three primary benefits: outliers can be rejected (e.g., stars that only one pipeline identifies as an oscillator can be considered dubious); the accuracy and precision of the parameters can be improved by averaging results for the same star from different pipelines; and the spread in values for a given star can be translated into an uncertainty estimate. To perform this averaging, however, the possibility that there are systematic differences in pipeline results needs to be taken into account. \cite{pinsonneault+2018} demonstrated that such systematics existed in their multi-pipeline analysis of {\it Kepler} data, finding that the relative zero-points of the measurements returned by different pipelines do not scale in the same way as their solar measurements do \cite{pinsonneault+2018}. Figures~\ref{fig:numax_twod_c4}~\&~\ref{fig:dnu_twod_c4} demonstrate these systematics in our {\it K2} $\numax$ and $\dnu$ values. In what follows, we describe an approach that re-scales each pipeline's $\numax$ and $\dnu$ values such that they are on average on the same scale as the other pipelines. These rescaled values are used to perform outlier rejection with sigma clipping, and the values are averaged for each star to compute more accurate and precise asteroseismic values, and to define empirical uncertainties on $\numax$ and $\dnu$.
  
To reduce the
systematic uncertainty across
pipelines for a
single star, we follow the approach taken by \cite{pinsonneault+2018}:
under the
assumption that each pipeline's $\numax$ and $\dnu$ values are
distributed around the true values, we can fit
for near-unity scale factors that effectively modify each
pipeline's solar reference values so that the mean $\numax$
and $\dnu$ returned by pipelines are all on the same scale. This
therefore reduces the systematic uncertainty due to the choice of a
single pipeline value on a single star's
$\numax$ or $\dnu$, and it allows for averaging values across pipelines on a star-by-star basis.

The above procedure to re-scale the pipeline-specific solar reference values
is performed by initially assuming that the pipeline values are already on the same
system, so that the re-scaled $\numax$, $\nu_{\mathrm{max,s,p}}'$, of a star, $s$, for
pipeline, $p$, equals
the raw value returned by the pipeline: $\nu_{\mathrm{max,s,p}}'
= \nu_{\mathrm{max,s,p}}$. An average value $\langle \nu_{\mathrm{max,s}}'
\rangle \equiv \sum_p \nu_{\mathrm{max,s,p}}'/N_p$ is then calculated for each star that has at least two pipelines returning a
value. The sum is over those $N_p$ reporting pipelines. A scalar factor for each pipeline, $X_{\nu_{\mathrm{max}},p}
\equiv \sum_{s}\left[\nu_{\mathrm{max,s,p}}/\langle
  \nu_{\mathrm{max,s}}' \rangle\right]/N_s$, is calculated using
the $N_s$ stars for which the pipeline returned a raw value,
  $\nu_{\mathrm{max,s,p}}$, and
  which had a defined mean value $\langle \nu_{\mathrm{max,s}}'
  \rangle$. The pipeline values are re-scaled by this factor
  so that $\nu_{\mathrm{max,s,p}}' =
  \nu_{\mathrm{max,s,p}}/X_{\nu_{\mathrm{max}},p}$. For each star, a $3\sigma$ clipping is performed to reject re-scaled values returned by a pipeline that are highly discordant with the results from other pipelines. This whole process is
  repeated until convergence in the re-scaled value,
  $\nu_{\mathrm{max,s,p}}'$. The same procedure is done for
  $\dnu$. Following the observation by
  \cite{pinsonneault+2018} that the RC exhibits significantly
  different structure in the pipeline $\numax$ and $\dnu$ zero-point
  differences, the same procedure is done for RGB \& RGB/AGB stars
  and RC stars, separately. In the end, four scale factors are
  derived for each pipeline: a $\numax$ scale factor for RGB
  stars; a $\dnu$ scale factor for RGB stars; a $\numax$
  scale factor for RC stars; and a $\dnu$ scale factor for
  RC stars. The resulting factors are at the per cent level or below. For each pipeline, we provide the re-scaled values, $\numax'$ and $\dnu'$, in addition to the raw values. We also provide
  mean values $\numaxmean$ and $\dnumean$, which have been
  averaged across pipelines.  The root mean square across all pipelines for each star are taken as the uncertainty on the $\numaxmean$ and $\dnumean$ values, $\sigma_{\numaxmean}$ and $\sigma_{\dnumean}$ (see \S\ref{sec:unc}). The sample with $\numaxmean$ is shown in Figure~\ref{fig:selection}.

Unless otherwise noted, the mean $\dnu$ values for each star,
$\langle \dnu' \rangle$, as well as the pipeline-specific re-scaled
$\dnu$ values, $\dnu'$, have been multiplied by a factor as described in \cite{sharma+2016} that depends on the star's temperature, metallicity, surface gravity, mass, and evolutionary
state. This factor is provided in Table~\ref{tab:vals} as
$X_{\mathrm{Sharma}}$, and is a theoretically-motivated factor
that improves the homology assumption in the $\dnu$ scaling relation (Equation~\ref{eq:scaling2}). We calculated these factors using
temperatures and metallicities from the EPIC \citep{huber+2016};
surface gravity from Equation~\ref{eq:scaling1} using $\numax$; mass
estimates from Equation \ref{eq:mass} using $\numax$ and
$\dnu$, before the factor in question has been applied; and
evolutionary states derived from the neural network approach laid out
in \S\ref{sec:evo}. The correction is applied multiplicatively such that $\dnu'$ is multiplied by $X_{\mathrm{Sharma}}$. Pipeline values rejected by the above sigma-clipping process will not have a corrected pipeline value populated in Table~\ref{tab:vals}.

Table~\ref{tab:refs} shows the resulting solar reference scale factors
compared to those from \cite{pinsonneault+2018}. The agreement is
mixed, depending on the pipeline. As noted in \S\ref{sec:extraction}, the pipelines as implemented for this analysis have been modified to account for {\it K2} data, and so may well differ slightly in the way they perform compared to their implementation for \cite{pinsonneault+2018}. Additionally, we consider results from BAM \citep{zinn+2019bam}, which was not considered for the \cite{pinsonneault+2018} analysis. The uncertainties on the scale factors derived in
this work are also larger than the ones found by
\cite{pinsonneault+2018}, in part due to the increased scatter in the
$\numax$ and $\dnu$ derived with {\it K2} data compared to {\it Kepler} data. These uncertainties are calculated assuming $\sigma_{\numax}$ and $\sigma_{\dnu}$ are exact (i.e., with no uncertainty on the standard deviation used to calculate them [see \S\ref{sec:unc}]). Given this assumption, the uncertainties on the scale factors are indicative and not definitive. We also provide the re-scaled, pipeline-specific solar reference values themselves in Table~\ref{tab:refs} (attained by multiplying the pipeline-specific solar reference value by the pipeline's solar reference scale factor derived here).

Even with these re-scalings, residual differences between pipelines
as a function of $\numax$ and $\dnu$ will remain. These
trends are shown in the bottom panels of Figures~\ref{fig:numax_rgb_mean}
-~\ref{fig:dnu_rc_mean}, where fractional differences between
$\numaxmean$~\&~$\dnumean$ and individual pipeline re-scaled values,
$\numax'$~\&~$\dnu'$ are shown. By definition, the mean of $\numax'$
and $\dnu'$ across the pipelines are $\numaxmean$ and $\dnumean$,
though what these figures demonstrate is sub-structure as a function of $\numax$
and $\dnu$, indicating $\numax$- and $\dnu$-dependent systematic errors, not
zero-point errors. We turn to a more robust estimate of the systematic errors on $\numaxmean$ and $\dnumean$ in \S\ref{sec:tgas}.

After these re-scalings are applied for each pipeline
  $\numax$ and $\dnu$, the absolute value of the solar reference
  values are free to be chosen, which we take to be $\numaxsun = 3076\muhz$ and
  $\dnusun = 135.146\muhz$ \citep{pinsonneault+2018}. The effect of this choice
  is evaluated in \S~\ref{sec:tgas}.

  The main {\it K2} GAP DR2 sample is defined to be that with a valid
  $\langle \nu_{\mathrm{max}}' \rangle$ and $\langle \Delta
  \nu' \rangle$ (i.e., stars for which at least two pipelines
  returned both $\numax$ and $\dnu$ that agree to within $3\sigma$), and its contents are summarized in
Table~\ref{tab:vals}. The sample contains 4395 stars.

\subsubsection{Evolutionary states}
\label{sec:evo}
Recent results have shown that highly accurate evolutionary
state classification between RGB and RC can be achieved even on short (\textit{K2}-like) time series using machine
learning approaches \citep{hon+2017a,hon+2018a,kuszlewicz_hekker_bell2020} where `classical' asteroseismic-based classification is not possible \citep{bedding+2011,mosser+2019,elsworth+2019}. We note also that determining evolutionary state for short times series may be possible by measuring $\epsilon_c$, the radial order position for the central radial mode of a power spectrum, to the extent that mode identification is possible with \textit{K2} light curves; see \cite{kallinger+2012}. In this work, we determine evolutionary state using the method described in \cite{hon+2017a,hon+2018a}, which has a similar accuracy to that of the recent machine learning approach from \cite{kuszlewicz_hekker_bell2020} ($95\%$ and $91\%$ for \textit{K2}-like data, respectively). The chosen technique requires
that the granulation background of the stellar power spectrum be removed, and that $\numax$ and $\dnu$ be provided in order to search the appropriate part of the spectrum for evolutionary state diagnostics. We remove the background by subtracting a smoothed version
of each power spectrum
in log space. This
approach avoids removing low-frequency modes, as can happen when smoothing with a
window with fixed size in linear frequency, by using the fact that solar-like
oscillators with different $\numax$ have similar granulation
background shapes in log
space.
We then apply the machine-learning technique from
\cite{hon+2017a,hon+2018a} using the background-subtracted power spectrum, $\numaxmean$, and $\dnumean$ as inputs. For details on how the neural network is trained, including how the probability of being RGB or RC is calibrated, see \cite{hon+2017a,hon+2018a}. Because the machine learning algorithm was not trained on stars with $\dnu < 3.2 \muhz$ (corresponding to stars with radii larger than those in the RC), we assign stars in this $\dnu$ regime to be ambiguous RGB/AGB stars. The spectra of these stars from \textit{K2} cannot be used to distinguish between RGB and AGB because both types of stars are shell burning stars, and so we consider them for the purposes of the following analysis to be equivalent to RGB stars.

\subsubsection{Uncertainties on $\numaxmean$ and $\dnumean$, $\sigma_{\numaxmean}$ and $\sigma_{\dnumean}$}
\label{sec:unc}
\begin{figure*}
\centering
  \subfloat{\includegraphics[width=0.37\textwidth]{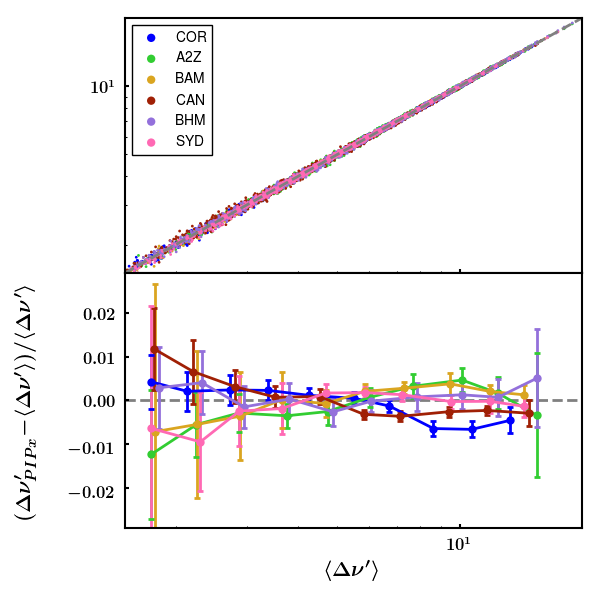}}
  \subfloat{\includegraphics[width=0.37\textwidth]{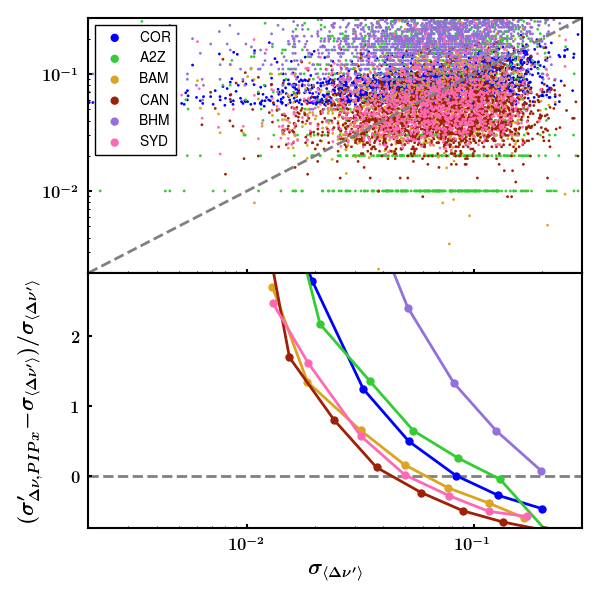}}
\caption{Same as Figure~\protect\ref{fig:numax_rgb_mean}, but for $\dnu$.}
\label{fig:dnu_rgb_mean}
  \end{figure*}

\begin{figure*}
\centering
  \subfloat{\includegraphics[width=0.37\textwidth]{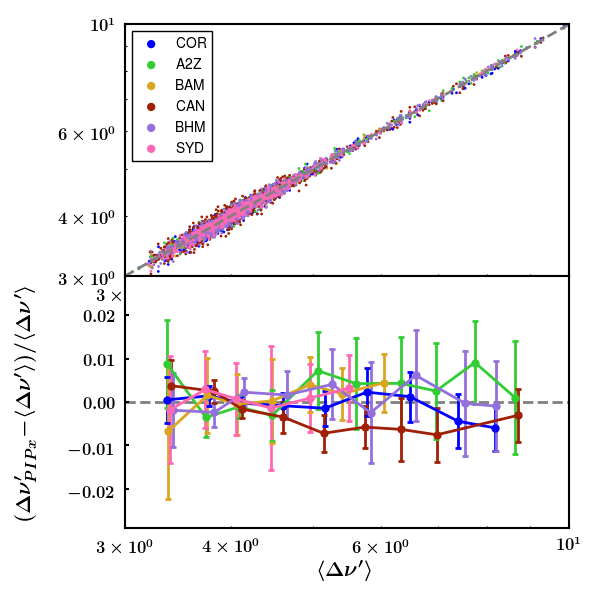}}
  \subfloat{\includegraphics[width=0.37\textwidth]{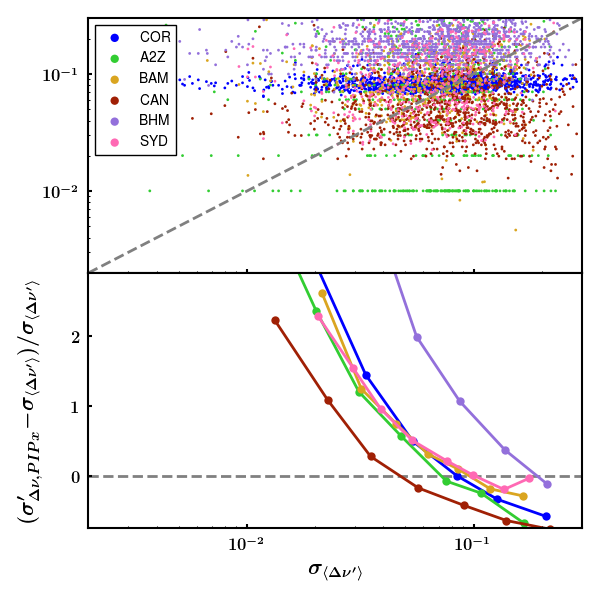}}
\caption{Same as Figure~\protect\ref{fig:dnu_rgb_mean}, but for RC stars.}
\label{fig:dnu_rc_mean}
  \end{figure*}

The root mean square of the re-scaled $\numax'$ and $\dnu'$ values for each
star across pipelines can be thought of as --- and in this work are taken to be --- the statistical uncertainties on
$\numaxmean$ and $\dnumean$, denoted $\sigma_{\numaxmean}$ and
$\sigma_{\dnumean}$. These values are listed in Table~\ref{tab:vals}. We compare $\sigma_{\numaxmean}$ and
$\sigma_{\dnumean}$ to the reported statistical
uncertainties on the parameters returned by each pipeline in the bottom panels of
Figures~\ref{fig:numax_rgb_mean}~-~\ref{fig:dnu_rc_mean}. These panels show the fractional differences in the pipeline-reported uncertainty and the root mean square of $\numax'$ and $\dnu'$ across pipelines, and
indicates whether or not the pipeline values are over- (above the grey
dashed line) or under-estimated (below the grey dashed line), assuming
the pipeline values are distributed like normal variables around
$\numaxmean$ and $\dnumean$. The uncertainties in $\numax$ and $\dnu$ appear to be over-estimated for stars with smaller $\sigma_{\numaxmean}$ and $\sigma_{\dnumean}$. This over-estimation worsens with decreasing $\sigma_{\numaxmean}$ and $\sigma_{\dnumean}$, and does so more rapidly for RC stars than RGB stars. Part of this over-estimation trend may well be a selection effect: $\sigma_{\numaxmean}$  and $\sigma_{\dnumean}$ are computed after a sigma-clipping procedure, which will tend to make a smaller root mean square. We evaluate the accuracy of these uncertainties further in \S\ref{sec:unc2}.

\subsubsection{Radius and mass coefficients, $\kapparmean$ and $\kappammean$}
\label{sec:derived}
Given Equations~\ref{eq:scaling1}~\&~\ref{eq:scaling2}, the radius of a star may be derived according to:
\begin{align}
\frac{R}{\rsun} &\approx \left(\frac{\numax}{\numaxsun}\right)
\left(\frac{\dnu}{\dnusun}\right)^{-2}
\left(\frac{\teff}{\teffsun}\right)^{1/2}\\
&\equiv \kappa_R \left(\frac{\teff}{\teffsun}\right)^{1/2},
\label{eq:radius}
\end{align}
and its radius from
\begin{align}
\frac{M}{\msun} &\approx \left(\frac{\numax}{\numaxsun}\right)^3
\left(\frac{\dnu}{\dnusun}\right)^{-4}
\left(\frac{\teff}{\teffsun}\right)^{3/2}\\
&\equiv \kappa_M \left(\frac{\teff}{\teffsun}\right)^{3/2}.
\label{eq:mass}
\end{align}

It is our wish not to impose a choice of temperature on the user when providing asteroseismic radii and masses, which is why we provide $\kappa_R$ and $\kappa_M$ instead of direct radius and mass. With the re-scaled asteroseismic values, $\numax'$ and
$\dnu'$, in hand, we can construct the radius and mass
coefficients, $\kappa_R'$ and $\kappa_M'$ for each star and each pipeline, which
correspond to the asteroseismic component of the scaling relations for
stellar mass and radius in solar units. We show the pairwise comparisons between pipeline values of $\kappa_R'$ and $\kappa_M'$ for all stars in the DR2 sample in Figures~\ref{fig:kappa_r_err}~\&~\ref{fig:kappa_m_err}. As with the pairwise $\numax$ and $\dnu$ comparisons (Figures~\ref{fig:numax_twod_c4}~\&~\ref{fig:dnu_twod_c4}), we see systematic differences between pipeline $\kappa_R'$ and $\kappa_M'$ values, which we quantify in \S\ref{sec:syst_mean}. We also
construct the average for each star, $\kapparmean$ and $\kappammean$, based on $\numaxmean$ and
$\dnumean$. Uncertainties on $\kapparmean$ and $\kappammean$ are
calculated according to standard propagation of error, using
$\sigma_{\numaxmean}$ and $\sigma_{\dnumean}$. All of these coefficients are reported in
Table~\ref{tab:kappas}.

    \subsection{Statistical uncertainties in $\numaxmean$, $\dnumean$, $\kapparmean$, and $\kappammean$}
    \label{sec:unc2}
We show in Figures~\ref{fig:numaxrgb}-\ref{fig:dnurc} the distribution
of the fractional uncertainties, $\sigma_{\numaxmean}/\numaxmean$ and
$ \sigma_{\dnumean}/\dnumean $ as a function of evolutionary state
(RGB or RGB/AGB, RC). The curves over-plotted on the distributions
represent models of the uncertainty distributions assuming Gaussian
statistics. Under the assumption that all stars in \textit{K2} have
the same, true fractional uncertainty, $\sigma$, these distributions
would be described by generalized gamma distributions with probability
density function $f(x, a, d, p) = \frac{(p/a^d) x^{d-1}
  e^{-(x/a)^p}}{\Gamma(d/p)}$, where $\Gamma(z)$ denotes the gamma
function, $p = 2$, $d=dof-1$, and $a=\sqrt{2\sigma^2/(dof-1)}$. We
consider two models: one with $\sigma$ fixed to be the observed median
fractional uncertainty for each $dof$, and one with two generalized
gamma distributions for which both $\sigma$ and $dof$ are allowed to
vary. The grey (black) curves in
Figures~\ref{fig:numaxrgb}-\ref{fig:dnurc} are weighted sums of the
best-fitting single-component (two-component) models across all $dof$
(see the Appendix for details). It is clear that the observed
distributions of $\sigma_{\numaxmean}/\numaxmean$ and $
\sigma_{\dnumean}/\dnumean $ are not perfectly described by a
generalized gamma distribution with a unique $\sigma$, as they would
be if all stars had the same fractional uncertainty in $\numaxmean$
and $\dnumean$. Nevertheless, allowing $\sigma$ to be a function of
the number of reporting pipelines appears to be a surprisingly good
approximation. This indicates that 1) the fractional uncertainties for
all stars are not a strong function of $\numax$ or $\dnu$ (but rather
have a fractional uncertainty that varies modestly according to the
number of reporting pipelines), and 2) our uncertainty estimates are
distributed like they should be according to $\chi$ statistics. As we
note in the Appendix, the fractional uncertainties vary mostly
according to the evolutionary state, with RC parameters less precisely
measured, and it happens that the typical uncertainties are the same
for $\numax$ and $\dnu$ for RGB stars. Our typical
uncertainties for $\numaxmean$ and $\dnumean$ are listed in
Table~\ref{tab:kepk2}, along with the corresponding median fractional
uncertainties from APOKASC-2 \citep{pinsonneault+2018} (``APOKASC-2''
in the table). We also include a comparison to the median fractional
uncertainties from the analysis of \cite{yu+2018} (``Y18" in the
table). The latter analysis uses only the SYD pipeline, as opposed to
APOKASC-2, which reports parameters averaged across five different
asteroseismic pipelines. The methodology in this work is much the same
as that of \cite{pinsonneault+2018}, but we include a comparison to
\cite{yu+2018} to give an indication of the variation in uncertainties
resulting from an aggregated pipeline versus individual pipeline
approach. The most significance difference between the uncertainties
of the two \textit{Kepler} analyses is in RC $\numax$, for which
uncertainties from \cite{yu+2018} are larger than those from
APOKASC-2; the larger $\numax$ uncertainties from \cite{yu+2018} map
into correspondingly larger RC $\kappa_R$ and $\kappa_M$
uncertainties. Comparing our results to those of APOKASC-2, we see
that the uncertainties in $\numax$ in \textit{K2} are up to a factor
of two larger than in \textit{Kepler}, and the uncertainties in $\dnu$
are larger by up to a factor of four for RGB stars. These differences
between \textit{Kepler} and \textit{K2} come from differences in photometric precision and the differences in dwell time, and will be further explored in the next \textit{K2} GAP data release.

The analogous uncertainty distributions for $\kapparmean$ and $\kappammean$ are shown in Figures~\ref{fig:kapparrgb}-\ref{fig:kappamrc}. Note that the number of stars plotted, $N$, is not 4395 (the total number of stars we provide with $\kapparmean$ and $\kappammean$). This is because in this treatment, we require the number of pipelines reporting values for $\numax$ and $\dnu$ be the same in order to fulfill the generalized gamma distribution requirements; most stars do not have the same number of $\numax$ measurements as $\dnu$: there are 1030 such RGB stars and 257 such RC stars. Considering the excellent match of the fitted generalized gamma distributions (black), we adopt statistical uncertainties for $\kapparmean$ and $\kappammean$ as listed in Table~\ref{tab:kepk2}. We also list the corresponding median fractional uncertainties in \textit{Kepler}, computed using the evolutionary states and $\numax$ \& $\dnu$ values from \cite{pinsonneault+2018} (``APOKASC-2'') or those from \cite{yu+2018} (``Y18"). We find that the \textit{K2} radius (mass) uncertainties are larger by up to a factor of three (two) compared to that of \textit{Kepler} for RGB stars from APOKASC-2. The uncertainties in RC stellar parameters are more comparable between the two data sets, with the increase in uncertainty from \textit{Kepler} to \textit{K2} not being larger than a factor of two. The uncertainties are more comparable between \cite{yu+2018} results and \textit{K2} because of the larger uncertainty in RC $\numax$ from \cite{yu+2018} compared to APOKASC-2.

\subsection{Systematic uncertainties in $\numaxmean$, $\dnumean$, $\kapparmean$, and $\kappammean$}
\label{sec:syst_mean}
\begin{figure*}[htp]
\centering
  \includegraphics[width=0.8\textwidth]{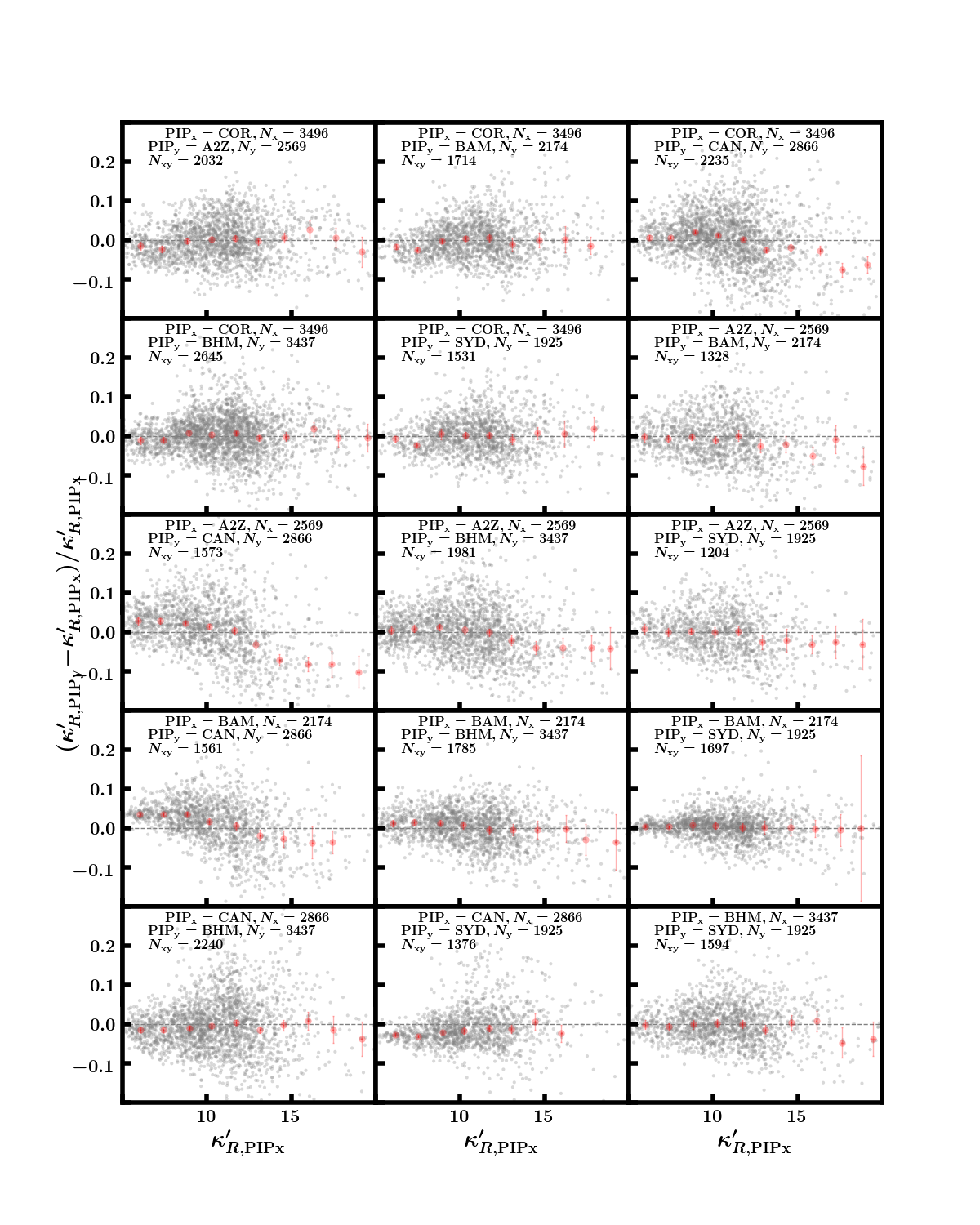}
    \caption{Pairwise comparisons between asteroseismology pipeline
      $\kappa_R'$ values.}
    \label{fig:kappa_r_err}        
\end{figure*}

\begin{figure*}[htp]
\centering
  \includegraphics[width=0.8\textwidth]{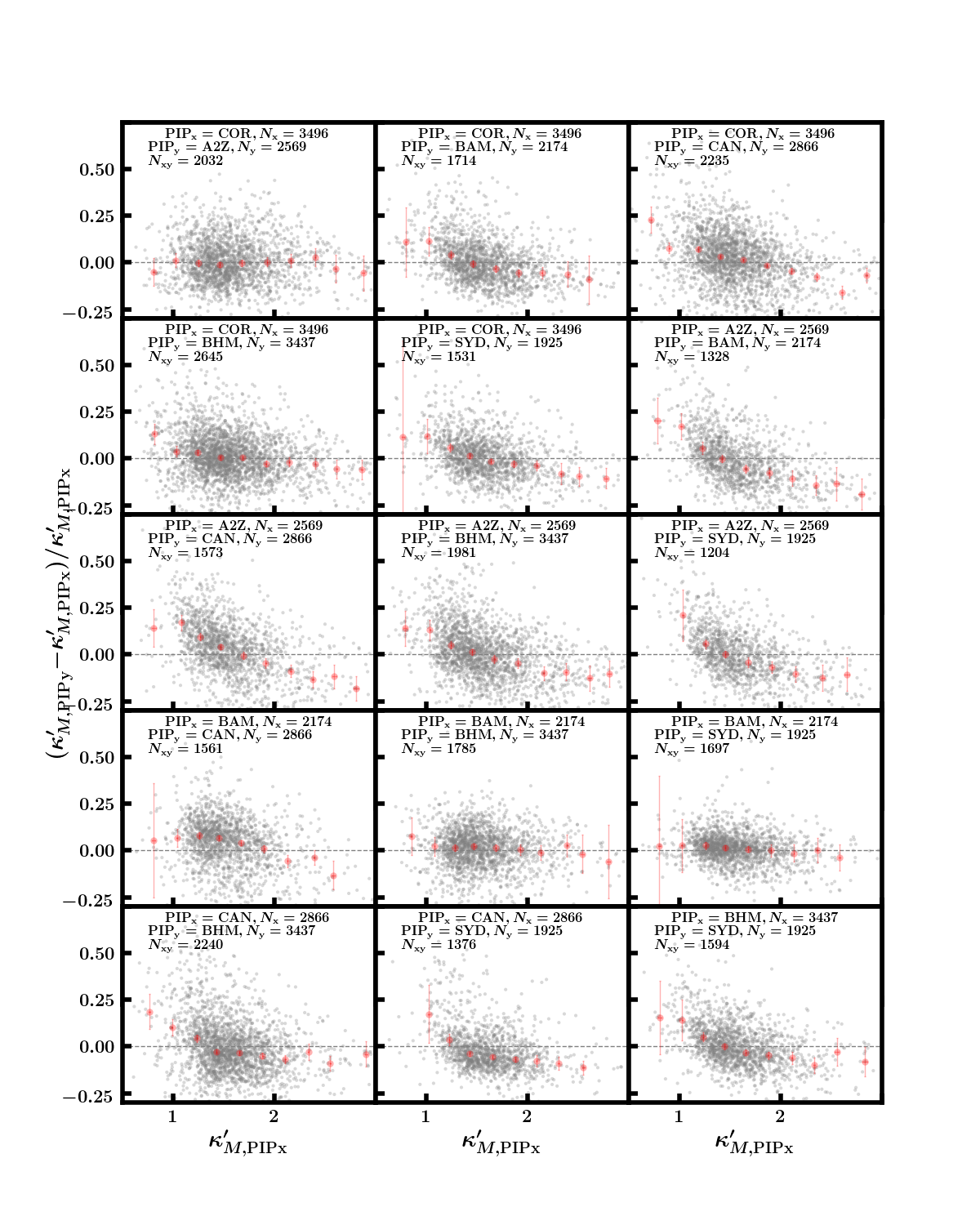}
    \caption{Pairwise comparisons between asteroseismology pipeline
      $\kappa_M'$ values.}
    \label{fig:kappa_m_err}        
\end{figure*}

We now turn to systematic uncertainties in $\numaxmean$, $\dnumean$, $\kapparmean$, and $\kappammean$ that take the form of $\numax$- and $\dnu$-dependent offsets among the pipelines. By definition, the rescaling process described in \S\ref{sec:zeropoint} removes offsets among the pipelines by averaging over all $\numax$ and $\dnu$. However, there are $\numax$- and $\dnu$-dependent trends seen in the fractional differences in $\numaxmean$ and $\dnumean$ shown in Figures~\ref{fig:numax_rgb_mean}~-~\ref{fig:dnu_rc_mean}. These trends do not affect the statistical uncertainty discussion above, as they are removed when computing $\sigma_{\numaxmean}$ and $\sigma_{\dnumean}$. Our approach to account for these systematics is therefore to adopt the largest excursion of any pipeline from $\numaxmean$ and $\dnumean$ for each bin plotted in the bottom panel of Figures~\ref{fig:numax_rgb_mean}~-~\ref{fig:dnu_rc_mean}, add to that the uncertainty on the median, and adopting the result as $2\sigma$ systematic uncertainties. The resulting ($1\sigma$) uncertainties are listed in Table~\ref{tab:syst_mean}. For most stars, the typical systematic uncertainty is less than the statistical uncertainty: for a typical RGB star with $(\numax, \dnu) \sim (75 \muhz, 7.5 \muhz)$ or RC star with $(\numax, \dnu) \sim (30 \muhz, 4.0\muhz)$, the systematic uncertainties are similar, at $\sim0.6\%$ in $\numaxmean$ and  $\sim0.3\%$ in $\dnumean$. We adopt these numbers as typical systematic uncertainties for $\numaxmean$ and $\dnumean$, though they are clearly a function of $\numax$ and $\dnu$. Generally, though, the uncertainty is larger at smaller $\numax$ and $\dnu$. This is a result of both intrinsically fewer stars in this regime (as they are more evolved, and therefore shorter-lived) as well as difficulties in measuring low-frequency $\numax$ and $\dnu$ due to the {\it K2} frequency resolution. We take the systematic uncertainties in $\kapparmean$ and $\kappammean$ to be $1\%$ and $2\%$, from propagation of the systematic uncertainties in $\numaxmean$ and $\dnumean$. As with $\numaxmean$ and $\dnumean$, in detail, these systematic uncertainties are a function of $\kapparmean$ and $\kappammean$ (see Figures~\ref{fig:kappa_r_err}~\&~\ref{fig:kappa_m_err}).

We note that the $\dnu$ correction applied to $\dnumean$ and therefore $\kapparmean$ and $\kappammean$, $X_{\mathrm{Sharma}}$, is computed using the EPIC temperature and metallicity scale.  We acknowledge the user may wish to use their own temperatures, and therefore we caution that using a different temperature scale will introduce systematics. For example, using a temperature scale 100K hotter (cooler) than the EPIC temperature will make radii 1\% lower (higher) if $X_{\mathrm{Sharma}}$ is not also recomputed with the user's adopted temperatures. In order to give the user as much convenience as possible, we provide EPIC temperatures in Table~\ref{tab:vals}, should the user wish to compute consistent radii/masses; we also include the EPIC metallicities used for computing $X_{\mathrm{Sharma}}$. In the event the user wishes to use a different temperature scale and does not wish to sustain additional $\sim 1\%$ systematic uncertainties, we encourage re-computing $X_{\mathrm{Sharma}}$ with the user's own temperatures and/or metallicities using \texttt{
asfgrid} \citep{sharma+2016, asfgrid}, which is available at \url{http://www.physics.usyd.edu.au/k2gap/Asfgrid/}.

\begin{figure*}[htp]
\centering
  \includegraphics[width=0.6\textwidth]{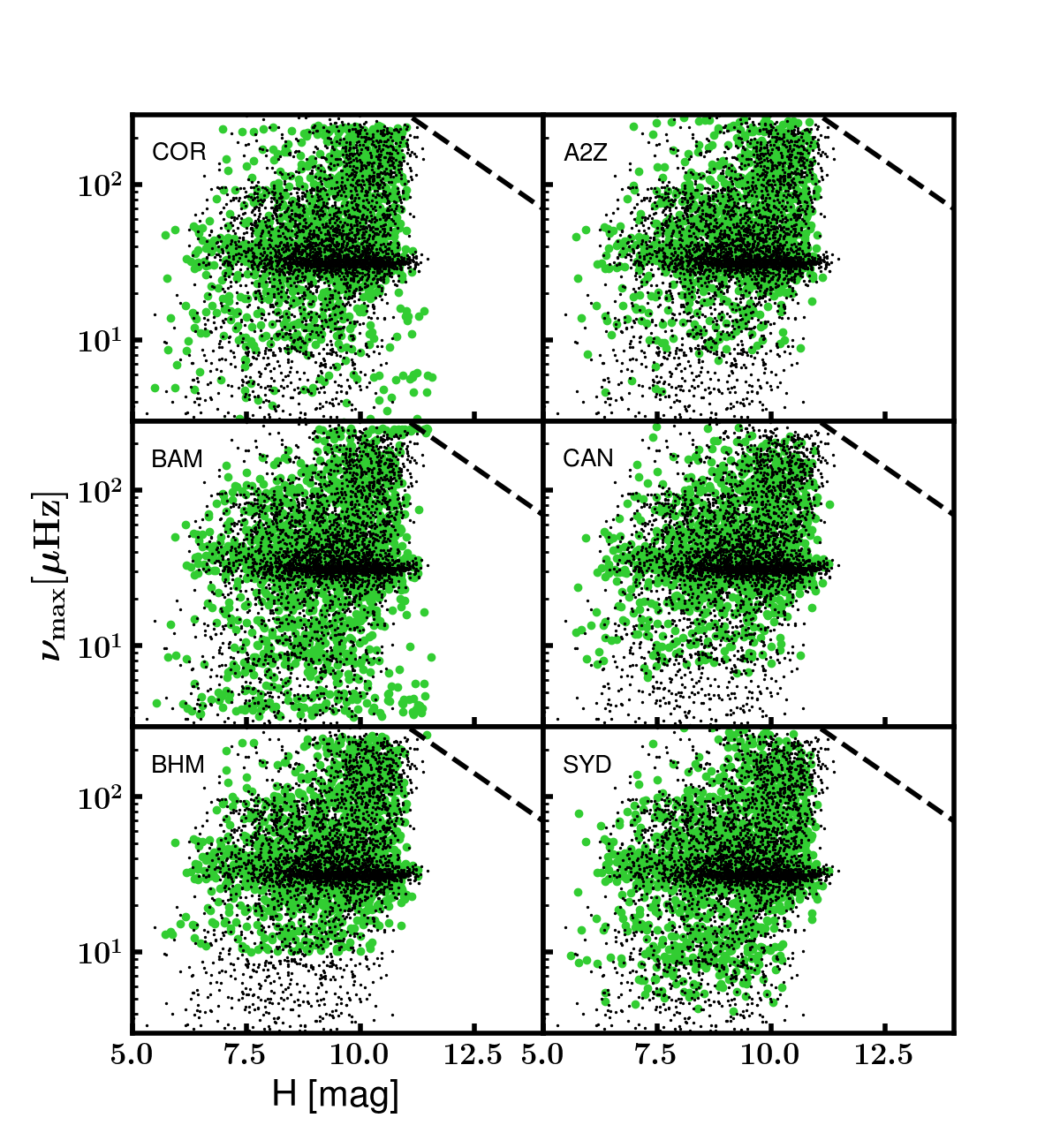}
  \caption{Distributions of $\numax$ as a function of magnitude, as predicted by \texttt{Galaxia} (black) and observed (green), for {\it K2} C4. The dashed lines represent our adopted detectability threshold of $\numaxdetect < 5\times10^4 1.6^{-H}$.}
\label{fig:mag_numax_c4}
\end{figure*}

\begin{figure*}[htp]
\centering
  \includegraphics[width=0.6\textwidth]{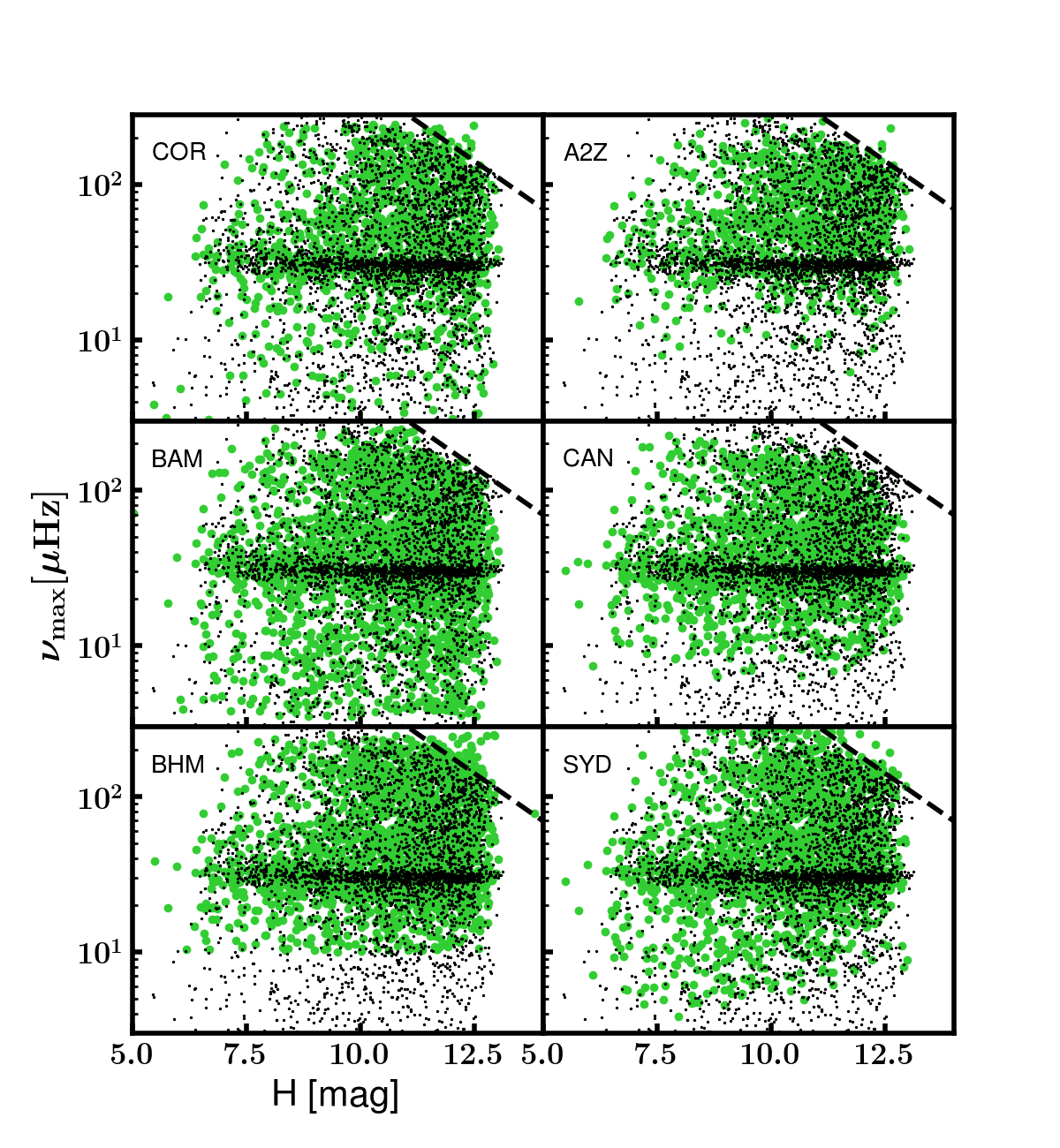}
  \caption{Distributions of $\numax$ as a function of magnitude, as predicted by \texttt{Galaxia} (black) and observed (green), for {\it K2} C6. The dashed lines represent our adopted detectability threshold of $\numaxdetect < 5\times10^4 1.6^{-H}$.}
 \label{fig:mag_numax_c6}
\end{figure*}

\begin{figure*}[htp]
\centering
\includegraphics[width=0.6\textwidth]{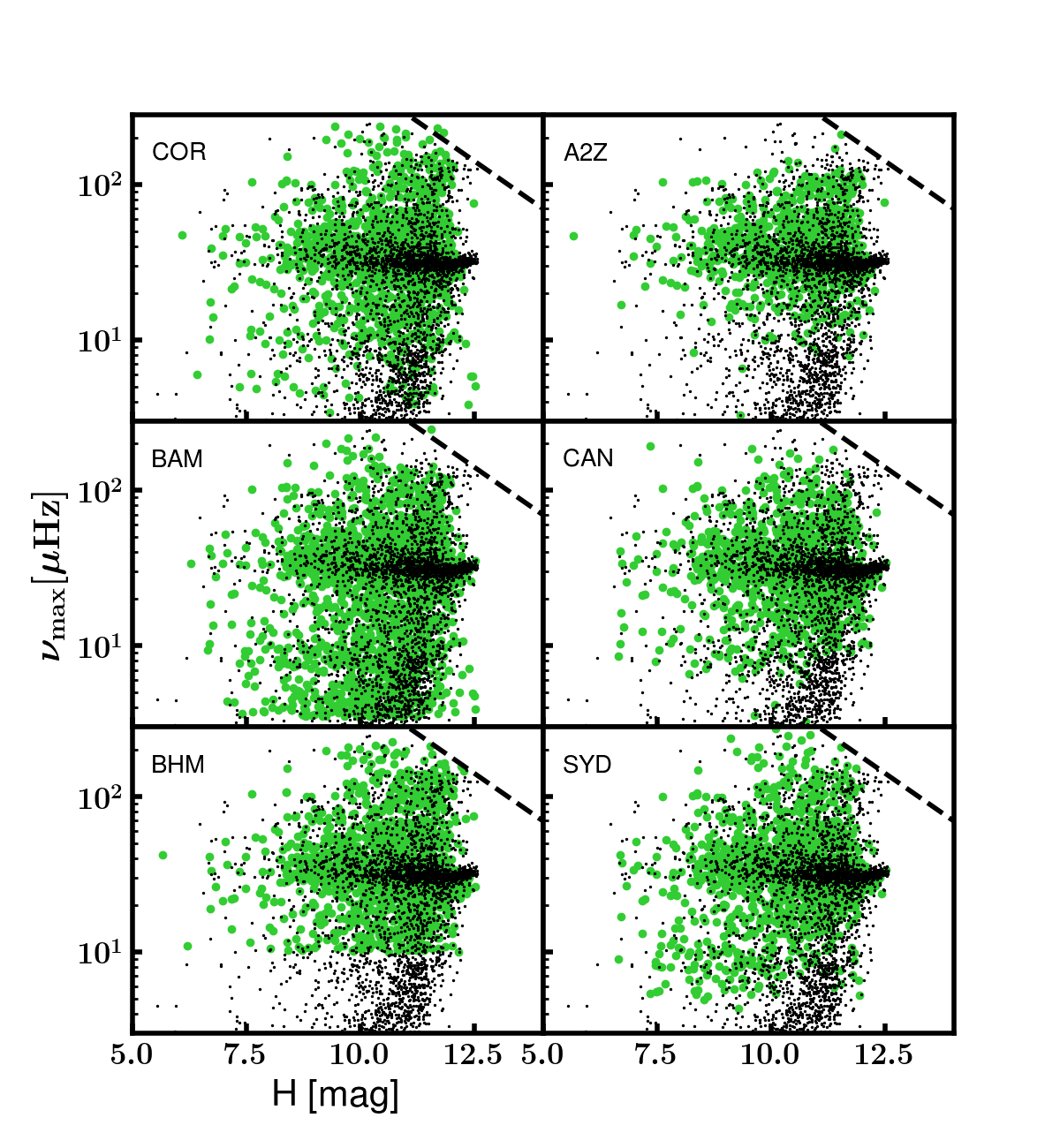}
  \caption{Distributions of $\numax$ as a function of magnitude, as predicted by \texttt{Galaxia} (black) and observed (green), for {\it K2} C7. The dashed lines represent our adopted detectability threshold of $\numaxdetect < 5\times10^4 1.6^{-H}$.}
\label{fig:mag_numax_c7}
\end{figure*}

\section{Results}
\label{sec:results}
\subsection{Comparison to \texttt{Galaxia}}
\label{sec:galaxia}
We start by comparing properties of the {\it K2} GAP DR2 sample to
those of a \texttt{Galaxia} simulation \citep{sharma+2011}, with corrections made to the simulated metallicity scale described in \cite{sharma+2019}. Each
campaign has been modeled separately because they each probe different
regions of the Galaxy. To make the simulated populations comparable to
the data, we select only the simulated stars that would be seismically detected. We defined the detectable sample to be stars with
$3 \muhz < \numax < 280 \muhz$ and signal-to-noise ratio yielding a
probability of detection greater than $95\%$ (calculated according to the
procedure used in \cite{chaplin+2011}). We impose the same selection of the simulated and observed stars (see \S\ref{sec:selection} and also \citealt{sharma+2019}), using a synthetic $V$-band magnitude that depends on $J-K_{\mathrm{s}}$ color
according to $V =
K_{\mathrm{s}}+2.0((J-K_{\mathrm{s}})+0.14)+0.382 e^{2(J-K_{\mathrm{s}}-0.2)}$ \citep{sharma+2018a}.

\begin{figure*}[htp]
\centering
  \subfloat{\includegraphics[width=0.31\textwidth]{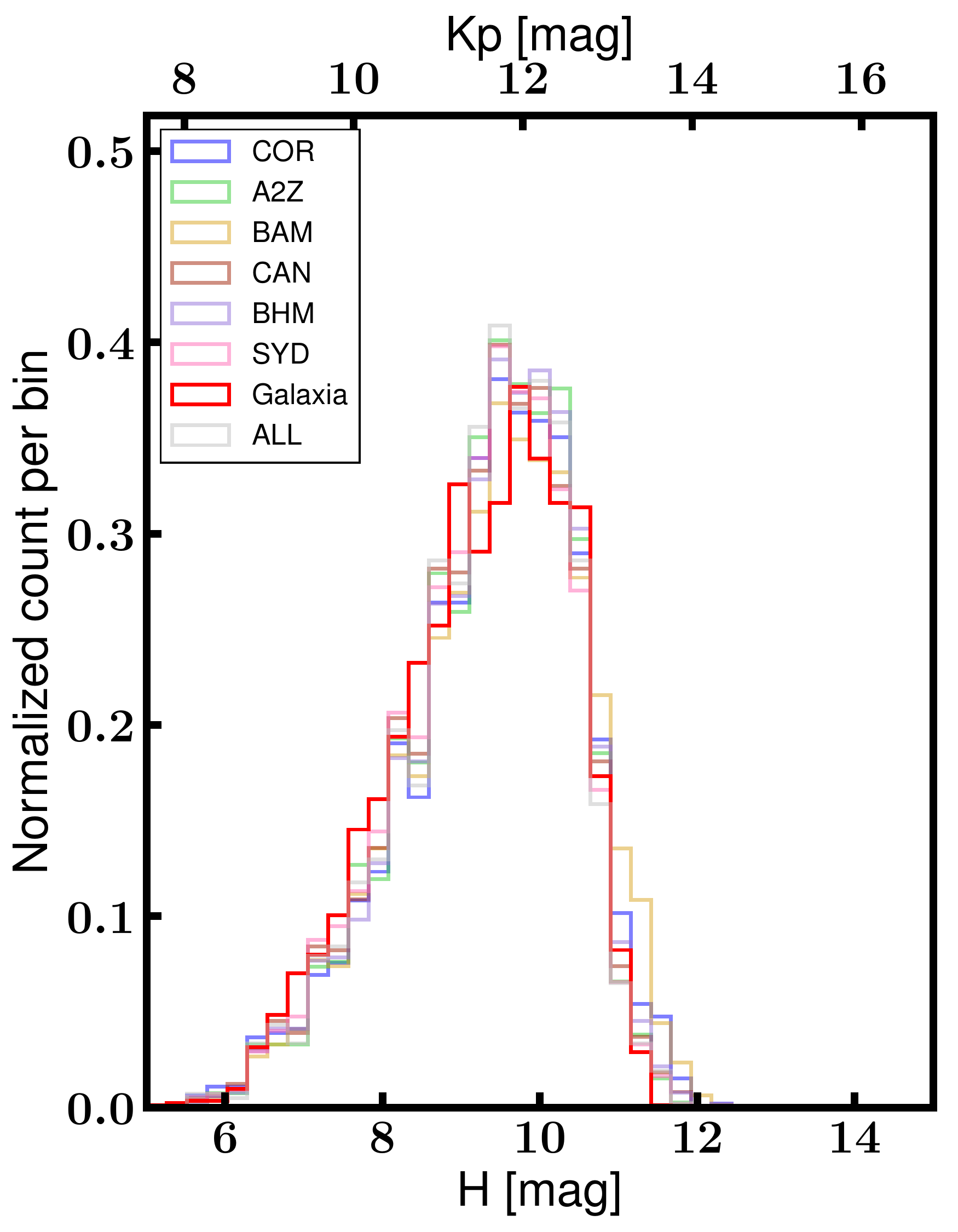}}
  \subfloat{\includegraphics[width=0.31\textwidth]{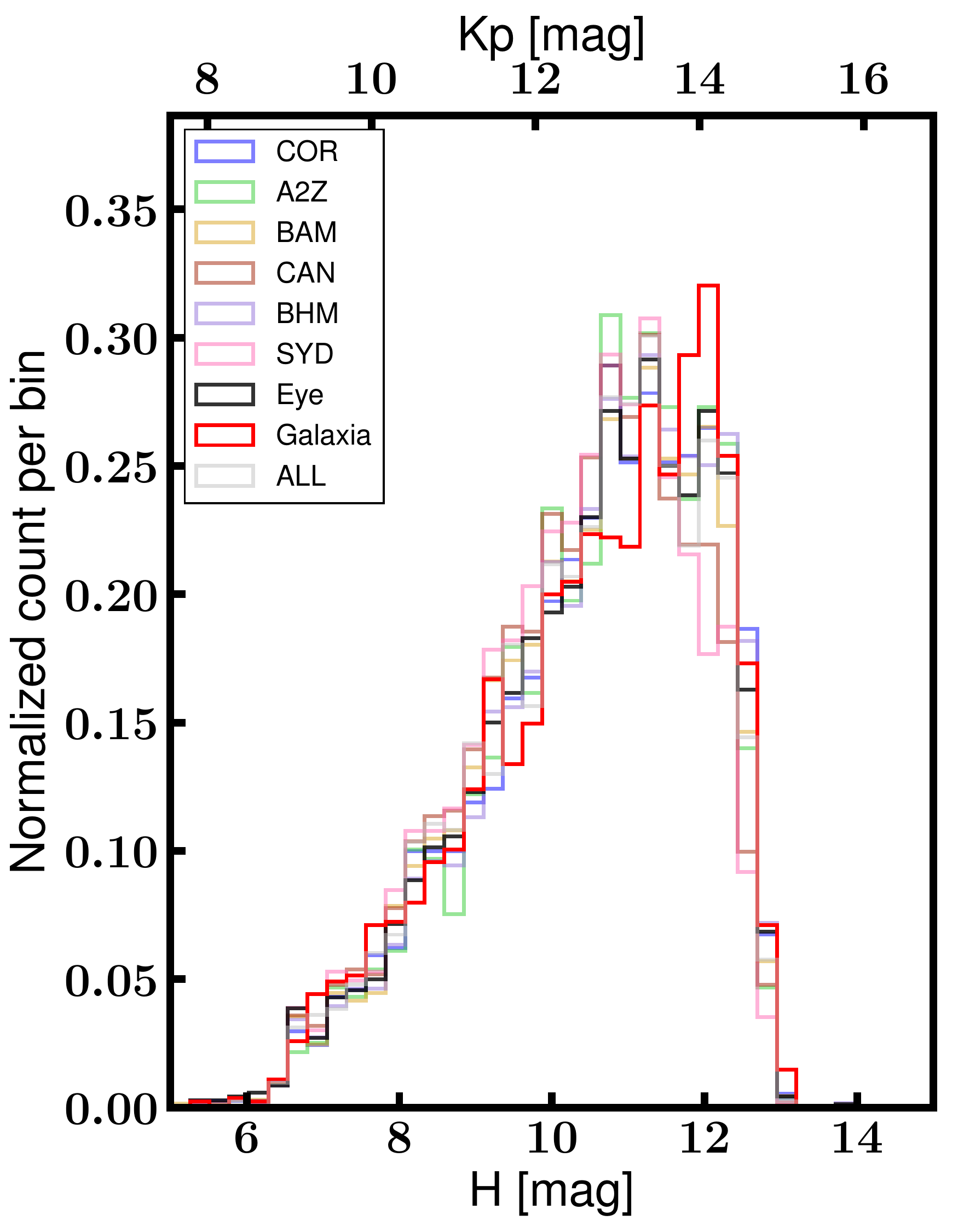}}
    \subfloat{\includegraphics[width=0.31\textwidth]{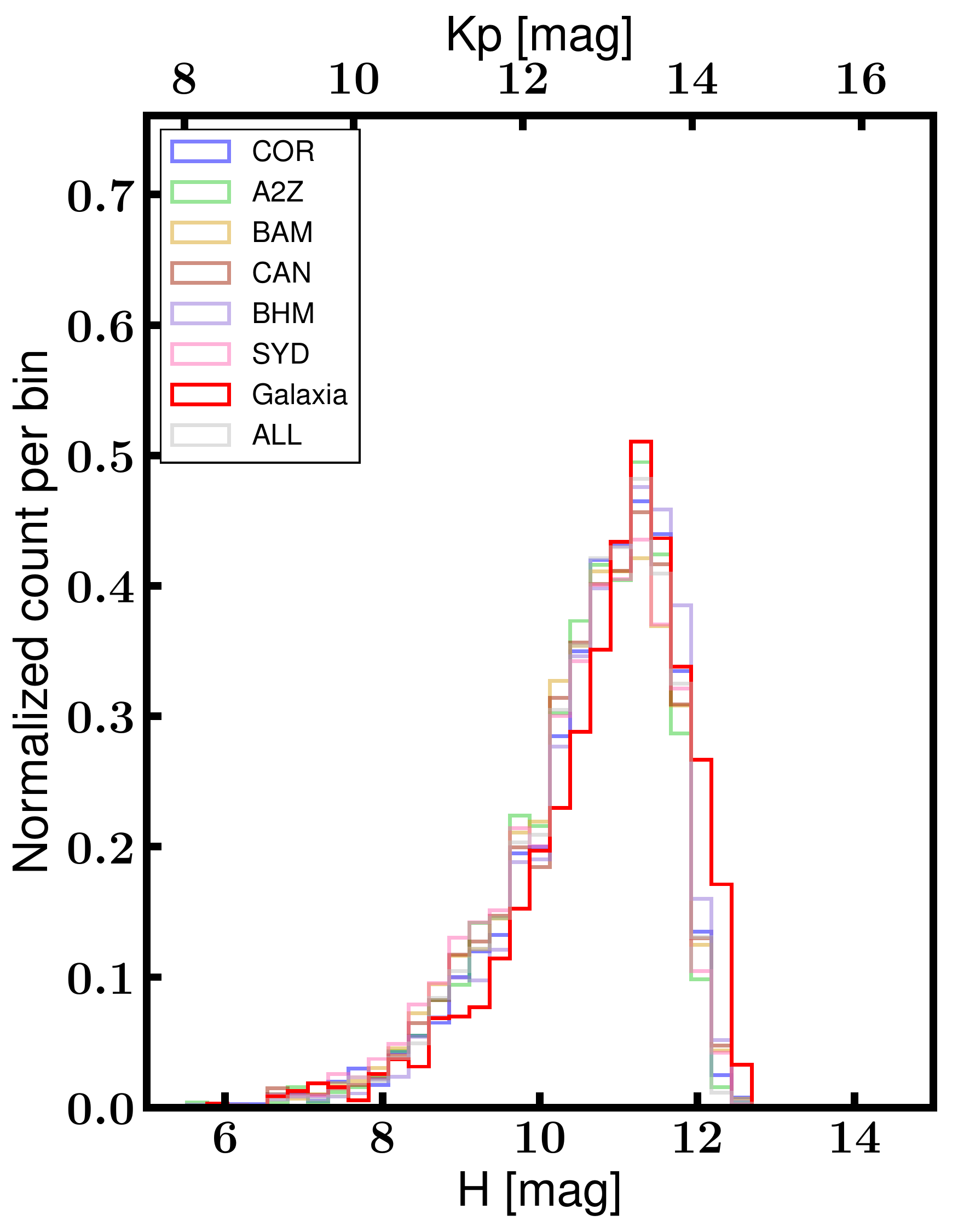}}
  \caption{Distributions of $H$-band magnitude for C4 (left), C6 (middle), and C7 (right), as predicted by \texttt{Galaxia} (red) and observed for each pipeline, according to the legend. The distribution of stars with $\numaxmean$ is labelled as ``ALL" --- this is not the same as summing the individual pipeline histograms because not every star will have a $\numaxmean$ since that requires at least two pipelines reporting values. The approximate $Kp$-band scale is indicated on the top x-axis.}
\label{fig:mag}
\end{figure*}

\texttt{Galaxia} models of the magnitude-$\numax$ distributions show good
agreement with the observations, as shown in
Figures~\ref{fig:mag_numax_c4}--\ref{fig:mag_numax_c7}. The most obvious feature in these plots is the RC, which, because red giants spend a relatively long amount of time in this phase, results in a ``clump" of stars at $\numax \sim 30 \muhz$. Lower-gravity
(lower-$\numax$) red giants oscillate with larger power than higher-gravity red giants,
and so at a fixed magnitude, it is easier to measure oscillations in
lower-gravity red giants. The amplitude of oscillations is primarily a function of surface gravity \citep{kallinger+2014}, and so the diagonal cutoff in the top right corner of both \texttt{Galaxia} predictions and (for the most part) observations is a result of the signal-to-noise ratio from the surface gravity--dependent oscillation amplitude compared to the magnitude-dependent white noise. To demonstrate this $\numax$-dependent white
noise limit, we assume a
detectability threshold of 
$\numaxdetect < 5\times10^4 1.6^{-H}$ in
Figures~\ref{fig:mag_numax_c4}--\ref{fig:mag_numax_c7} (dashed lines). This detectability threshold describes
the observed data well, and scales like a
flux-dependent white noise would.  In $Kp$-band space, this threshold would mean a detectability limit of $Kp \approx 15$,
fainter than which the white noise is too large to detect
high-$\numax$ oscillators. Compared to the detectability
threshold of
C1 described in \cite{stello+2017}, the dependence on magnitude is less steep, and likely
reflects the improved noise qualities following improved pointing control starting with C3. 
The reason the faint and bright limits do not form straight, vertical trends in Figures~\ref{fig:mag_numax_c4}--\ref{fig:mag_numax_c7}, which show $H$-band, is because we selected stars in $V$-band (see \S\ref{sec:selection}). For convenience, we also show the detection distributions as a function of only $H$-band magnitude in Figure~\ref{fig:mag}.

\begin{figure*}
\centering
  \includegraphics[width=0.45\textwidth]{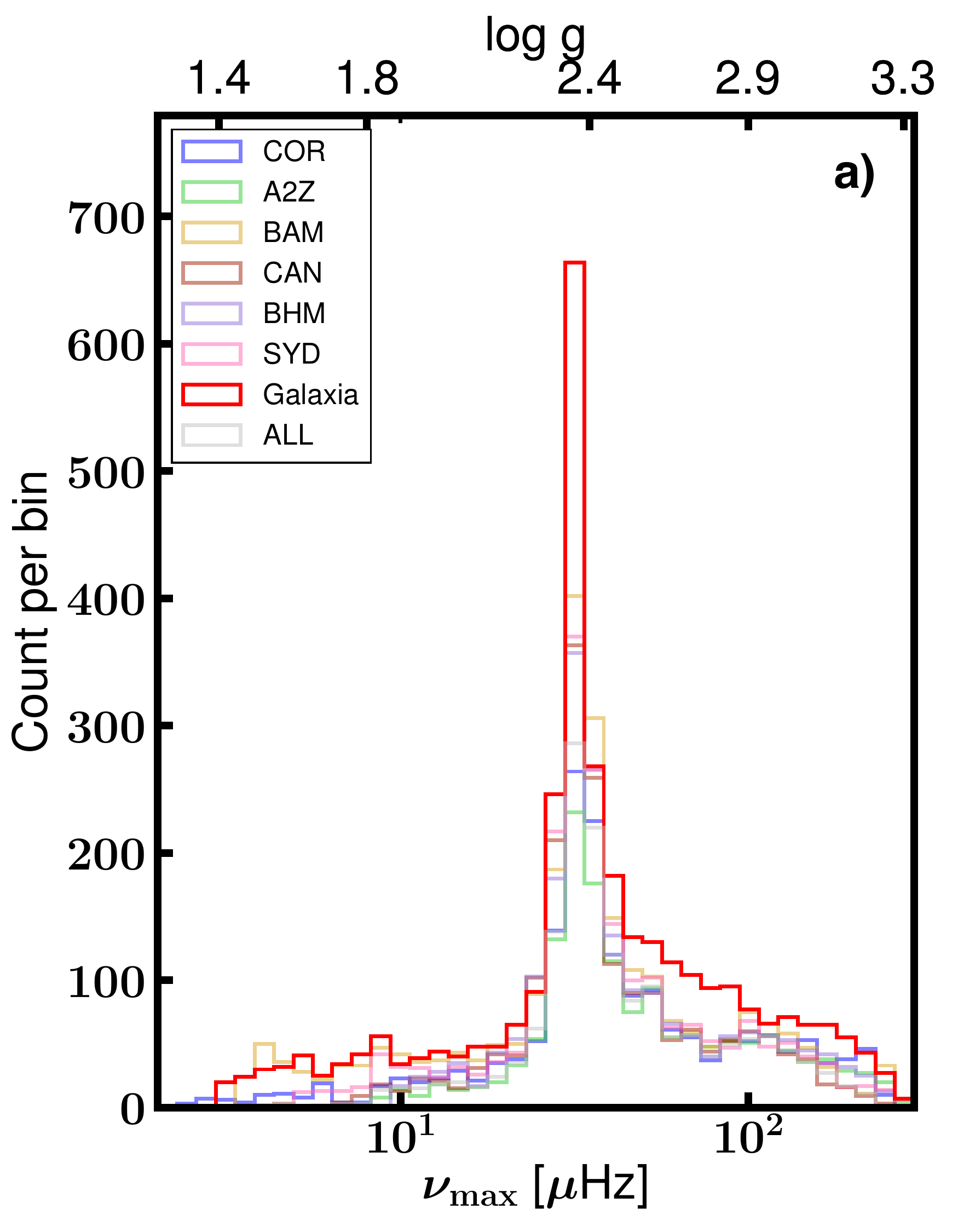}
  \includegraphics[width=0.45\textwidth]{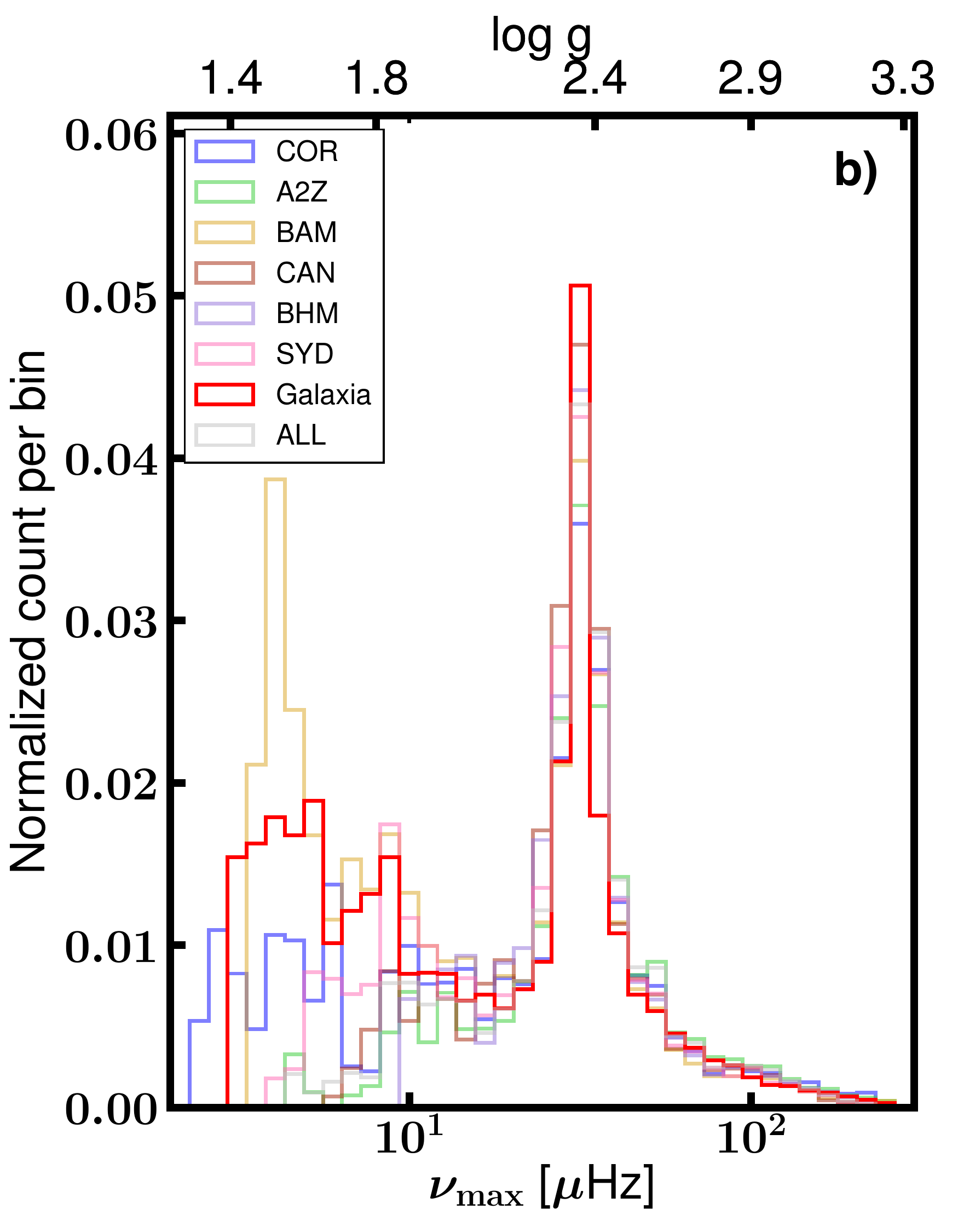}
  \caption{Left: Distributions of $\numax$ for stars that have been
    recovered by a particular pipeline (i.e., observed; colored according to the legend) compared to the \texttt{Galaxia} simulation of predicted detections for {\it K2}
    C4 (red).  The approximate
    asteroseismic surface gravity scale was computed with scaling
    relations according to Equation~\protect\ref{eq:scaling1}, and
    assuming a temperature of $4500$K. Right: Same as left, but showing normalized counts such that the distributions represent probability density.}
  \label{fig:c4_eye}
\end{figure*}

\begin{figure*}
\centering
    \includegraphics[width=0.45\textwidth]{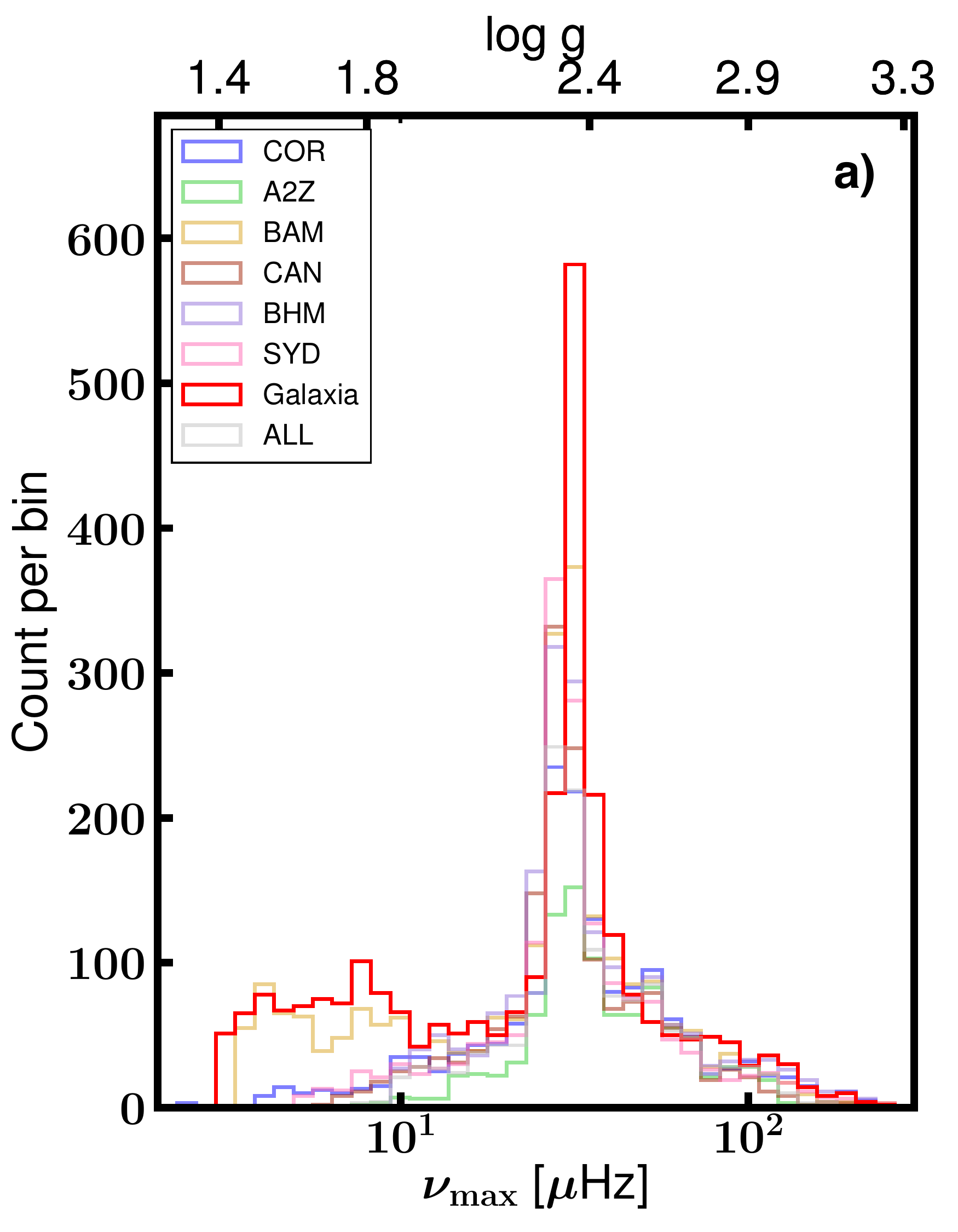}
  \includegraphics[width=0.45\textwidth]{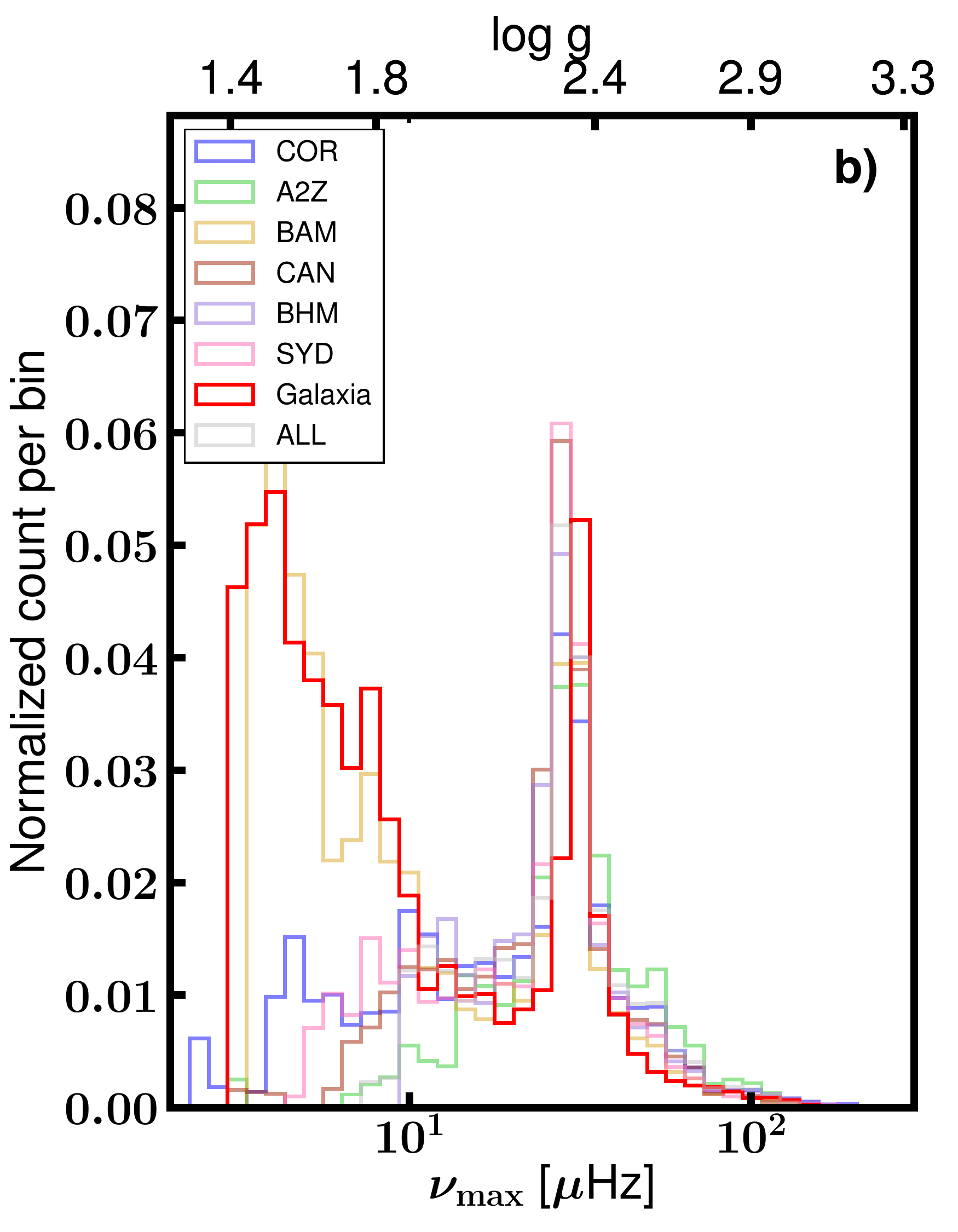}
  \caption{Same as Figure~\protect\ref{fig:c4_eye}, but for C7.}
  \label{fig:c7_eye}
\end{figure*}

We condense the comparison between observed and simulated asteroseismic values to just the $\numax$ dimension for
C4 and C7 in Figures~\ref{fig:c4_eye}~\&~\ref{fig:c7_eye}. The agreement is
generally good. There are two main discrepancies, however. First, the number of predicted oscillators in C4 does not agree with the observations (Figure~\ref{fig:c4_eye}a). However, plotting the normalized distribution such that each bin is divided by the bin size (representing probability density) results in agreement, except for the low-$\numax$ regime (Figure~\ref{fig:c4_eye}b). We discuss the discrepancy in Figure~\ref{fig:c4_eye}a below. Second, there are fewer observed low $\numax$ stars ($\numax \lesssim
10-20\muhz$) than predicted in both C4 and C7. Although \cite{stello+2017} found this same bias in {\it K2} GAP
DR1, we attempt to verify that it is not due to a bias in the
\texttt{Galaxia} models themselves by manually inspecting all
{\it K2} GAP spectra in C6 for evidence of solar-like oscillations. Two
experts did the exercise separately, classifying
the {\it K2} spectra according to whether or not they had a detectable
$\numax$ (yes/maybe/no), additionally assigning a visual estimate of
$\numax$. Objects that showed evidence of solar-like
oscillations below the effective high-pass cutoff of $3 \muhz$ were
classified as no. Note that this exercise did not require the detection of $\dnu$ for
classification as a solar-like oscillator, which is consistent with
our definition of a pipeline detection as a valid $\numax$ measurement
(see \S\ref{sec:extraction}). The results from each person were nearly
identical, with any contested classifications discussed individually
and a final consensus classification agreed upon.

\begin{figure*}
\centering
  \includegraphics[width=0.5\textwidth]{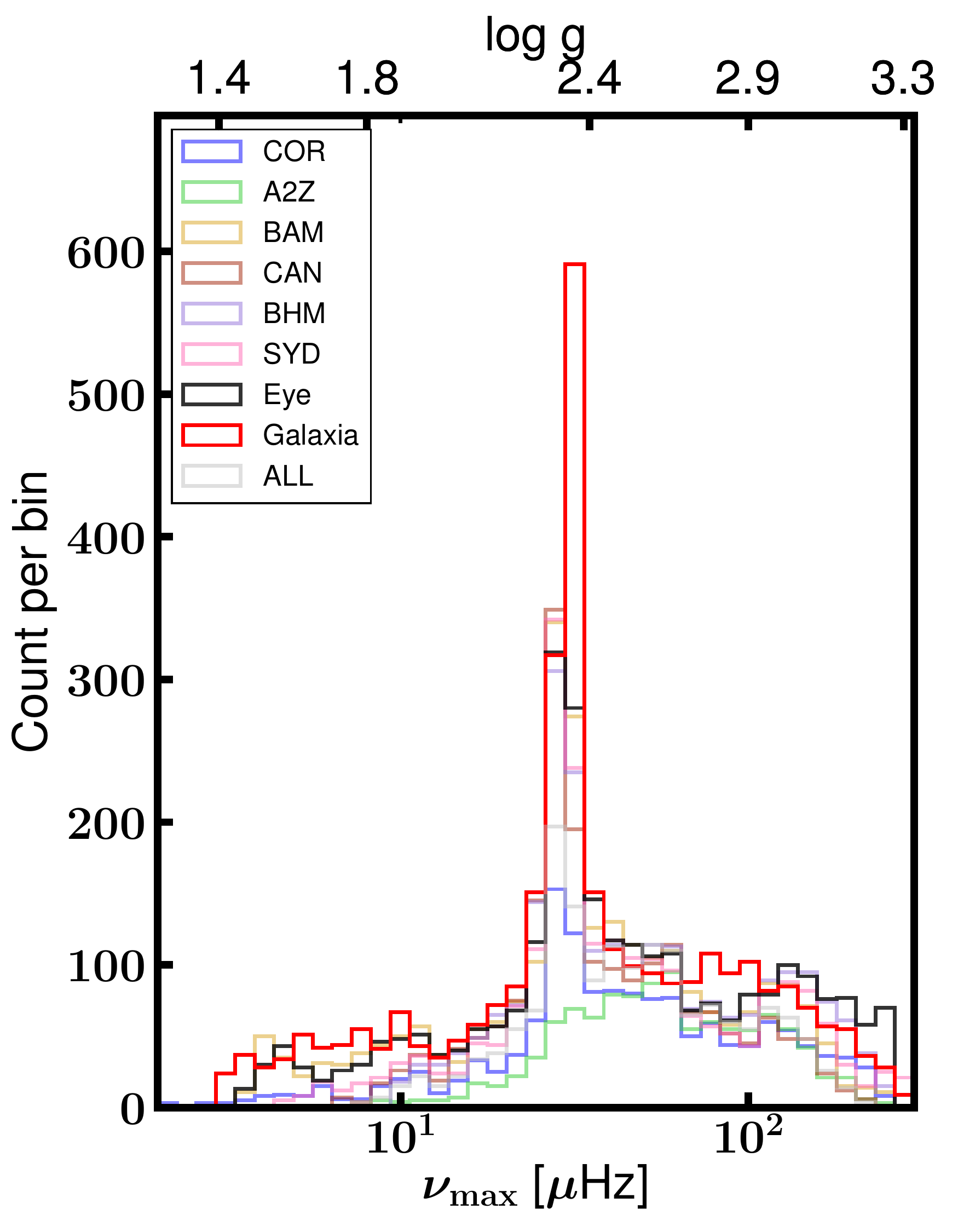}
  \caption{Distributions of $\numax$ for stars that have been
    recovered by a particular pipeline (i.e., observed; colored according to the legend); that have a
    visible $\numax$ as determined by eye (black); and that are expected
    to have a detectable $\numax$ according to the \texttt{Galaxia} simulation (red) for {\it K2}
    C6. The approximate asteroseismic surface gravity scale was computed with scaling
    relations according to Equation~\protect\ref{eq:scaling1}, and
    assuming a temperature of $4500$K.}
  \label{fig:c6_eye}
\end{figure*}

As evident in
Figure~\ref{fig:c6_eye}, the predicted $\numax$ distribution from \texttt{Galaxia} is in good agreement with the visually confirmed
distribution (``yes"s are plotted, not ``maybe"s), though they are formally inconsistent with being drawn from
the same distribution, according to a Kolmogorov-Smirnov
test. By going through all of the observed
targets by eye and not just ones that were returned by pipelines as being red
giants, we are able to be
more confident that the \texttt{Galaxia} predictions are robust and
not subject to obvious biases.
The distributions of $\numax$ returned by individual pipelines all
fall short of being formally consistent with either the visually-confirmed distribution or the \texttt{Galaxia} distribution, though
some of the pipeline $\numax$ distributions qualitatively show good agreement with the predicted and visually-confirmed distributions.

As noted above, there is nonetheless a bias against
detecting stars with $\numax$, below $\sim 10-20\muhz$. This is true for all three of the campaigns, which
indicates that this detection bias was not solely a function of the
particular DR1 sample, which were all in C1. With the data we have in
hand, we cannot definitively say what causes this bias. At these low frequencies, there are relatively few modes observable. This may hinder detecting these oscillations for pipelines that rely on finding $\dnu$ as part of its $\numax$ detection step, and can also hamper fits to the power excess using a Gaussian since a Gaussian does not very well describe a few discrete modes. It is also possible that low-$\numax$ oscillators are harder to recover at lower frequencies because of the relatively short \textit{K2} dwell time, which leads to a smaller ratio of the frequency resolution to the mode width; the result is that the spectra of these stars can sometimes be hard to distinguish from a pure granulation background. The degree to which these oscillators are detected or not is highly pipeline-dependent, as one can see from Figures~\ref{fig:c4_eye}-\ref{fig:c6_eye}.

Keeping this low-$\numax$ detection bias in mind, we can go on to test the detection rate for $\numax > 20 \muhz$, using \texttt{Galaxia} and visual inspection as ground truths. In C7, there are 1740 such stars that have asteroseismic values from at least one pipeline. This number is consistent with the 1758 expected stars from the \texttt{Galaxia} simulation for C7. The 2312 stars recovered by at least one pipeline in C6 is between the number found by visual inspection, 2214, and that predicted by Galaxia, 2511.  In C4, however, there are significantly fewer stars observed by at least one pipeline than predicted by \texttt{Galaxia} (2177 versus 2670). These `missing' stars are at magnitudes $Kp < 13$, and therefore should yield observable asteroseismic parameters, given our empirical detectability threshold. The \texttt{Galaxia} model seems to under-predict reddening in this campaign, which may explain this discrepancy. Indeed, the predicted fraction of dwarfs (and therefore the fraction of stars that would not be detected as oscillators in long-cadence \textit{K2} data) depends on the assumed reddening. For \textit{K2} GAP, the selection `draws a line' in $J-K_{\mathrm{s}}$ color space to separate the sample of predominantly blue dwarfs from the sample containing the red giants (and some red dwarfs), which are the targets with $J-K_{\mathrm{s}} > 0.5$. The same cut is applied to the synthetic stellar population in the \texttt{Galaxia} simulation, using an assumed reddening. An under-estimated reddening in \texttt{Galaxia} means that fewer of the blue dwarfs are reddened enough to fall on the giant side of the $J-K_{\mathrm{s}} > 0.5$ dwarf/giant dividing line, increasing the number of predicted oscillators compared to reality.

\subsection{Absolute radius calibration}
\label{sec:tgas}

The derived $\langle \dnu' \rangle$ and
$\langle \numax' \rangle$ have not necessarily been placed on an absolute scale. While we
scaled the asteroseismic values to a common mean scale, we
were free to impose solar reference values for $\numaxmean$ and
$\dnumean$ to be $\numaxsun = 3076\muhz$ and $\dnusun = 135.146\muhz$, which are relatively close to the average pipeline-specific solar reference values, and which are the same as determined by the absolute asteroseismology re-scaling done in the APOKASC-2 analysis. Were we working with \textit{Kepler} data and using the same pipelines as in the APOKASC-2 analysis, this choice of solar reference values would be valid and would put the asteroseismic radii on a scale that is consistent with open cluster masses and \textit{Gaia} radii \citep{pinsonneault+2018,zinn+2019a}. There are at least two reasons why choosing our solar reference values as $\numaxsun = 3076\muhz$ and $\dnusun = 135.146\muhz$ may not result in $\kapparmean$ and $\kappammean$ being on an absolute scale. First, although our re-scaling procedure to derive $\langle \dnu' \rangle$ and $\langle \numax' \rangle$ is nearly the same as in \cite{pinsonneault+2018}, we have added BAM as one of the pipelines that contributes to the re-scaling procedure: APOKASC-2 used results from A2Z, CAN, COR, SYD, and OCT, and in this work, we have used results from A2Z, CAN, COR, SYD, BHM (based on OCT), and BAM (see \S\ref{sec:extraction} for summaries of the pipeline methodologies). The addition of BAM in this work to the pipelines used for aggregating asteroseismic results may result in a slightly different mean scale for $\numaxmean$ and $\dnumean$. This is because the APOAKSC-2 solar reference values were chosen such that the aggregated stellar masses agreed with open cluster masses --- using a different set of pipelines to average over may have required a different set of solar reference values to achieve agreement with open cluster masses. We see this in the differences between the re-scaling values from our analysis and from that of \cite{pinsonneault+2018}, shown in Table~\ref{tab:refs}. Second, there is evidence to suggest that there are systematic biases in asteroseismic parameters based on the dwell time of the data (Zinn et al., in prep.), which could be suggestive of a need to modify the $\numaxsun$ and/or $\dnusun$ for {\it K2} data compared to \textit{Kepler} data. We now test this choice of zero-points by comparing the derived radii to radii using parallaxes from \textit{Gaia} Data Release 2 \citep{gaia2018,lindegren+2018}. 

We populated the overlap sample of stars in \textit{K2} GAP DR2 with both $\dnumean$ and $\numaxmean$
and {\it Gaia} DR2 by matching on 2MASS ID using the \textit{Gaia}
Archive\footnote{\url{https://gea.esac.esa.int/archive/}}. We also
required APOGEE metallicities and temperatures from DR16 \citep{apogeedr16_2019}. Because of a known, position, magnitude, and color-dependent zero-point in the \textit{Gaia} parallaxes \citep[e.g.,][]{lindegren+2018,zinn+2019}, we did not work directly with the \textit{Gaia} DR2 parallaxes. Instead, we followed the methodology of \cite{schoenrich+2019a} to derive distance estimates for stars in our {\it K2} GAP sample. We did this separately for RC and RGB stars, with the understanding that RGB and RC populations will have different selection functions, which is an important consideration in the Bayesian distance estimates in the \cite{schoenrich+2019a} framework \citep[see also][]{schoenrich_aumer2017}. 

For the purposes of establishing a \textit{Gaia} calibration of $\numaxsun$ and $\dnusun$, we used only stars with more than two pipelines returning results for $\dnu$, and only considered stars with $\pi > 0.4$ and \textit{Gaia} $G$-band $< 13$mag to ensure that the results are less sensitive to any residual \textit{Gaia} parallax zero-points that may not be accounted for in the \cite{schoenrich+2019a} method; [Fe/H]$ > -1$ to ensure that there are no metallicity-dependent asteroseismic radius systematics (see \citealt{zinn+2019a}); and $R < 30\rsun$ to ensure there are no radius-dependent asteroseismic radius systematics (see \citealt{zinn+2019a}). For this sample, which has spectroscopic information, we re-compute $\dnu$ correction factors using metallicities that are adjusted to account for non-solar alpha abundances according to the \cite{salaris+1993} prescription: 
$\mathrm{[Fe/H]}' = \mathrm{[Fe/H]} + \log_{10}(0.638 \times 10^{\mathrm{[\alpha/M]}} + 0.362)$. 
We computed a \textit{Gaia} radius for the 261
resulting stars following a Monte Carlo
procedure of the sort detailed in \cite{zinn+2017b}. The method uses the Stefan-Boltzmann law to translate the flux, temperature, and distance of a star into a radius. To do so, we
computed bolometric fluxes with a $K_{\mathrm{s}}$-band bolometric
correction \citep{gonzalezhernandez&bonifacio2009}, APOGEE effective temperatures, and
metallicities, combined with asteroseismic surface gravities from $\numaxmean$. Extinctions were computed using the three-dimensional dust map of \protect\cite{green+2015}, as
implemented in \texttt{mwdust}\footnote{\url{https://github.com/jobovy/mwdust}} \citep{bovy+2016}. Five stars with asteroseismic and \textit{Gaia} radii discrepant at more than the 3$\sigma$ level were removed from subsequent analysis.

\begin{figure*}
\centering
\includegraphics[width=0.5\textwidth]{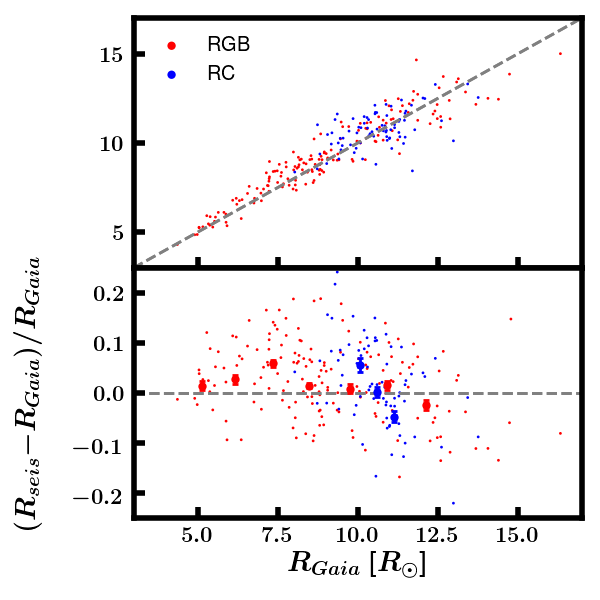}
\caption{Comparison of asteroseismic scaling relation radii
  (Eq.~\ref{eq:radius}) using {\it K2} GAP DR2 $\kapparmean$ in combination with APOGEE effective temperatures and {\it Gaia} DR2 radii derived based on corrected \textit{Gaia} parallaxes (see text). Grey
  dashed lines show one-to-one relations. In the bottom panel, blue
  error bars indicate weighted averages and standard errors on the weighted averages for
  RC stars (blue) and RGB stars (red).}
\label{fig:tgas}
  \end{figure*}

We compare the {\it Gaia} radius scale to our {\it K2} asteroseismic radius scale in Figure~\ref{fig:tgas}. The top panel shows the points colored by
evolutionary state (RGB in red and RC in blue), and the bottom panels shows the fractional
agreement of the two radius scales. Figure~\ref{fig:tgas} indicates that the radius scale of $\kapparmean$ is consistent
with the {\it Gaia} radius scale for both RC and RGB stars to within $\sim 3\%$. We find that the RGB stars are in more disagreement than the RC stars: while the median agreement is $3.0\% \pm 0.4\%$ for RGB stars, it is $1.6\% \pm 0.8\%$ for RC stars. This median statistic is computed as the median radius ratio for all RGB or RC stars, with the uncertainty in the median taken to be $\sigma_{\rm{med}} = \sqrt{ \frac{\pi}{2} \sum \sigma_R^2/N^2}$, where $\sigma_R = \frac{R_{\rm{seis}}}{R_{Gaia}} \sqrt{ \left( \frac{\sigma_{R, Gaia}}{R_{Gaia}} \right )^2 + \left( \frac{\sigma_{R,\rm{seis}}}{R_{\rm{seis}}}\right)^2}$. Using the EPIC extinctions instead of those from \cite{green+2015} leads to insignificant variations in the radius agreement. However, the agreement is discrepant at the $\sim 3\sigma$ level between stars in C4 versus those in C7. This could be an indication of \textit{Gaia} parallax zero-point issues, given we do not expect such variations in the asteroseismic data by campaign. We therefore also consider the \textit{Gaia} zero-point from \cite{khan+2019}, who compared asteroseismic distances to \textit{Gaia} distances in \textit{K2} C3 and C6. For this exercise, we restricted our sample to the stars in C6, and adopted their derived zero-point of $-17\mu as$. The result is that the asteroseismic radii are consistent with \textit{Gaia} radii to within $\approx 1\%$ (RGB) and $\approx 5\%$ (RC). 

Because the median agreement between the radius scales could be biased by underlying skewed distributions of the individual radii, we finally evaluate the agreement using a weighted mean: $\langle R_{\rm{seis}}/R_{Gaia} \rangle = \frac{\sum \frac{R_{\rm{seis}}}{R_{Gaia}} / \sigma^2_R}{\sum{1/\sigma^2_R}}$, where $\sigma_R = \frac{R_{\rm{seis}}}{R_{Gaia}} \sqrt{ \left( \frac{\sigma_{R, Gaia}}{R_{Gaia}} \right )^2 + \left( \frac{\sigma_{R,\rm{seis}}}{R_{\rm{seis}}}\right)^2}$. We calculate this for those stars that have fractional parallax uncertainties less than $10\%$, in order to mitigate potential biases due to parallax systematics. According to this metric, the agreement becomes $2.2 \pm 0.3\%$ for RGB stars and $2.0 \pm 0.6\%$ for RC stars.

Due to the variation in this agreement based on the tests described above, we opt not to re-scale our $\numaxmean$ or $\dnumean$ values, and instead allow for a systematic zero-point uncertainty in our derived $\kapparmean$ values. Acknowledging these uncertainties in the \textit{K2}-\textit{Gaia} agreement, we use the weighted mean estimate of the radius scales using the \cite{schoenrich+2019a} \textit{Gaia} distances. Our asteroseismic radius coefficients could therefore be over-estimated by up to $2.2 \pm 0.3\%$ for RGB stars and up to $2.0 \pm 0.6\%$ for RC stars. The uncertainty on this agreement is solely due to the standard uncertainty on the mean, and does not account for intrinsic scatter or trends in the radius agreement. Indeed, this should be thought of as being in addition to the systematic uncertainty from pipeline-to-pipeline variation in $\kapparmean$ as a function of $\numax$ and $\dnumean$ discussed in \S\ref{sec:syst_mean}. We also note that this agreement does not account for systematic variation in the radius ratio due to choice of temperature, bolometric correction, or \textit{Gaia} zero-point, which may contribute to a systematic uncertainty of about $\pm 2\%$ \citep{zinn+2019a}. 

Broadly speaking, the excellent level of agreement between asteroseismology and \textit{Gaia} corroborates findings of the accuracy of the scaling relations in this regime from previous work based on asteroseismology-\textit{Gaia} comparisons \citep{huber+2017,zinn+2019a}. In detail, there do appear to be trends with radius evident in Figure~\ref{fig:tgas}: the RGB radii appear to inflate compared to \textit{Gaia} radii at around $R \sim 7.5\rsun$ (red error bars), while the red clump radii (blue error bars) seem to deflate compared to the \textit{Gaia} radii with increasing radius at all radii. At least for the RC stars, systematics in the tracks used to generate the $\dnu$ corrections could be to blame, particularly given the disagreement among RC models from the literature \citep{an+2019a}. Indeed, it appears that the RC radius trend is mostly a trend in $\dnu$, with some metallicity dependence, as well. At the population level, the RC radii have been found to agree within $5\%$ with \textit{Gaia} radii \citep{hall+2019a}. However, to our knowledge, the scaling relations for RC stars have not been tested as a function of radius as we do here. We note also that a $\approx 1\%$ relative difference between the zero-point for RGB versus RC stars is not ruled out by \cite{pinsonneault+2018} \citep[see also][]{khan+2019}.

Regarding the mean agreement of RGB and RC stars, the most likely culprit for the (small) radius disagreement is a systematic in the solar reference value combination $\numaxsun/\dnusun^2$. That the absolute \textit{K2} asteroseismic radius scale is consistent to within $\sim
2\%$ with the {\it Gaia} radius scale naively implies that our $\kappammean$ are within
$\sim 6\%$ of an absolute mass scale, according to standard
propagation of error from Equation~\ref{eq:radius} to \ref{eq:mass}. However, this is only approximate, because we can test only $\numaxsun/\dnusun^2$ against \textit{Gaia}, whereas the mass coefficient goes as $\numaxsun^3/\dnusun^4$. 

In summary, because we are delivering asteroseismic values $\numaxmean$ and $\dnumean$ that are averaged values from several pipelines, we needed to evaluate to what extent the resulting asteroseismic scale is on an absolute scale. We did this by comparing to \textit{Gaia} radii, and we found a systematic offset between the {\it K2} GAP DR2 asteroseismic and \textit{Gaia} radius scales of about $2\%$. This translates roughly to a $20\%$ systematic uncertainty in age. Given the typical uncertainty in mass listed in Table~\ref{tab:kepk2}, ages based on our RGB asteroseismic masses would be expected to have statistical uncertainties of $\approx 20\%$, with potential scale shifts by $\approx 20\%$ due to the level of systematics we identify in this section. This anticipated $20\%$ statistical age uncertainty makes the data particularly interesting for potentially identifying the history of minor mergers in the Galaxy based on their impact on the age--velocity dispersion relation \citep{martig_minchev_flynn2014}. For this and other Galactic archaeology applications (e.g., age-abundance patterns), the $20\%$ systematic uncertainty should not be significant, given that the differential age relationship between stellar populations would be preserved. Regarding mass- or magnitude-dependent systematics among RGB asteroseismic parameters, the small inflation of RGB asteroseismic radii at $R\sim 7.5\rsun$ seems to map onto a corresponding trend in $\numax$, and so this may introduce an inflation in the RGB mass scale by perhaps up to $15\%$ for the minority of stars with $50 \muhz \lesssim \numaxmean \lesssim 80 \muhz$. RC stars in our sample, on the other hand, appear to suffer from strong radius-dependent trends that seem to be related to the $\dnu$ and not $\numax$ scaling relation: there is a strong trend of RC agreement with $\dnu$, which suggests that the $\dnu$ scaling relation for RC stars is not well-calibrated using our $\dnu$ corrections. Despite the concerning magnitude of the RC systematic, RC ages are not in popular use because of uncertainties in modeling mass loss \citep{casagrande+2016}. We will nonetheless explore the RC systematic further in the next and final {\it K2} GAP data release, as having accurate red clump masses and radii is important for reckoning red clump models with observed red clump properties \citep[e.g.,][]{an+2019a}.

\section{Conclusion}
\label{sec:conclusion}
We have described the second data release of {\it K2} GAP, containing red giants
for campaigns 4, 6, and 7. We have derived evolutionary state
classifications for our sample, and have placed the raw asteroseismic
observations on a self-consistent scale, resulting in 4395 stars
with mean asteroseismic parameters. We have also provided ready-to-use derived
quantities, $\kappammean$ and $\kapparmean$, for these stars, which yield masses and radii
when combined with a weakly temperature-dependent factor that users
may compute with their preferred effective temperature. We conclude
the following:

\begin{enumerate}
  \item The observed {\it K2} GAP targets in campaigns 4, 6, and 7 have reproducible selection functions, which enable them for use in Galactic archaeology studies.
    \item $\numax$- and $\dnu$- dependent trends among pipelines have
      been improved by bringing the pipelines onto a common
      scale. This re-scaling process effectively changes
      pipeline-specific solar reference values at or below the 1\% level, in
      different measures, depending on the pipeline and the
      evolutionary state of the star.
      \item We provide empirical uncertainties in $\numax$ and $\dnu$ values for stars that have results from at least two pipelines, which have statistically reasonable distributions, and which indicate that fractional uncertainties are not strong functions of $\numax$ or $\dnu$ or the number of pipelines reporting, but rather vary mostly according to evolutionary state: RGB stars have better-measured parameters than do RC stars. Systematic uncertainties for $\numax$ and $\dnu$ values are similar across evolutionary state, at $\sim 0.6\%$ and $\sim 0.3\%$ for both RGB and RC stars.
      \item The distributions of our mean $\numax$ are in good
        agreement with those predicted by theoretical stellar population synthesis models. Crucially, both the
        observed and predicted $\numax$ distributions globally agree with an
        unbiased estimate of the $\numax$ distributions from manual inspection of the data, which
        indicates our model predictions are accurate, and our observed
        samples are largely complete. A notable exception to the asteroseismic detection completeness is for red giants with $\numax \lesssim
          10-20\muhz$, where pipelines may report lower-than-expected
          numbers of oscillating stars.
          \item The radius and mass coefficients that we provide, $\kapparmean$ and $\kappammean$, have typical uncertainties of $\sigma_{\kapparmean} = 3.3\%$ (RGB or RGB/AGB) \& $\sigma_{\kapparmean} = 5.0\%$ (RC) and $\sigma_{\kappammean} = 7.7\%$ (RGB or RGB/AGB) \& $\sigma_{\kappammean} = 10.5\%$ (RC). These uncertainties are a factor of two to three higher than the uncertainties from \textit{Kepler} radii and masses.
    \item Our asteroseismic radii have been validated to be on the {\it Gaia} radius scale, to within $2.2 \pm 0.3\%$ for RGB stars and $2.0\% \pm 0.6\%$ for RC stars.
\end{enumerate}

{\it K2} GAP Data Release 2 successfully builds upon {\it K2} GAP DR1 in
providing evolutionary state information; re-scaled asteroseismic parameters and uncertainties that take advantage of
the information from multiple asteroseismic pipelines;
radius and mass coefficients; and placing the radius coefficients on
an absolute scale. Future work will focus on calibrating the mass
coefficients, which at this point cannot be definitively placed on an
absolute scale, for lack of convenient mass calibrators in the
sample. In the next {\it K2} GAP data release, we may be able to place
masses on an absolute scale, as we have done for radii in this
work, by appealing to the red giant branch mass of open clusters observed by {\it K2}.

\newpage

\newpage

\begin{longrotatetable}
\begin{splitdeluxetable*}{lccccccccccccccBccccccccccccl}
  \tablecaption{Raw asteroseismic $\numax$ and $\dnu$ values for {\it K2} GAP
  DR2 for each pipeline, with evolutionary states \label{tab:raw}}
  \tabletypesize{\footnotesize}
\tablehead{EPIC & Campaign & Priority & Evo. State & $\nu_{\mathrm{max,A2Z}}$ &
  $\sigma_{\nu\mathrm{max,A2Z}}$ & $\dnu_{\mathrm{A2Z}}$ &
  $\sigma_{\dnu,\mathrm{A2Z}}$ & $\nu_{\mathrm{max,BAM}}$ &
  $\sigma_{\nu\mathrm{max,BAM}}$ & $\dnu_{\mathrm{BAM}}$ &
  $\sigma_{\dnu,\mathrm{BAM}}$ & $\nu_{\mathrm{max,BHM}}$ &
  $\sigma_{\nu\mathrm{max,BHM}}$ & $\dnu_{\mathrm{BHM}}$ &
  $\sigma_{\dnu,\mathrm{BHM}}$ & $\nu_{\mathrm{max,CAN}}$ &
  $\sigma_{\nu\mathrm{max,CAN}}$ & $\dnu_{\mathrm{CAN}}$ &
  $\sigma_{\dnu,\mathrm{CAN}}$ & $\nu_{\mathrm{max,COR}}$ &
  $\sigma_{\nu\mathrm{max,COR}}$ & $\dnu_{\mathrm{COR}}$ &
  $\sigma_{\dnu,\mathrm{COR}}$ & $\nu_{\mathrm{max,SYD}}$ &
  $\sigma_{\nu\mathrm{max,SYD}}$ & $\dnu_{\mathrm{SYD}}$ &
  $\sigma_{\dnu,\mathrm{SYD}}$}
\startdata
& & & &$\muhz$ & $\muhz$ & $\muhz$ & $\muhz$ & $\muhz$ & $\muhz$ & $\muhz$
& $\muhz$ & $\muhz$ & $\muhz$ & $\muhz$ & $\muhz$  & $\muhz$ & $\muhz$
& $\muhz$ & $\muhz$ & $\muhz$ & $\muhz$ & $\muhz$ & $\muhz$ & $\muhz$
& $\muhz$ & $\muhz$ & $\muhz$ \\
210306475 &         4 &       903 & RGB  &  28.550 & 2.25 &   3.620 &   0.010 &  30.135 & 0.808 &   3.277 &   0.197 &  26.900 & 1.4 &   3.470 &   0.160 &  27.670 & 1.27 &     --- &     --- &     --- & --- &     --- &     --- &  29.016 & 0.63245 &   3.450 &   0.170\\
210307958 &         4 &      2771 & RGB  &  27.940 & 2.47 &   3.890 &   0.000 &  30.316 & 0.723 &   3.723 &   0.124 &     --- & --- &     --- &     --- &  29.260 & 1.03 &   4.004 &   0.094 &  28.350 & 0.83 &   3.805 &   0.083 &  28.789 & 0.62378 &   3.740 &   0.203\\
210314854 &         4 &      1141 & RGB  &  29.030 & 2.03 &   4.100 &   0.120 &  31.863 & 0.743 &   4.115 &   0.083 &  31.200 & 1.0 &   4.090 &   0.190 &  30.600 & 1.49 &   4.066 &   0.056 &  31.640 & 0.89 &   4.204 &   0.079 &  31.044 & 0.88666 &   4.033 &   0.238\\
210315825 &         4 &      1651 & RGB  &  59.580 & 3.63 &   5.910 &   0.160 &  59.253 & 1.379 &   5.899 &   0.060 &  60.000 & 1.3 &   6.010 &   0.190 &  59.390 & 1.28 &   5.960 &   0.084 &  58.960 & 1.26 &   5.975 &   0.087 &  59.213 & 1.21725 &   5.873 &   0.058\\
210318976 &         4 &       988 & RGB  &  24.800 & 1.88 &   3.270 &   0.040 &  25.080 & 0.841 &     --- &     --- &  21.700 & 1.1 &   3.080 &   0.100 &  23.660 & 1.22 &   3.482 &   0.090 &  24.540 & 0.73 &   3.402 &   0.076 &  24.144 & 0.77871 &     --- &     ---\\
\enddata
\tablecomments{These are the parameters returned by
  a given pipeline, along with their uncertainties, without any
  of the re-scaling described in \S\ref{sec:zeropoint}
  applied. Evolutionary states are also included in this table, which
  have been derived in this work (see \S\ref{sec:evo}). If classified, a
  star's evolutionary state is assigned as either ``RGB'', ``RGB/AGB", or ``RC''. ``Priority" refers to the {\it K2} GAP target priority discussed in \S\ref{sec:data} (a smaller numerical value corresponds to higher priority); serendipitous targets do not have a populated priority entry. A full version of this table is available in the online journal.}
\end{splitdeluxetable*}
\end{longrotatetable}

\begin{deluxetable*}{cccccccc}
  \tablecaption{Numbers of stars with raw asteroseismic values ($\numax$, $\dnu$), re-scaled asteroseismic values ($\numax'$, $\dnu'$), and radius~\&~mass coefficients ($\kappa_R'$, $\kappa_M'$), as a function of pipeline and campaign \label{tab:num}}
  \tabletypesize{\footnotesize}
  \tablehead{ & &  $\numax$  &  $\numax'$ & $\dnu$ & $\dnu'$ & $\kappa_R'$ & $\kappa_M'$ }
  \startdata
C4 & A2Z & 1536 & 1375 & 1536 & 1331 & 1331 & 1331 \\ \hline 
C6 & A2Z & 1086 & 1018 & 1086 & 985 & 985 & 985 \\ \hline 
C7 & A2Z & 993 & 912 & 293 & 279 & 279 & 279 \\ \hline 
Total & A2Z & 3615 & 3305 & 2915 & 2595 & 2595 & 2595 \\ \hline 
C4 & BAM & 2478 & 1480 & 844 & 741 & 741 & 741 \\ \hline 
C6 & BAM & 2529 & 1515 & 955 & 844 & 844 & 844 \\ \hline 
C7 & BAM & 2315 & 1267 & 677 & 589 & 589 & 589 \\ \hline 
Total & BAM & 7322 & 4262 & 2476 & 2174 & 2174 & 2174 \\ \hline 
C4 & BHM & 1984 & 1414 & 1529 & 1189 & 1189 & 1189 \\ \hline 
C6 & BHM & 2275 & 1482 & 1702 & 1229 & 1229 & 1229 \\ \hline 
C7 & BHM & 1803 & 1231 & 1238 & 1019 & 1019 & 1019 \\ \hline 
Total & BHM & 6062 & 4127 & 4469 & 3437 & 3437 & 3437 \\ \hline 
C4 & CAN & 1897 & 1395 & 968 & 788 & 788 & 788 \\ \hline 
C6 & CAN & 1956 & 1420 & 1455 & 1189 & 1189 & 1189 \\ \hline 
C7 & CAN & 1564 & 1137 & 1048 & 889 & 889 & 889 \\ \hline 
Total & CAN & 5417 & 3952 & 3471 & 2866 & 2866 & 2866 \\ \hline 
C4 & COR & 1803 & 1374 & 1803 & 1304 & 1304 & 1304 \\ \hline 
C6 & COR & 1443 & 1118 & 1443 & 1043 & 1043 & 1043 \\ \hline 
C7 & COR & 1561 & 1188 & 1561 & 1149 & 1149 & 1149 \\ \hline 
Total & COR & 4807 & 3680 & 4807 & 3496 & 3496 & 3496 \\ \hline 
C4 & SYD & 2136 & 1416 & 853 & 695 & 695 & 695 \\ \hline 
C6 & SYD & 2207 & 1335 & 868 & 727 & 727 & 727 \\ \hline 
C7 & SYD & 1675 & 1089 & 584 & 503 & 503 & 503 \\ \hline 
Total & SYD & 6018 & 3840 & 2305 & 1925 & 1925 & 1925 \\ \hline 
\enddata
\end{deluxetable*}

\begin{longrotatetable}
\begin{splitdeluxetable*}{lcccccccccccBcccccccccccccl}
  \tablecaption{Derived asteroseismic $\numax$ and $\dnu$ values for {\it K2} GAP
  DR2 \label{tab:vals}}
  \tabletypesize{\footnotesize}
  \tablehead{EPIC & $\langle \numax' \rangle$ & $\sigma_{\langle \numax'
      \rangle}$ & $N_{\nu\mathrm{max}}$ & $\langle\dnu'\rangle$ &
    $\sigma_{\langle\dnu'\rangle}$ & $N_{\dnu}$ & $X_{\mathrm{Sharma}}$ & $\sigma_{X{\mathrm{Sharma}}}$
    & $\langle\dnu\rangle$ & $\nu_{\mathrm{max,A2Z}}'$ & $\dnu_{\mathrm{A2Z}}'$ &
    $\nu_{\mathrm{max,BAM}}'$ & $\dnu_{\mathrm{BAM}}'$ &
    $\nu_{\mathrm{max,BHM}}'$ & $\dnu_{\mathrm{BHM}}'$ &
    $\nu_{\mathrm{max,CAN}}'$ & $\dnu_{\mathrm{CAN}}'$ &
    $\nu_{\mathrm{max,COR}}'$ & $\dnu_{\mathrm{COR}}'$ &
    $\nu_{\mathrm{max,SYD}}'$ & $\dnu_{\mathrm{SYD}}'$ & EPIC $T_{\rm{eff}}$ & $\sigma_{T}$ & EPIC [Fe/H] & $\sigma_{\rm{[Fe/H]}}$}
\startdata 
 & $\muhz$ & $\muhz$ & & $\muhz$ & $\muhz$ & & & & $\muhz$ & $\muhz$ & $\muhz$ & $\muhz$ & $\muhz$ & $\muhz$ & $\muhz$ & $\muhz$ & $\muhz$ & $\muhz$ & $\muhz$ & $\muhz$ & $\muhz$ & K & K & & \\  
210306475 &  28.487 &   1.235 &     5 &   3.610 &   0.099 &     3 &   1.027 &   0.010 &   3.517 &  28.502 &   3.723 &  30.004 &     --- &  26.764 &   3.558 &  27.948 &     --- &     --- &     --- &  29.218 &   3.547 & 4797 &  134 & -0.27 & 0.30 \\
210307958 &  28.974 &   0.935 &     5 &   3.957 &   0.123 &     5 &   1.032 &   0.013 &   3.834 &  27.893 &   4.090 &  30.184 &   3.853 &     --- &     --- &  29.554 &   4.110 &  28.249 &   3.926 &  28.990 &   3.867 & 4750 &  138 & -0.36 & 0.26 \\
210314854 &  30.907 &   0.990 &     6 &   4.170 &   0.057 &     6 &   1.016 &   0.018 &   4.102 &  28.981 &   4.167 &  31.724 &   4.202 &  31.042 &   4.157 &  30.907 &   4.118 &  31.527 &   4.268 &  31.260 &   4.115 & 4953 &  174 & -0.51 & 0.33 \\
210315825 &  59.422 &   0.463 &     6 &   6.088 &   0.045 &     6 &   1.025 &   0.010 &   5.939 &  59.479 &   6.071 &  58.995 &   6.067 &  59.696 &   6.156 &  59.986 &   6.077 &  58.749 &   6.125 &  59.626 &   6.029 & 4827 &  180 & -0.30 & 0.30 \\
210318976 &  24.478 &   0.414 &     5 &   3.487 &   0.094 &     3 &   1.031 &   0.013 &   3.381 &  24.758 &   3.381 &  24.971 &     --- &     --- &     --- &  23.897 &   3.565 &  24.452 &   3.508 &  24.312 &     --- & 4680 &  140 & -0.20 & 0.26 \\
\enddata
\tablecomments{Asteroseismic values re-scaled for scalar offsets among
  pipelines are denoted by a prime (the pipeline-specific solar
  reference scale factors are listed in Table~\ref{tab:refs}); mean $\numax$ and $\dnu$ values for each star across all pipelines are denoted by
  $\numaxmean$ and $\dnumean$; the standard deviation of these values for each star
  across all pipelines are denoted by $\sigma_{<\nu\mathrm{max}'>}$ and
  $\sigma_{<\dnu'>}$. $\dnumean$ is adjusted using theoretically-motivated correction factors, $X_{\mathrm{Sharma}}$ \protect\cite{sharma+2016}, for use in asteroseismic scaling relations; an uncorrected version of $\dnumean$ for each star is provided, $\langle \dnu \rangle  = \dnumean/X_{\mathrm{Sharma}}$, should the user wish to compute custom $\dnu$ corrections. EPIC temperatures and metallicities are provided for this purpose, though these are relatively uncertain estimates of the true temperatures and metallicities (these uncertainties are also provided for convenience). The uncertainties in $X_{\mathrm{Sharma}}$,  $\sigma_{X{\mathrm{Sharma}}}$, are computed by perturbing the EPIC temperature and metallicities in a Monte Carlo procedure. Pipeline-specific re-scaled values, $\numax'$ and
  $\dnu'$, are only provided for targets
  for which at least two pipelines returned concordant results, and otherwise
  have a blank entry; the numbers of
  pipelines returning valid results for $\numax$ or $\dnu$ are denoted by
  $N_{\nu \mathrm{max}}$ and $N_{\dnu}$.
  See text for details. A full version of this table is available in the online journal.}
\end{splitdeluxetable*}
\end{longrotatetable}

\begin{longrotatetable}
\begin{deluxetable}{ccccccc}
  \tablecaption{Derived solar reference value scale factors and solar reference values from this work (see \S\ref{sec:zeropoint}), 
    compared to those computed for some of the same pipelines using a
    similar method with {\it Kepler} data (\citealt{pinsonneault+2018}; ``APOKASC-2''). \label{tab:refs}}
  \tabletypesize{\footnotesize}
  \tablehead{ & A2Z &  CAN &  COR &  SYD &  BAM &  BHM }
  \startdata
$X_{\numax,{\mathrm{\ RGB,\ APOKASC2}}}$ & 1.0023 $\pm$ 0.00002 & 1.0082 $\pm$ 0.00002 & 0.9989 $\pm$ 0.00002 & 1.0006 $\pm$ 0.00002 &    --- &    ---\\  
$X_{\numax,\mathrm{\ RGB}}$ & 1.0017 $\pm$ 0.00150 & 0.9901 $\pm$ 0.00072 & 1.0036 $\pm$ 0.00062 & 0.9931 $\pm$ 0.00072 & 1.0044 $\pm$ 0.00056 & 1.0051 $\pm$ 0.00063 \\ 
 ${\numaxsun}_{\mathrm{RGB}}$ & 3102.6  & 3108.8  & 3061.0  & 3068.6  & 3107.5  & 3065.6 \\  
$X_{\dnu,\mathrm{\ RGB,\ APOKASC2}}$ & 0.9993 $\pm$ 0.00001 & 1.0007 $\pm$ 0.00001 & 1.0051 $\pm$ 0.00001 & 0.9995 $\pm$ 0.00001 &    --- &    ---\\  
$X_{\dnu,\mathrm{\ RGB}}$ & 0.9977 $\pm$ 0.00090 & 1.0050 $\pm$ 0.00048 & 1.0001 $\pm$ 0.00061 & 0.9982 $\pm$ 0.00120 & 0.9966 $\pm$ 0.00159 & 1.0008 $\pm$ 0.00099 \\ 
 ${\dnusun}_{\mathrm{RGB}}$ & 134.61  & 135.59  & 134.93  & 134.86  & 134.38  & 135.03 \\
 $X_{\numax,\mathrm{\ RC,\ APOKASC2}}$ & 1.0035 $\pm$ 0.00003 & 1.0067 $\pm$ 0.00002 & 0.9909 $\pm$ 0.00002 & 1.0010 $\pm$ 0.00003 &    --- &    ---\\  
$X_{\numax,\mathrm{\ RC}}$ & 0.9931 $\pm$ 0.00260 & 0.9830 $\pm$ 0.00115 & 1.0056 $\pm$ 0.00090 & 0.9971 $\pm$ 0.00113 & 1.0160 $\pm$ 0.00083 & 0.9989 $\pm$ 0.00101 \\ 
$ {\numaxsun}_{\mathrm{RC}}$ & 3075.9  & 3086.5  & 3067.1  & 3080.9  & 3143.4  & 3046.8 \\
 $X_{\dnu,\mathrm{\ RC,\ APOKASC2}}$ & 0.9965 $\pm$ 0.00003 & 1.0108 $\pm$ 0.00002 & 0.9960 $\pm$ 0.00001 & 1.0032 $\pm$ 0.00002 &    --- &    ---\\  
$X_{\dnu,\mathrm{\ RC}}$ & 0.9960 $\pm$ 0.00151 & 1.0070 $\pm$ 0.00094 & 1.0006 $\pm$ 0.00086 & 0.9927 $\pm$ 0.00390 & 0.9945 $\pm$ 0.00328 & 1.0032 $\pm$ 0.00154 \\ 
 ${\dnusun}_{\mathrm{RC}}$ & 134.38  & 135.86  & 135.00  & 134.11  & 134.09  & 135.35 \\  
\enddata
\end{deluxetable}
\end{longrotatetable}

\begin{longrotatetable}
\begin{splitdeluxetable*}{lcccccccccccccBccccccccccccccl}
  \tablecaption{Radius and mass coefficients \label{tab:kappas}}
  \tabletypesize{\footnotesize}
  \tablehead{EPIC & $\kapparmean$ & $\sigma_{\kapparmean}$ &
    $\kappammean$ & $\sigma_{\kappammean}$ &
    $\kappa_{R,\mathrm{A2Z}}^{\prime}$ & $\sigma_{\kappa
      R',\mathrm{A2Z}}$ & $\kappa_{M,\mathrm{A2Z}}^{\prime}$ &
    $\sigma_{\kappa M',\mathrm{A2Z}}$ &
    $\kappa_{R,\mathrm{BAM}}^{\prime}$ & $\sigma_{\kappa
      R',\mathrm{BAM}}$ & $\kappa_{M,\mathrm{BAM}}^{\prime}$ &
    $\sigma_{\kappa M',\mathrm{BAM}}$ &
    $\kappa_{R,\mathrm{BHM}}^{\prime}$ & $\sigma_{\kappa
      R',\mathrm{BHM}}$ & $\kappa_{M,\mathrm{BHM}}^{\prime}$ &
    $\sigma_{\kappa M',\mathrm{BHM}}$ &
    $\kappa_{R,\mathrm{CAN}}^{\prime}$ & $\sigma_{\kappa
      R',\mathrm{CAN}}$ & $\kappa_{M,\mathrm{CAN}}^{\prime}$ &
    $\sigma_{\kappa M',\mathrm{CAN}}$ &
    $\kappa_{R,\mathrm{COR}}^{\prime}$ & $\sigma_{\kappa
      R',\mathrm{COR}}$ & $\kappa_{M,\mathrm{COR}}^{\prime}$ &
    $\sigma_{\kappa M',\mathrm{COR}}$ &
    $\kappa_{R,\mathrm{SYD}}^{\prime}$ & $\sigma_{\kappa
      R',\mathrm{SYD}}$ & $\kappa_{M,\mathrm{SYD}}^{\prime}$ &
    $\sigma_{\kappa M',\mathrm{SYD}}$}
  \startdata
  & $$ & $$ & $$ & $$ & $$ & $$ & $$ & $$ & $$ & $$ & $$ & $$ & $$
  & $$ & $$ & $$ \\
210306475  &  12.976 &   0.906 &   1.559 &   0.265 &  12.209 &   0.964 &   1.381 &   0.327 &     --- &     --- &     --- &     --- &  12.556 &   1.304 &   1.372 &   0.327 &     --- &     --- &     --- &     --- &     --- &     --- &     --- &     --- &  13.793 &   1.360 &   1.807 &   0.367\\
210307958  &  10.987 &   0.767 &   1.137 &   0.179 &   9.900 &     --- &   0.889 &     --- &  12.072 &   0.834 &   1.430 &   0.212 &     --- &     --- &     --- &     --- &  10.390 &   0.598 &   1.037 &   0.145 &  10.883 &   0.560 &   1.088 &   0.133 &  11.509 &   1.234 &   1.248 &   0.274\\
210314854  &  10.555 &   0.446 &   1.119 &   0.124 &   9.909 &   0.898 &   0.925 &   0.222 &  10.668 &   0.492 &   1.174 &   0.124 &  10.668 &   1.033 &   1.148 &   0.237 &  10.823 &   0.603 &   1.177 &   0.183 &  10.277 &   0.478 &   1.083 &   0.122 &  10.961 &   1.309 &   1.221 &   0.302\\
210315825  &   9.521 &   0.160 &   1.751 &   0.066 &   9.582 &   0.773 &   1.776 &   0.375 &   9.517 &   0.291 &   1.737 &   0.139 &   9.354 &   0.612 &   1.698 &   0.237 &   9.644 &   0.337 &   1.814 &   0.154 &   9.299 &   0.331 &   1.652 &   0.141 &   9.740 &   0.275 &   1.839 &   0.134\\
210318976  &  11.953 &   0.677 &   1.137 &   0.136 &  12.862 &   1.022 &   1.331 &   0.309 &     --- &     --- &     --- &     --- &     --- &     --- &     --- &     --- &  11.166 &   0.804 &   0.969 &   0.179 &  11.797 &   0.620 &   1.106 &   0.138 &     --- &     --- &     --- &     ---\\
\enddata
\tablecomments{$\kapparmean$ and
    $\kappammean$, and their uncertainties, are computed based on $\langle \dnu' \rangle$ and
$\langle \numax' \rangle$, according to Equations~\ref{eq:radius}~\&~\ref{eq:mass}, and represent pipeline-averaged radius and mass coefficients. Pipeline-specific radius and mass coefficients, $\kappa_R'$ and $\kappa_M'$, are
    computed with pipeline-specific
    asteroseismic parameters, $\dnu'$ and
$\numax'$. See \S\ref{sec:derived} for details. A full version of this table is available in the online journal.}
\end{splitdeluxetable*}
\end{longrotatetable}

\begin{deluxetable*}{cccc|ccc}
  \tablecaption{Median fractional uncertainties of \textit{Kepler} and \textit{K2} asteroseismic quantities (in percent) \label{tab:kepk2}}
  \tabletypesize{\footnotesize}
  \tablehead{& & RGB or & RGB/AGB &  & RC &}
  \startdata
  & APOKASC-2 & Y18 & \textit{K2} & APOKASC-2 & Y18 & \textit{K2} \\ \hline
  $\sigma_{\nu_{\rm{max}}}$ & 0.9 & 1.0 & 1.7 & 1.3 & 2.1 & 2.4 \\ \hline
  $\sigma_{\Delta{\nu}}$ &0.4 & 0.3 & 1.7 & 1.1 & 1.1 & 2.3 \\ \hline
  $\sigma_{\kappa_R}$ & 1.3 & 1.1 & 3.3 & 2.7 & 3.3 & 5.0 \\ \hline
  $\sigma_{\kappa_M}$  & 3.4 & 3.1 & 7.7 & 6.2 & 8.4 & 10.5 \\ \hline
\enddata
\tablecomments{``APOKASC-2" indicates median fractional uncertainties from the analysis of \cite{pinsonneault+2018}, while ``Y18" refers to the analysis of \cite{yu+2018}.}
\end{deluxetable*}

\begin{longrotatetable}
\begin{deluxetable}{cccccccc}
\tablecaption{Systematic uncertainties of $\numaxmean$ and $\dnumean$, as a function of $\numaxmean$ and $\dnumean$ \label{tab:syst_mean}}
\tablehead{ $\numaxmean_{\mathrm{RGB}}$ & $\sigma_{\numaxmean, \mathrm{RGB}}$ & $\numaxmean_{\mathrm{RC}}$ & $\sigma_{\numaxmean, \mathrm{RC}}$ & $\dnumean_{\mathrm{RGB}}$ & $\sigma_{\dnumean, \mathrm{RGB}}$ & $\dnumean_{\mathrm{RC}}$ & $\sigma_{\dnumean, \mathrm{RC}}$}
\startdata
$\muhz$ & \% & $\muhz$ & \% & $\muhz$ & \% & $\muhz$ & \% \\
12  &  0.67 &  23  &  1.2 &    1.7  &  1.3 &  3.4  &  0.85\\ \hline
17  &  0.91 &  28  &  0.83 &    2.3  &  1.0 &  3.7  &  0.52\\ \hline
23  &  0.64 &  31  &  0.24 &    2.9  &  0.68 &  4.1  &  0.24\\ \hline
30  &  0.40 &  38  &  0.68 &    3.8  &  0.31 &  4.5  &  0.47\\ \hline
42  &  0.29 &  43  &  0.74 &    4.9  &  0.28 &  5.1  &  0.66\\ \hline
56  &  0.55 &  53  &  1.1 &    5.8  &  0.21 &  5.7  &  0.75\\ \hline
76  &  0.68 &  62  &  0.69 &    7.1  &  0.24 &  6.3  &  0.59\\ \hline
110  &  0.73 &  73  &  0.92 &    9.3  &  0.40 &  7.0  &  0.68\\ \hline
140  &  0.70 &  84  &  1.2 &    11  &  0.43 &  7.7  &  0.92\\ \hline
190  &  0.50 &  93  &  1.4 &    15  &  0.81 &  8.7  &  0.43\\ \hline
\enddata
\tablecomments{Systematic uncertainties of $\numaxmean$ and $\dnumean$, listed as per cent. The binned medians of the fractional difference between an individual pipeline's asteroseismic values and the mean values, $\numaxmean$ and $\dnumean$ shown in the bottom panels of Figures~\ref{fig:numax_rgb_mean}~-~\ref{fig:dnu_rc_mean} are taken to be indications of systematic uncertainty in $\numaxmean$ and $\dnumean$. See \S\ref{sec:syst_mean} for details. Entries that are not populated contain ten or fewer stars in a given $\numaxmean$ or $\dnumean$ bin, and are also not plotted in Figures~\ref{fig:numax_rgb_mean}~-~\ref{fig:dnu_rc_mean}.}
\end{deluxetable}
\end{longrotatetable}

\acknowledgments
JCZ and MHP acknowledge support from NASA grants
80NSSC18K0391 and NNX17AJ40G. YE and CJ acknowledge the support of the
UK Science and Technology Facilities Council (STFC). SM would like to
acknowledge support from the Spanish Ministry with the Ramon y Cajal
fellowship number RYC-2015-17697. RAG acknowledges funding received
from the PLATO CNES grant. RS acknowledges funding via a Royal Society
University Research Fellowship. D.H. acknowledges support from the
Alfred P. Sloan Foundation and the National Aeronautics and Space
Administration (80NSSC19K0108). V.S.A. acknowledges support from the
Independent Research Fund Denmark (Research grant 7027-00096B), and
the Carlsberg foundation (grant agreement CF19-0649). This research
was supported in part by the National Science Foundation under Grant
No. NSF PHY-1748958.

Funding for the Stellar Astrophysics Centre (SAC) is provided by The Danish National Research Foundation (Grant agreement no. DNRF106). 

The {\it K2} Galactic Archaeology Program is supported by the National Aeronautics and Space Administration under Grant NNX16AJ17G issued through the {\it K2} Guest Observer Program. This publication makes use of data products from the Two Micron All Sky Survey, which is a joint project of the University of Massachusetts and the Infrared Processing and Analysis Center/California Institute of Technology, funded by the National Aeronautics and Space Administration and the National Science Foundation.

This work has made use of data from the European Space Agency (ESA)
mission
{\it Gaia} (\url{https://www.cosmos.esa.int/gaia}), processed by the
{\it Gaia}
Data Processing and Analysis Consortium (DPAC,
\url{https://www.cosmos.esa.int/web/gaia/dpac/consortium}). Funding
for the DPAC
has been provided by national institutions, in particular the
institutions
participating in the {\it Gaia} Multilateral Agreement.

Funding for the Sloan Digital Sky Survey IV has been provided by the Alfred P. Sloan Foundation, the U.S. Department of Energy Office of Science, and the Participating Institutions. SDSS-IV acknowledges
support and resources from the Center for High-Performance Computing at
the University of Utah. The SDSS web site is www.sdss.org.

SDSS-IV is managed by the Astrophysical Research Consortium for the 
Participating Institutions of the SDSS Collaboration including the 
Brazilian Participation Group, the Carnegie Institution for Science, 
Carnegie Mellon University, the Chilean Participation Group, the French Participation Group, Harvard-Smithsonian Center for Astrophysics, 
Instituto de Astrof\'isica de Canarias, The Johns Hopkins University, Kavli Institute for the Physics and Mathematics of the Universe (IPMU) / 
University of Tokyo, the Korean Participation Group, Lawrence Berkeley National Laboratory, 
Leibniz Institut f\"ur Astrophysik Potsdam (AIP),  
Max-Planck-Institut f\"ur Astronomie (MPIA Heidelberg), 
Max-Planck-Institut f\"ur Astrophysik (MPA Garching), 
Max-Planck-Institut f\"ur Extraterrestrische Physik (MPE), 
National Astronomical Observatories of China, New Mexico State University, 
New York University, University of Notre Dame, 
Observat\'ario Nacional / MCTI, The Ohio State University, 
Pennsylvania State University, Shanghai Astronomical Observatory, 
United Kingdom Participation Group,
Universidad Nacional Aut\'onoma de M\'exico, University of Arizona, 
University of Colorado Boulder, University of Oxford, University of Portsmouth, 
University of Utah, University of Virginia, University of Washington, University of Wisconsin, 
Vanderbilt University, and Yale University.

\software{asfgrid \citep{asfgrid}, emcee \citep{foreman-mackey+2013}, NumPy \citep{numpy}, pandas \citep{pandas}, Matplotlib \citep{matplotlib}, IPython \citep{ipython}, SciPy \citep{scipy}}

\appendix
\renewcommand{\theprogram}{A\arabic{program}}  
We model the distribution of the fractional uncertainties, $\sigma_{\numaxmean}/\numaxmean$ and $ \sigma_{\dnumean}/\dnumean $ as a function of the number of pipelines used to compute the uncertainties, $dof$, and evolutionary state (RGB or RGB/AGB versus RC). As described in the text, we use two generalized gamma distributions to model each distribution of stars with a given $dof$ (either 2, 3, 4, 5, or 6). When summed, the two fitted generalized gamma distributions are solutions to a least-squares minimization problem to describe the data, with Poisson uncertainties assumed for each bin in the observed distribution. To arrive at the fit, each component is weighted using a free parameter to describe the relative contribution of each component, and the degrees of freedom for each component are required to be less than or equal to the nominal degrees of freedom for the observed distribution (i.e., the number of reporting pipelines, $dof$). The solution is found using the Trust Region Reflective method as implemented in the \textit{scipy} function, \textit{curve\_fit}. As a reference, we also fit each of the six distributions for both RGB or RGB/AGB and RC stars using a generalized gamma distribution with the uncertainty taken to be the median observed fractional uncertainty, and with the $dof$ fixed to be the number of reporting pipelines. We do the same for the fractional uncertainty distributions for $\kapparmean$ and $\kappammean$.

For $\numaxmean$ among RGB or RGB/AGB stars, the fractional uncertainties inferred from fitting the two-component generalized gamma distribution model vary according to the number of pipelines that contribute to the scatter estimate, from $1.1-2.1\%$. For RC stars, the fitted uncertainties are larger, and have a range of $2.2-3.9\%$. The fitted fractional uncertainties on $\dnumean$ have marginally smaller fractional uncertainties, and can range from  $1.4\%$ to $1.9\%$ for RGB or RGB/AGB stars and from $1.9\%$ to $2.8\%$ for RC stars. For both $\numaxmean$ and $\dnumean$, therefore, the uncertainties vary more as a function of evolutionary state than number of pipelines reporting. 

\begin{figure*}[htp!]
\subfloat{\includegraphics[width=0.5\textwidth]{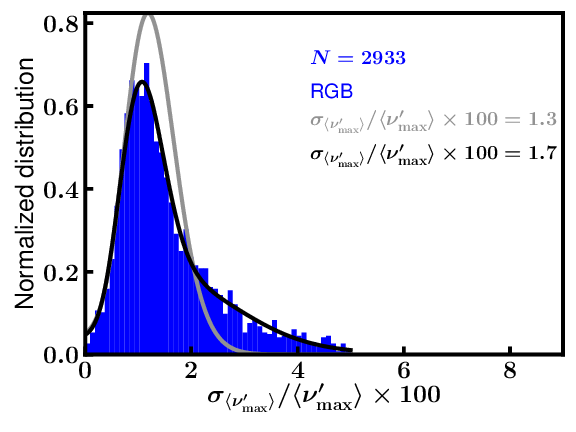}}
\subfloat{\includegraphics[width=0.5\textwidth]{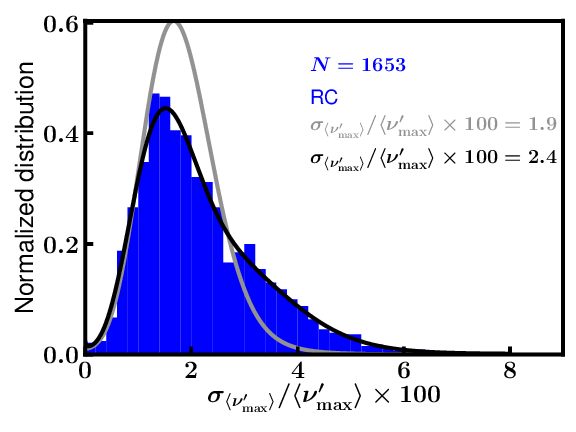}}
\caption{Left: The distribution of uncertainties in $\numaxmean$ for RGB stars, with curves showing models for the distributions assuming the median (grey) and best-fitting uncertainties (black) --- the characteristic uncertainty according to each of these two models is shown in grey and black in the legend. These distributions and models are the results of summing the distributions and models for stars with $dof=\{2, 3,4,5,6\}$ pipelines reporting (see Appendix for details). The number of stars contributing to the observed distribution is listed as $N$. Right: Same as Left, but for RC stars.}
\label{fig:numaxrgb}     
\end{figure*}

\begin{figure*}[htp!]
\subfloat{\includegraphics[width=0.5\textwidth]{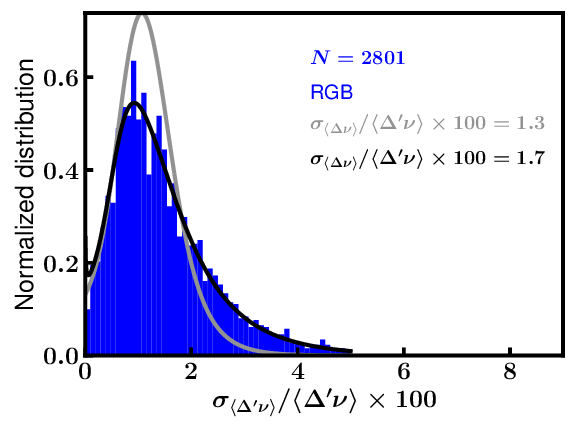}}
\subfloat{\includegraphics[width=0.5\textwidth]{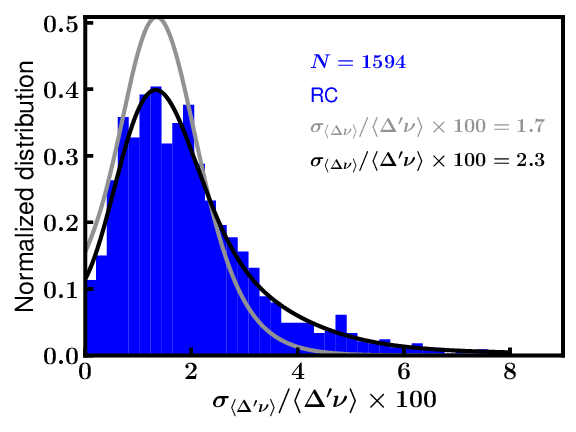}}
\caption{Left: Same as Figure~\protect\ref{fig:numaxrgb}, but for $\dnumean$. Right: Same as Left, but for RC stars.}
\label{fig:dnurc}     
\end{figure*}

The uncertainty distributions for all of the RGB and RC stars are shown in Figures~\ref{fig:numaxrgb}-\ref{fig:dnurc}. We show both the expected distribution according to the observed median uncertainties using a fixed $dof$ (grey curve), as well as the expected distribution from the two-component model (black curve), where we sum the uncertainties from each $dof$, weighting by the number of stars with a given $dof$. The agreement between the model for the uncertainties and the observed uncertainty distributions indicates that the uncertainties are largely not a function of $\numax$ or $\dnu$. Nevertheless, the approximation is not completely accurate: lower values of $\numax$ and $\dnu$ tend to have marginally larger fractional uncertainties --- up to 1\% larger across the entire observed parameter range for $\numax$ and $\dnu$ among RGB stars, and up to $3\%$ larger for $\numax$ among $dof=2$ and $dof=3$ RC stars with $\numax < 30\muhz$. (These latter low-$\numax$ RC stars contribute to the extra bump at $\sigma_{\numaxmean}/\numaxmean \approx 3$ in Figure~\ref{fig:numaxrgb}.) The uncertainties are not strong functions of magnitude, which reflects the fact that the uncertainties are not dominated by white noise, but rather the length of the light curve, intrinsic properties of the star (e.g., evolutionary state), and pipeline agreement.

\begin{figure*}[htp!]
\subfloat{\includegraphics[width=0.5\textwidth]{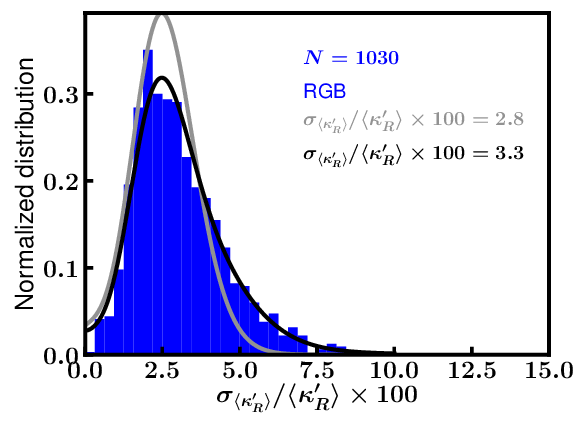}}
\subfloat{\includegraphics[width=0.5\textwidth]{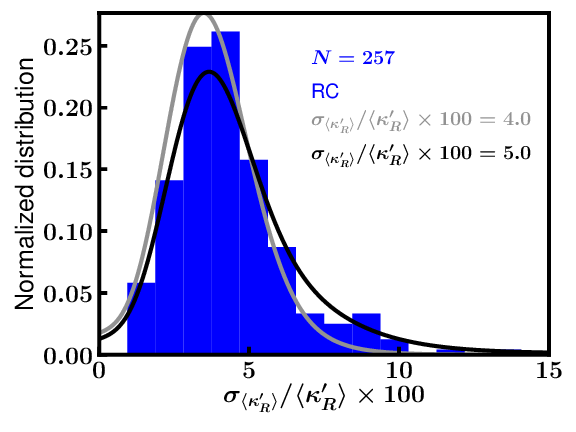}}
\caption{Left: Same as Figure~\protect\ref{fig:numaxrgb}, but for $\kapparmean$. Right: Same as Left, but for RC stars.}
\label{fig:kapparrgb}     
\end{figure*}

\begin{figure*}[htp!]
\subfloat{\includegraphics[width=0.5\textwidth]{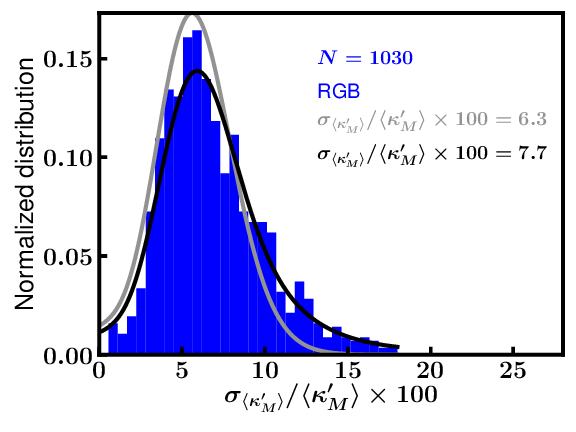}}
\subfloat{\includegraphics[width=0.5\textwidth]{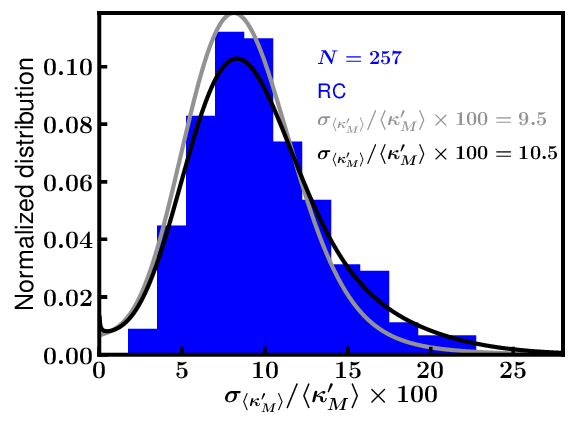}}
\caption{Left: Same as Figure~\protect\ref{fig:kapparrgb}, but for $\kappammean$. Right: Same as Left but for RC stars.}
\label{fig:kappamrc}     
\end{figure*}

$\kappammean$ fractional uncertainties reinforce the trend for fractional uncertainties to vary more as a function of evolutionary state than a function of number of pipelines reporting: for RGB and RGB/AGB, the range is $6.2-9.0\%$ and for RC stars, the range is $9.9-12.1\%$. The uncertainty distributions of $\kapparmean$ and $\kappammean$ for all of the RGB and RC stars are shown in Figures~\ref{fig:kapparrgb}-\ref{fig:kappamrc}, where the grey and black curves correspond to the expected distribution according to the sample median uncertainties and the best-fitting uncertainties, and which are obtained by appropriately weighted sums of the uncertainties at each $dof$.

\bibliography{bib}

\begin{thebibliography}{}
\expandafter\ifx\csname natexlab\endcsname\relax\def\natexlab#1{#1}\fi
\providecommand{\url}[1]{\href{#1}{#1}}

\bibitem[{{Ahumada} {et~al.}(2020){Ahumada}, {Prieto}, {Almeida}, {Anders},
  {Anderson}, {Andrews}, {Anguiano}, {Arcodia}, {Armengaud}, {Aubert}, {Avila},
  {Avila-Reese}, {Badenes}, {Balland }, {Barger}, {Barrera-Ballesteros},
  {Basu}, {Bautista}, {Beaton}, {Beers}, {Benavides}, {Bender}, {Bernardi},
  {Bershady}, {Beutler}, {Bidin}, {Bird}, {Bizyaev}, {Blanc}, {Blanton},
  {Boquien}, {Borissova}, {Bovy}, {Brand t}, {Brinkmann}, {Brownstein},
  {Bundy}, {Bureau}, {Burgasser}, {Burtin}, {Cano-D{\'\i}az}, {Capasso},
  {Cappellari}, {Carrera}, {Chabanier}, {Chaplin}, {Chapman}, {Cherinka},
  {Chiappini}, {Doohyun Choi}, {Chojnowski}, {Chung}, {Clerc}, {Coffey},
  {Comerford}, {Comparat}, {da Costa}, {Cousinou}, {Covey}, {Crane}, {Cunha},
  {Ilha}, {Dai}, {Damsted}, {Darling}, {Davidson}, {Davies}, {Dawson}, {De},
  {de la Macorra}, {De Lee}, {Queiroz}, {Deconto Machado}, {de la Torre},
  {Dell'Agli}, {du Mas des Bourboux}, {Diamond-Stanic}, {Dillon}, {Donor},
  {Drory}, {Duckworth}, {Dwelly}, {Ebelke}, {Eftekharzadeh}, {Davis Eigenbrot},
  {Elsworth}, {Eracleous}, {Erfanianfar}, {Escoffier}, {Fan}, {Farr},
  {Fern{\'a}ndez-Trincado}, {Feuillet}, {Finoguenov}, {Fofie},
  {Fraser-McKelvie}, {Frinchaboy}, {Fromenteau}, {Fu}, {Galbany}, {Garcia},
  {Garc{\'\i}a-Hern{\'a}ndez}, {Oehmichen}, {Ge}, {Maia}, {Geisler}, {Gelfand
  }, {Goddy}, {Gonzalez-Perez}, {Grabowski}, {Green}, {Grier}, {Guo}, {Guy},
  {Harding}, {Hasselquist}, {Hawken}, {Hayes}, {Hearty}, {Hekker}, {Hogg},
  {Holtzman}, {Horta}, {Hou}, {Hsieh}, {Huber}, {Hunt}, {Chitham}, {Imig},
  {Jaber}, {Angel}, {Johnson}, {Jones}, {J{\"o}nsson}, {Jullo}, {Kim},
  {Kinemuchi}, {Kirkpatrick}, {Kite}, {Klaene}, {Kneib}, {Kollmeier}, {Kong},
  {Kounkel}, {Krishnarao}, {Lacerna}, {Lan}, {Lane}, {Law}, {Le Goff}, {Leung},
  {Lewis}, {Li}, {Lian}, {Lin}, {Long}, {Longa-Pe{\~n}a}, {Lundgren}, {Lyke},
  {Ted Mackereth}, {MacLeod}, {Majewski}, {Manchado}, {Maraston}, {Martini},
  {Masseron}, {Masters}, {Mathur}, {McDermid}, {Merloni}, {Merrifield},
  {M{\'e}sz{\'a}ros}, {Miglio}, {Minniti}, {Minsley}, {Miyaji}, {Mohammad},
  {Mosser}, {Mueller}, {Muna}, {Mu{\~n}oz-Guti{\'e}rrez}, {Myers}, {Nadathur},
  {Nair}, {Nandra}, {do Nascimento}, {Nevin}, {Newman}, {Nidever}, {Nitschelm},
  {Noterdaeme}, {O'Connell}, {Olmstead}, {Oravetz}, {Oravetz}, {Osorio},
  {Pace}, {Padilla}, {Palanque-Delabrouille}, {Palicio}, {Pan}, {Pan},
  {Parker}, {Paviot}, {Peirani}, {Ram{\'r}ez}, {Penny}, {Percival},
  {Perez-Fournon}, {P{\'e}rez-R{\`a}fols}, {Petitjean}, {Pieri},
  {Pinsonneault}, {Poovelil}, {Povick}, {Prakash}, {Price-Whelan}, {Raddick},
  {Raichoor}, {Ray}, {Rembold}, {Rezaie}, {Riffel}, {Riffel}, {Rix}, {Robin},
  {Roman-Lopes}, {Rom{\'a}n-Z{\'u}{\~n}iga}, {Rose}, {Ross}, {Rossi}, {Rowland
  s}, {Rubin}, {Salvato}, {S{\'a}nchez}, {S{\'a}nchez-Menguiano},
  {S{\'a}nchez-Gallego}, {Sayres}, {Schaefer}, {Schiavon}, {Schimoia},
  {Schlafly}, {Schlegel}, {Schneider}, {Schultheis}, {Schwope}, {Seo},
  {Serenelli}, {Shafieloo}, {Shamsi}, {Shao}, {Shen}, {Shetrone}, {Shirley},
  {Aguirre}, {Simon}, {Skrutskie}, {Slosar}, {Smethurst}, {Sobeck}, {Sodi},
  {Souto}, {Stark}, {Stassun}, {Steinmetz}, {Stello}, {Stermer},
  {Storchi-Bergmann}, {Streblyanska}, {Stringfellow}, {Stutz}, {Su{\'a}rez},
  {Sun}, {Taghizadeh-Popp}, {Talbot}, {Tayar}, {Thakar}, {Theriault}, {Thomas},
  {Thomas}, {Tinker}, {Tojeiro}, {Toledo}, {Tremonti}, {Troup}, {Tuttle},
  {Unda-Sanzana}, {Valentini}, {Vargas-Gonz{\'a}lez}, {Vargas-Maga{\~n}a},
  {V{\'a}zquez-Mata}, {Vivek}, {Wake}, {Wang}, {Weaver}, {Weijmans}, {Wild},
  {Wilson}, {Wilson}, {Wolthuis}, {Wood-Vasey}, {Yan}, {Yang}, {Y{\`e}che},
  {Zamora}, {Zarrouk}, {Zasowski}, {Zhang}, {Zhao}, {Zhao}, {Zheng}, {Zheng},
  {Zhu}, \& {Zou}}]{apogeedr16_2019}
{Ahumada}, R., {Prieto}, C.~A., {Almeida}, A., {et~al.} 2020, \apjs, 249, 3

\bibitem[{{An} {et~al.}(2019){An}, {Pinsonneault}, {Terndrup}, \&
  {Chung}}]{an+2019a}
{An}, D., {Pinsonneault}, M.~H., {Terndrup}, D.~M., \& {Chung}, C. 2019, \apj,
  879, 81

\bibitem[{{Bedding} {et~al.}(2011){Bedding}, {Mosser}, {Huber},
  {Montalb{\'a}n}, {Beck}, {Christensen-Dalsgaard}, {Elsworth}, {Garc{\'\i}a},
  {Miglio}, {Stello}, {White}, {De Ridder}, {Hekker}, {Aerts}, {Barban},
  {Belkacem}, {Broomhall}, {Brown}, {Buzasi}, {Carrier}, {Chaplin}, {di Mauro},
  {Dupret}, {Frandsen}, {Gilliland }, {Goupil}, {Jenkins}, {Kallinger},
  {Kawaler}, {Kjeldsen}, {Mathur}, {Noels}, {Silva Aguirre}, \&
  {Ventura}}]{bedding+2011}
{Bedding}, T.~R., {Mosser}, B., {Huber}, D., {et~al.} 2011, \nat, 471, 608

\bibitem[{{Belkacem} {et~al.}(2011){Belkacem}, {Goupil}, {Dupret}, {Samadi},
  {Baudin}, {Noels}, \& {Mosser}}]{belkacem+2011}
{Belkacem}, K., {Goupil}, M.~J., {Dupret}, M.~A., {et~al.} 2011, \aap, 530,
  A142

\bibitem[{{Borucki} {et~al.}(2008){Borucki}, {Koch}, {Basri}, {Batalha},
  {Brown}, {Caldwell}, {Christensen-Dalsgaard}, {Cochran}, {Dunham}, {Gautier},
  {Geary}, {Gilliland}, {Jenkins}, {Kondo}, {Latham}, {Lissauer}, \&
  {Monet}}]{borucki+2008}
{Borucki}, W., {Koch}, D., {Basri}, G., {et~al.} 2008, in IAU Symposium, Vol.
  249, Exoplanets: Detection, Formation and Dynamics, ed. Y.-S. {Sun},
  S.~{Ferraz-Mello}, \& J.-L. {Zhou}, 17--24

\bibitem[{{Bovy} {et~al.}(2016){Bovy}, {Rix}, {Green}, {Schlafly}, \&
  {Finkbeiner}}]{bovy+2016}
{Bovy}, J., {Rix}, H.-W., {Green}, G.~M., {Schlafly}, E.~F., \& {Finkbeiner},
  D.~P. 2016, \apj, 818, 130

\bibitem[{{Brown} {et~al.}(1991){Brown}, {Gilliland}, {Noyes}, \&
  {Ramsey}}]{brown+1991}
{Brown}, T.~M., {Gilliland}, R.~L., {Noyes}, R.~W., \& {Ramsey}, L.~W. 1991,
  \apj, 368, 599

\bibitem[{{Bugnet} {et~al.}(2018){Bugnet}, {Garc{\'\i}a}, {Davies}, {Mathur},
  {Corsaro}, {Hall}, \& {Rendle}}]{bugnet+2018}
{Bugnet}, L., {Garc{\'\i}a}, R.~A., {Davies}, G.~R., {et~al.} 2018, \aap, 620,
  A38

\bibitem[{{Casagrande} {et~al.}(2016){Casagrande}, {Silva Aguirre},
  {Schlesinger}, {Stello}, {Huber}, {Serenelli}, {Sch{\"o}nrich}, {Cassisi},
  {Pietrinferni}, {Hodgkin}, {Milone}, {Feltzing}, \&
  {Asplund}}]{casagrande+2016}
{Casagrande}, L., {Silva Aguirre}, V., {Schlesinger}, K.~J., {et~al.} 2016,
  \mnras, 455, 987

\bibitem[{{Chaplin} {et~al.}(2008){Chaplin}, {Houdek}, {Appourchaux},
  {Elsworth}, {New}, \& {Toutain}}]{chaplin+2008}
{Chaplin}, W.~J., {Houdek}, G., {Appourchaux}, T., {et~al.} 2008, \aap, 485,
  813

\bibitem[{{Chaplin} {et~al.}(2011){Chaplin}, {Kjeldsen}, {Bedding},
  {Christensen-Dalsgaard}, {Gilliland}, {Kawaler}, {Appourchaux}, {Elsworth},
  {Garc{\'{\i}}a}, {Houdek}, {Karoff}, {Metcalfe}, {Molenda-{\.Z}akowicz},
  {Monteiro}, {Thompson}, {Verner}, {Batalha}, {Borucki}, {Brown}, {Bryson},
  {Christiansen}, {Clarke}, {Jenkins}, {Klaus}, {Koch}, {An}, {Ballot}, {Basu},
  {Benomar}, {Bonanno}, {Broomhall}, {Campante}, {Corsaro}, {Creevey}, {Esch},
  {Gai}, {Gaulme}, {Hale}, {Handberg}, {Hekker}, {Huber}, {Mathur}, {Mosser},
  {New}, {Pinsonneault}, {Pricopi}, {Quirion}, {R{\'e}gulo}, {Roxburgh},
  {Salabert}, {Stello}, \& {Suran}}]{chaplin+2011}
{Chaplin}, W.~J., {Kjeldsen}, H., {Bedding}, T.~R., {et~al.} 2011, \apj, 732,
  54

\bibitem[{{Elsworth} {et~al.}(2019){Elsworth}, {Hekker}, {Johnson},
  {Kallinger}, {Mosser}, {Pinsonneault}, {Hon}, {Kuszlewicz}, {Miglio},
  {Serenelli}, {Stello}, {Tayar}, \& {Vrard}}]{elsworth+2019}
{Elsworth}, Y., {Hekker}, S., {Johnson}, J.~A., {et~al.} 2019, \mnras, 489,
  4641

\bibitem[{{Foreman-Mackey} {et~al.}(2013){Foreman-Mackey}, {Hogg}, {Lang}, \&
  {Goodman}}]{foreman-mackey+2013}
{Foreman-Mackey}, D., {Hogg}, D.~W., {Lang}, D., \& {Goodman}, J. 2013, \pasp,
  125, 306

\bibitem[{{Gaia Collaboration} {et~al.}(2018){Gaia Collaboration}, {Brown},
  {Vallenari}, {Prusti}, {de Bruijne}, {Babusiaux}, {Bailer-Jones}, {Biermann},
  {Evans}, {Eyer}, \& et~al.}]{gaia2018}
{Gaia Collaboration}, {Brown}, A.~G.~A., {Vallenari}, A., {et~al.} 2018, \aap,
  616, A1

\bibitem[{{Gonz{\'a}lez Hern{\'a}ndez} \&
  {Bonifacio}(2009)}]{gonzalezhernandez&bonifacio2009}
{Gonz{\'a}lez Hern{\'a}ndez}, J.~I., \& {Bonifacio}, P. 2009, \aap, 497, 497

\bibitem[{{Green} {et~al.}(2015){Green}, {Schlafly}, {Finkbeiner}, {Rix},
  {Martin}, {Burgett}, {Draper}, {Flewelling}, {Hodapp}, {Kaiser}, {Kudritzki},
  {Magnier}, {Metcalfe}, {Price}, {Tonry}, \& {Wainscoat}}]{green+2015}
{Green}, G.~M., {Schlafly}, E.~F., {Finkbeiner}, D.~P., {et~al.} 2015, \apj,
  810, 25

\bibitem[{{Hall} {et~al.}(2019){Hall}, {Davies}, {Elsworth}, {Miglio},
  {Bedding}, {Brown}, {Khan}, {Hawkins}, {Garc{\'{\i}}a}, {Chaplin}, \&
  {North}}]{hall+2019a}
{Hall}, O.~J., {Davies}, G.~R., {Elsworth}, Y.~P., {et~al.} 2019, \mnras, 486,
  3569

\bibitem[{{Hekker} {et~al.}(2010){Hekker}, {Broomhall}, {Chaplin}, {Elsworth},
  {Fletcher}, {New}, {Arentoft}, {Quirion}, \& {Kjeldsen}}]{hekker+2010}
{Hekker}, S., {Broomhall}, A.-M., {Chaplin}, W.~J., {et~al.} 2010, \mnras, 402,
  2049

\bibitem[{{Hon} {et~al.}(2017){Hon}, {Stello}, \& {Yu}}]{hon+2017a}
{Hon}, M., {Stello}, D., \& {Yu}, J. 2017, \mnras, 469, 4578

\bibitem[{{Hon} {et~al.}(2018{\natexlab{a}}){Hon}, {Stello}, \&
  {Yu}}]{hon+2018a}
---. 2018{\natexlab{a}}, \mnras, 476, 3233

\bibitem[{{Hon} {et~al.}(2018{\natexlab{b}}){Hon}, {Stello}, \&
  {Zinn}}]{hon+2018b}
{Hon}, M., {Stello}, D., \& {Zinn}, J.~C. 2018{\natexlab{b}}, \apj, 859, 64

\bibitem[{{Howell} {et~al.}(2014){Howell}, {Sobeck}, {Haas}, {Still},
  {Barclay}, {Mullally}, {Troeltzsch}, {Aigrain}, {Bryson}, {Caldwell},
  {Chaplin}, {Cochran}, {Huber}, {Marcy}, {Miglio}, {Najita}, {Smith},
  {Twicken}, \& {Fortney}}]{howell+2014}
{Howell}, S.~B., {Sobeck}, C., {Haas}, M., {et~al.} 2014, \pasp, 126, 398

\bibitem[{{Huber} {et~al.}(2009){Huber}, {Stello}, {Bedding}, {Chaplin},
  {Arentoft}, {Quirion}, \& {Kjeldsen}}]{huber+2009}
{Huber}, D., {Stello}, D., {Bedding}, T.~R., {et~al.} 2009, Communications in
  Asteroseismology, 160, 74

\bibitem[{{Huber} {et~al.}(2012){Huber}, {Ireland}, {Bedding}, {Brand{\~a}o},
  {Piau}, {Maestro}, {White}, {Bruntt}, {Casagrande}, {Molenda-{\.Z}akowicz},
  {Silva Aguirre}, {Sousa}, {Barclay}, {Burke}, {Chaplin},
  {Christensen-Dalsgaard}, {Cunha}, {De Ridder}, {Farrington}, {Frasca},
  {Garc{\'{\i}}a}, {Gilliland}, {Goldfinger}, {Hekker}, {Kawaler}, {Kjeldsen},
  {McAlister}, {Metcalfe}, {Miglio}, {Monteiro}, {Pinsonneault}, {Schaefer},
  {Stello}, {Stumpe}, {Sturmann}, {Sturmann}, {ten Brummelaar}, {Thompson},
  {Turner}, \& {Uytterhoeven}}]{huber+2012}
{Huber}, D., {Ireland}, M.~J., {Bedding}, T.~R., {et~al.} 2012, \apj, 760, 32

\bibitem[{{Huber} {et~al.}(2016){Huber}, {Bryson}, {Haas}, {Barclay},
  {Barentsen}, {Howell}, {Sharma}, {Stello}, \& {Thompson}}]{huber+2016}
{Huber}, D., {Bryson}, S.~T., {Haas}, M.~R., {et~al.} 2016, \apjs, 224, 2

\bibitem[{{Huber} {et~al.}(2017){Huber}, {Zinn}, {Bojsen-Hansen},
  {Pinsonneault}, {Sahlholdt}, {Serenelli}, {Silva Aguirre}, {Stassun},
  {Stello}, {Tayar}, {Bastien}, {Bedding}, {Buchhave}, {Chaplin}, {Davies},
  {Garc{\'{\i}}a}, {Latham}, {Mathur}, {Mosser}, \& {Sharma}}]{huber+2017}
{Huber}, D., {Zinn}, J., {Bojsen-Hansen}, M., {et~al.} 2017, \apj, 844, 102

\bibitem[{Hunter(2007)}]{matplotlib}
Hunter, J.~D. 2007, Computing in Science \& Engineering, 9, 90.
\newblock \url{https://aip.scitation.org/doi/abs/10.1109/MCSE.2007.55}

\bibitem[{{Kallinger} {et~al.}(2016){Kallinger}, {Hekker}, {Garcia}, {Huber},
  \& {Matthews}}]{kallinger+2016}
{Kallinger}, T., {Hekker}, S., {Garcia}, R.~A., {Huber}, D., \& {Matthews},
  J.~M. 2016, Science Advances, 2, 1500654

\bibitem[{{Kallinger} {et~al.}(2012){Kallinger}, {Hekker}, {Mosser}, {De
  Ridder}, {Bedding}, {Elsworth}, {Gruberbauer}, {Guenther}, {Stello}, {Basu},
  {Garc{\'\i}a}, {Chaplin}, {Mullally}, {Still}, \&
  {Thompson}}]{kallinger+2012}
{Kallinger}, T., {Hekker}, S., {Mosser}, B., {et~al.} 2012, \aap, 541, A51

\bibitem[{{Kallinger} {et~al.}(2014){Kallinger}, {De Ridder}, {Hekker},
  {Mathur}, {Mosser}, {Gruberbauer}, {Garc{\'\i}a}, {Karoff}, \&
  {Ballot}}]{kallinger+2014}
{Kallinger}, T., {De Ridder}, J., {Hekker}, S., {et~al.} 2014, \aap, 570, A41

\bibitem[{{Khan} {et~al.}(2019){Khan}, {Miglio}, {Mosser}, {Arenou},
  {Belkacem}, {Brown}, {Katz}, {Casagrand e}, {Chaplin}, {Davies}, {Rendle},
  {Rodrigues}, {Bossini}, {Cantat-Gaudin}, {Elsworth}, {Girardi}, {North}, \&
  {Vallenari}}]{khan+2019}
{Khan}, S., {Miglio}, A., {Mosser}, B., {et~al.} 2019, \aap, 628, A35

\bibitem[{{Kjeldsen} \& {Bedding}(1995)}]{kjeldsen&bedding1995}
{Kjeldsen}, H., \& {Bedding}, T.~R. 1995, \aap, 293, 87

\bibitem[{{Kuszlewicz} {et~al.}(2020){Kuszlewicz}, {Hekker}, \&
  {Bell}}]{kuszlewicz_hekker_bell2020}
{Kuszlewicz}, J.~S., {Hekker}, S., \& {Bell}, K.~J. 2020, \mnras, 497, 4843

\bibitem[{{Lindegren} {et~al.}(2018){Lindegren}, {Hern{\'a}ndez}, {Bombrun},
  {Klioner}, {Bastian}, {Ramos-Lerate}, {de Torres}, {Steidelm{\"u}ller},
  {Stephenson}, {Hobbs}, {Lammers}, {Biermann}, {Geyer}, {Hilger}, {Michalik},
  {Stampa}, {McMillan}, {Casta{\~n}eda}, {Clotet}, {Comoretto}, {Davidson},
  {Fabricius}, {Gracia}, {Hambly}, {Hutton}, {Mora}, {Portell}, {van Leeuwen},
  {Abbas}, {Abreu}, {Altmann}, {Andrei}, {Anglada}, {Balaguer-N{\'u}{\~n}ez},
  {Barache}, {Becciani}, {Bertone}, {Bianchi}, {Bouquillon}, {Bourda},
  {Br{\"u}semeister}, {Bucciarelli}, {Busonero}, {Buzzi}, {Cancelliere},
  {Carlucci}, {Charlot}, {Cheek}, {Crosta}, {Crowley}, {de Bruijne}, {de
  Felice}, {Drimmel}, {Esquej}, {Fienga}, {Fraile}, {Gai}, {Garralda},
  {Gonz{\'a}lez-Vidal}, {Guerra}, {Hauser}, {Hofmann}, {Holl}, {Jordan},
  {Lattanzi}, {Lenhardt}, {Liao}, {Licata}, {Lister}, {L{\"o}ffler},
  {Marchant}, {Martin-Fleitas}, {Messineo}, {Mignard}, {Morbidelli}, {Poggio},
  {Riva}, {Rowell}, {Salguero}, {Sarasso}, {Sciacca}, {Siddiqui}, {Smart},
  {Spagna}, {Steele}, {Taris}, {Torra}, {van Elteren}, {van Reeven}, \&
  {Vecchiato}}]{lindegren+2018}
{Lindegren}, L., {Hern{\'a}ndez}, J., {Bombrun}, A., {et~al.} 2018, \aap, 616,
  A2

\bibitem[{{Luger} {et~al.}(2018){Luger}, {Kruse}, {Foreman-Mackey}, {Agol}, \&
  {Saunders}}]{luger+2018}
{Luger}, R., {Kruse}, E., {Foreman-Mackey}, D., {Agol}, E., \& {Saunders}, N.
  2018, \aj, 156, 99

\bibitem[{{Majewski} {et~al.}(2010){Majewski}, {Wilson}, {Hearty}, {Schiavon},
  \& {Skrutskie}}]{majewski+2010}
{Majewski}, S.~R., {Wilson}, J.~C., {Hearty}, F., {Schiavon}, R.~R., \&
  {Skrutskie}, M.~F. 2010, in IAU Symposium, Vol. 265, Chemical Abundances in
  the Universe: Connecting First Stars to Planets, ed. K.~{Cunha}, M.~{Spite},
  \& B.~{Barbuy}, 480--481

\bibitem[{{Mamajek} {et~al.}(2015){Mamajek}, {Prsa}, {Torres}, {Harmanec},
  {Asplund}, {Bennett}, {Capitaine}, {Christensen-Dalsgaard}, {Depagne},
  {Folkner}, {Haberreiter}, {Hekker}, {Hilton}, {Kostov}, {Kurtz}, {Laskar},
  {Mason}, {Milone}, {Montgomery}, {Richards}, {Schou}, \&
  {Stewart}}]{mamajek+2015a}
{Mamajek}, E.~E., {Prsa}, A., {Torres}, G., {et~al.} 2015, ArXiv e-prints

\bibitem[{{Martig} {et~al.}(2014){Martig}, {Minchev}, \&
  {Flynn}}]{martig_minchev_flynn2014}
{Martig}, M., {Minchev}, I., \& {Flynn}, C. 2014, \mnras, 443, 2452

\bibitem[{{Mathur} {et~al.}(2010){Mathur}, {Garc{\'\i}a}, {R{\'e}gulo},
  {Creevey}, {Ballot}, {Salabert}, {Arentoft}, {Quirion}, {Chaplin}, \&
  {Kjeldsen}}]{mathur+2010}
{Mathur}, S., {Garc{\'\i}a}, R.~A., {R{\'e}gulo}, C., {et~al.} 2010, \aap, 511,
  A46

\bibitem[{{Mathur} {et~al.}(2011){Mathur}, {Hekker}, {Trampedach}, {Ballot},
  {Kallinger}, {Buzasi}, {Garc{\'\i}a}, {Huber}, {Jim{\'e}nez}, {Mosser},
  {Bedding}, {Elsworth}, {R{\'e}gulo}, {Stello}, {Chaplin}, {De Ridder},
  {Hale}, {Kinemuchi}, {Kjeldsen}, {Mullally}, \& {Thompson}}]{mathur+2011}
{Mathur}, S., {Hekker}, S., {Trampedach}, R., {et~al.} 2011, \apj, 741, 119

\bibitem[{McKinney(2010)}]{pandas}
McKinney, W. 2010, in Proceedings of the 9th Python in Science Conference, ed.
  S.~van~der Walt \& J.~Millman, 51 -- 56

\bibitem[{{Miglio} {et~al.}(2013){Miglio}, {Chiappini}, {Morel}, {Barbieri},
  {Chaplin}, {Girardi}, {Montalb{\'a}n}, {Valentini}, {Mosser}, {Baudin},
  {Casagrande}, {Fossati}, {Silva Aguirre}, \& {Baglin}}]{miglio+2013}
{Miglio}, A., {Chiappini}, C., {Morel}, T., {et~al.} 2013, \mnras, 429, 423

\bibitem[{{Mosser} \& {Appourchaux}(2009)}]{mosser&appourchaux2009}
{Mosser}, B., \& {Appourchaux}, T. 2009, \aap, 508, 877

\bibitem[{{Mosser} {et~al.}(2019){Mosser}, {Michel}, {Samadi}, {Miglio},
  {Davies}, {Girardi}, \& {Goupil}}]{mosser+2019}
{Mosser}, B., {Michel}, E., {Samadi}, R., {et~al.} 2019, \aap, 622, A76

\bibitem[{{Mosser} {et~al.}(2010){Mosser}, {Belkacem}, {Goupil}, {Miglio},
  {Morel}, {Barban}, {Baudin}, {Hekker}, {Samadi}, {De Ridder}, {Weiss},
  {Auvergne}, \& {Baglin}}]{mosser+2010}
{Mosser}, B., {Belkacem}, K., {Goupil}, M.~J., {et~al.} 2010, \aap, 517, A22

\bibitem[{P{\'e}rez \& Granger(2007)}]{ipython}
P{\'e}rez, F., \& Granger, B.~E. 2007, Computing in Science \& Engineering, 9,
  21.
\newblock \url{https://aip.scitation.org/doi/abs/10.1109/MCSE.2007.53}

\bibitem[{{Pinsonneault} {et~al.}(2018){Pinsonneault}, {Elsworth}, {Tayar},
  {Serenelli}, {Stello}, {Zinn}, {Mathur}, {Garc{\'\i}a}, {Johnson}, {Hekker},
  {Huber}, {Kallinger}, {M{\'e}sz{\'a}ros}, {Mosser}, {Stassun}, {Girardi},
  {Rodrigues}, {Silva Aguirre}, {An}, {Basu}, {Chaplin}, {Corsaro}, {Cunha},
  {Garc{\'\i}a-Hern{\'a}ndez}, {Holtzman}, {J{\"o}nsson}, {Shetrone}, {Smith},
  {Sobeck}, {Stringfellow}, {Zamora}, {Beers}, {Fern{\'a}ndez-Trincado},
  {Frinchaboy}, {Hearty}, \& {Nitschelm}}]{pinsonneault+2018}
{Pinsonneault}, M.~H., {Elsworth}, Y.~P., {Tayar}, J., {et~al.} 2018, The
  Astrophysical Journal Supplement Series, 239, 32

\bibitem[{{Rendle} {et~al.}(2019){Rendle}, {Miglio}, {Chiappini}, {Valentini},
  {Davies}, {Mosser}, {Elsworth}, {Garc{\'\i}a}, {Mathur}, {Jofr{\'e}},
  {Worley}, {Casagrande}, {Girardi}, {Lund}, {Feuillet}, {Gavel}, {Magrini},
  {Khan}, {Rodrigues}, {Johnson}, {Cunha}, {Lane}, {Nitschelm}, \&
  {Chaplin}}]{rendle+2019}
{Rendle}, B.~M., {Miglio}, A., {Chiappini}, C., {et~al.} 2019, \mnras, 490,
  4465

\bibitem[{{Salaris} {et~al.}(1993){Salaris}, {Chieffi}, \&
  {Straniero}}]{salaris+1993}
{Salaris}, M., {Chieffi}, A., \& {Straniero}, O. 1993, \apj, 414, 580

\bibitem[{{Sch{\"o}nrich} \& {Aumer}(2017)}]{schoenrich_aumer2017}
{Sch{\"o}nrich}, R., \& {Aumer}, M. 2017, \mnras, 472, 3979

\bibitem[{{Sch{\"o}nrich} {et~al.}(2019){Sch{\"o}nrich}, {McMillan}, \&
  {Eyer}}]{schoenrich+2019a}
{Sch{\"o}nrich}, R., {McMillan}, P., \& {Eyer}, L. 2019, \mnras, 487, 3568

\bibitem[{{Sharma} {et~al.}(2011){Sharma}, {Bland-Hawthorn}, {Johnston}, \&
  {Binney}}]{sharma+2011}
{Sharma}, S., {Bland-Hawthorn}, J., {Johnston}, K.~V., \& {Binney}, J. 2011,
  \apj, 730, 3

\bibitem[{{Sharma} \& {Stello}(2016)}]{asfgrid}
{Sharma}, S., \& {Stello}, D. 2016, {Asfgrid: Asteroseismic parameters for a
  star}, , , ascl:1603.009

\bibitem[{{Sharma} {et~al.}(2016){Sharma}, {Stello}, {Bland-Hawthorn}, {Huber
  }, \& {Bedding}}]{sharma+2016}
{Sharma}, S., {Stello}, D., {Bland-Hawthorn}, J., {Huber }, D., \& {Bedding},
  T.~R. 2016, \apj, 822, 15

\bibitem[{{Sharma} {et~al.}(2018){Sharma}, {Stello}, {Buder}, {Kos},
  {Bland-Hawthorn}, {Asplund}, {Duong}, {Lin}, {Lind}, {Ness}, {Huber},
  {Zwitter}, {Traven}, {Hon}, {Kafle}, {Khanna}, {Saddon}, {Anguiano}, {Casey},
  {Freeman}, {Martell}, {De Silva}, {Simpson}, {Wittenmyer}, \&
  {Zucker}}]{sharma+2018a}
{Sharma}, S., {Stello}, D., {Buder}, S., {et~al.} 2018, \mnras, 473, 2004

\bibitem[{{Sharma} {et~al.}(2019){Sharma}, {Stello}, {Bland-Hawthorn},
  {Hayden}, {Zinn}, {Kallinger}, {Hon}, {Asplund}, {Buder}, {De Silva},
  {D'Orazi}, {Freeman}, {Kos}, {Lewis}, {Lin}, {Lind}, {Martell}, {Simpson},
  {Wittenmyer}, {Zucker}, {Zwitter}, {Bedding}, {Chen}, {Cotar}, {Esdaile},
  {Horner}, {Huber}, {Kafle}, {Khanna}, {Li}, {Ting}, {Nataf}, {Nordlander},
  {Saadon}, {Traven}, {Wright}, \& {Wyse}}]{sharma+2019}
{Sharma}, S., {Stello}, D., {Bland-Hawthorn}, J., {et~al.} 2019, \mnras, 490,
  5335

\bibitem[{{Silva Aguirre} {et~al.}(2012){Silva Aguirre}, {Casagrande}, {Basu},
  {Campante}, {Chaplin}, {Huber}, {Miglio}, {Serenelli}, {Ballot}, {Bedding},
  {Christensen-Dalsgaard}, {Creevey}, {Elsworth}, {Garc{\'{\i}}a}, {Gilliland},
  {Hekker}, {Kjeldsen}, {Mathur}, {Metcalfe}, {Monteiro}, {Mosser},
  {Pinsonneault}, {Stello}, {Weiss}, {Tenenbaum}, {Twicken}, \&
  {Uddin}}]{silva_aguirre+2012}
{Silva Aguirre}, V., {Casagrande}, L., {Basu}, S., {et~al.} 2012, \apj, 757, 99

\bibitem[{{Silva Aguirre} {et~al.}(2018){Silva Aguirre}, {Bojsen-Hansen},
  {Slumstrup}, {Casagrande}, {Kawata}, {Ciuc{\v a}}, {Handberg}, {Lund},
  {Mosumgaard}, {Huber}, {Johnson}, {Pinsonneault}, {Serenelli}, {Stello},
  {Tayar}, {Bird}, {Cassisi}, {Hon}, {Martig}, {Nissen}, {Rix},
  {Sch{\"o}nrich}, {Sahlholdt}, {Trick}, \& {Yu}}]{silva-aguirre+2018a}
{Silva Aguirre}, V., {Bojsen-Hansen}, M., {Slumstrup}, D., {et~al.} 2018,
  \mnras, 475, 5487

\bibitem[{{Skrutskie} {et~al.}(2006){Skrutskie}, {Cutri}, {Stiening},
  {Weinberg}, {Schneider}, {Carpenter}, {Beichman}, {Capps}, {Chester},
  {Elias}, {Huchra}, {Liebert}, {Lonsdale}, {Monet}, {Price}, {Seitzer},
  {Jarrett}, {Kirkpatrick}, {Gizis}, {Howard}, {Evans}, {Fowler}, {Fullmer},
  {Hurt}, {Light}, {Kopan}, {Marsh}, {McCallon}, {Tam}, {Van Dyk}, \&
  {Wheelock}}]{skrutskie+2006}
{Skrutskie}, M.~F., {Cutri}, R.~M., {Stiening}, R., {et~al.} 2006, \aj, 131,
  1163

\bibitem[{{Spitoni} {et~al.}(2020){Spitoni}, {Verma}, {Silva Aguirre}, \&
  {Calura}}]{spitoni+2020}
{Spitoni}, E., {Verma}, K., {Silva Aguirre}, V., \& {Calura}, F. 2020, \aap,
  635, A58

\bibitem[{{Stello} {et~al.}(2015){Stello}, {Huber}, {Sharma}, {Johnson},
  {Lund}, {Handberg}, {Buzasi}, {Silva Aguirre}, {Chaplin}, {Miglio},
  {Pinsonneault}, {Basu}, {Bedding}, {Bland-Hawthorn}, {Casagrande}, {Davies},
  {Elsworth}, {Garcia}, {Mathur}, {Di Mauro}, {Mosser}, {Schneider},
  {Serenelli}, \& {Valentini}}]{stello+2015}
{Stello}, D., {Huber}, D., {Sharma}, S., {et~al.} 2015, \apjl, 809, L3

\bibitem[{{Stello} {et~al.}(2017){Stello}, {Zinn}, {Elsworth}, {Garcia},
  {Kallinger}, {Mathur}, {Mosser}, {Sharma}, {Chaplin}, {Davies}, {Huber},
  {Jones}, {Miglio}, \& {Silva Aguirre}}]{stello+2017}
{Stello}, D., {Zinn}, J., {Elsworth}, Y., {et~al.} 2017, \apj, 835, 83

\bibitem[{{Ulrich}(1986)}]{ulrich1986}
{Ulrich}, R.~K. 1986, \apjl, 306, L37

\bibitem[{{Vanderburg} \& {Johnson}(2014)}]{vanderburg&johnson2014}
{Vanderburg}, A., \& {Johnson}, J.~A. 2014, \pasp, 126, 948

\bibitem[{{Virtanen} {et~al.}(2020){Virtanen}, {Gommers}, {Oliphant},
  {Haberland}, {Reddy}, {Cournapeau}, {Burovski}, {Peterson}, {Weckesser},
  {Bright}, {van der Walt}, {Brett}, {Wilson}, {Jarrod Millman}, {Mayorov},
  {Nelson}, {Jones}, {Kern}, {Larson}, {Carey}, {Polat}, {Feng}, {Moore}, {Vand
  erPlas}, {Laxalde}, {Perktold}, {Cimrman}, {Henriksen}, {Quintero}, {Harris},
  {Archibald}, {Ribeiro}, {Pedregosa}, {van Mulbregt}, \&
  {Contributors}}]{scipy}
{Virtanen}, P., {Gommers}, R., {Oliphant}, T.~E., {et~al.} 2020, Nature
  Methods, 17, 261

\bibitem[{Walt {et~al.}(2011)Walt, Colbert, \& Varoquaux}]{numpy}
Walt, S. v.~d., Colbert, S.~C., \& Varoquaux, G. 2011, Computing in Science \&
  Engineering, 13, 22.
\newblock \url{https://aip.scitation.org/doi/abs/10.1109/MCSE.2011.37}

\bibitem[{{Yu} {et~al.}(2018){Yu}, {Huber}, {Bedding}, {Stello}, {Hon},
  {Murphy}, \& {Khanna}}]{yu+2018}
{Yu}, J., {Huber}, D., {Bedding}, T.~R., {et~al.} 2018, \apjs, 236, 42

\bibitem[{{Zinn} {et~al.}(2017){Zinn}, {Huber}, {Pinsonneault}, \&
  {Stello}}]{zinn+2017b}
{Zinn}, J.~C., {Huber}, D., {Pinsonneault}, M.~H., \& {Stello}, D. 2017, \apj,
  844, 166

\bibitem[{{Zinn} {et~al.}(2019{\natexlab{a}}){Zinn}, {Pinsonneault}, {Huber},
  \& {Stello}}]{zinn+2019}
{Zinn}, J.~C., {Pinsonneault}, M.~H., {Huber}, D., \& {Stello}, D.
  2019{\natexlab{a}}, The Astrophysical Journal, 878, 136

\bibitem[{{Zinn} {et~al.}(2019{\natexlab{b}}){Zinn}, {Pinsonneault}, {Huber},
  {Stello}, {Stassun}, \& {Serenelli}}]{zinn+2019a}
{Zinn}, J.~C., {Pinsonneault}, M.~H., {Huber}, D., {et~al.} 2019{\natexlab{b}},
  \apj, 885, 166

\bibitem[{{Zinn} {et~al.}(2019{\natexlab{c}}){Zinn}, {Stello}, {Huber}, \&
  {Sharma}}]{zinn+2019bam}
{Zinn}, J.~C., {Stello}, D., {Huber}, D., \& {Sharma}, S. 2019{\natexlab{c}},
  \apj, 884, 107

\end{thebibliography}
%\begin{thebibliography}
%\end{thebibliography}
\end{document}